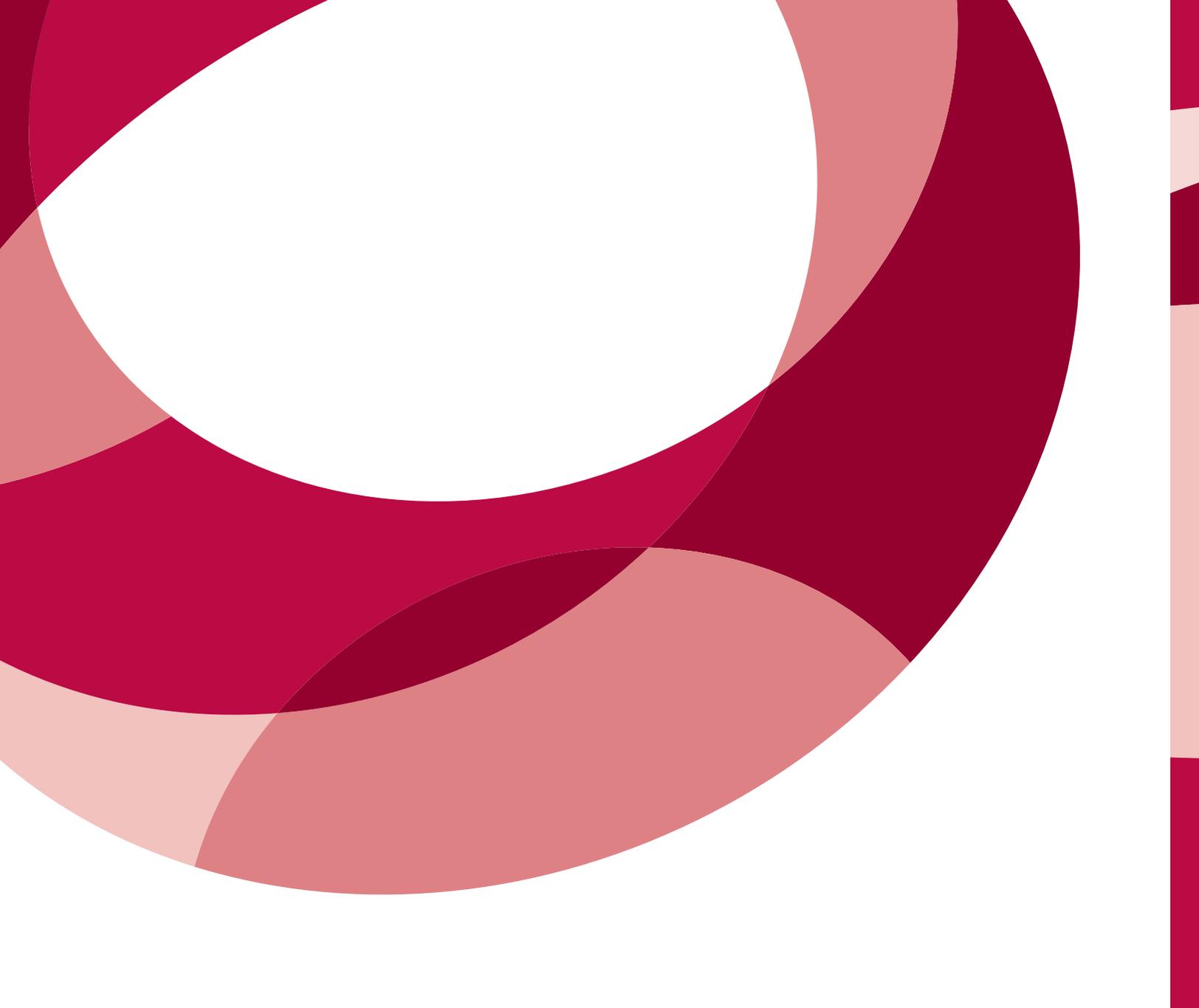

# A 20-Year Community Roadmap for Artificial Intelligence Research in the US

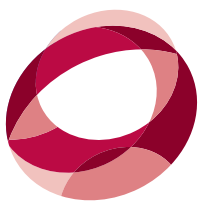
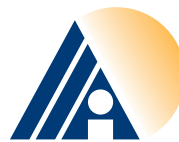

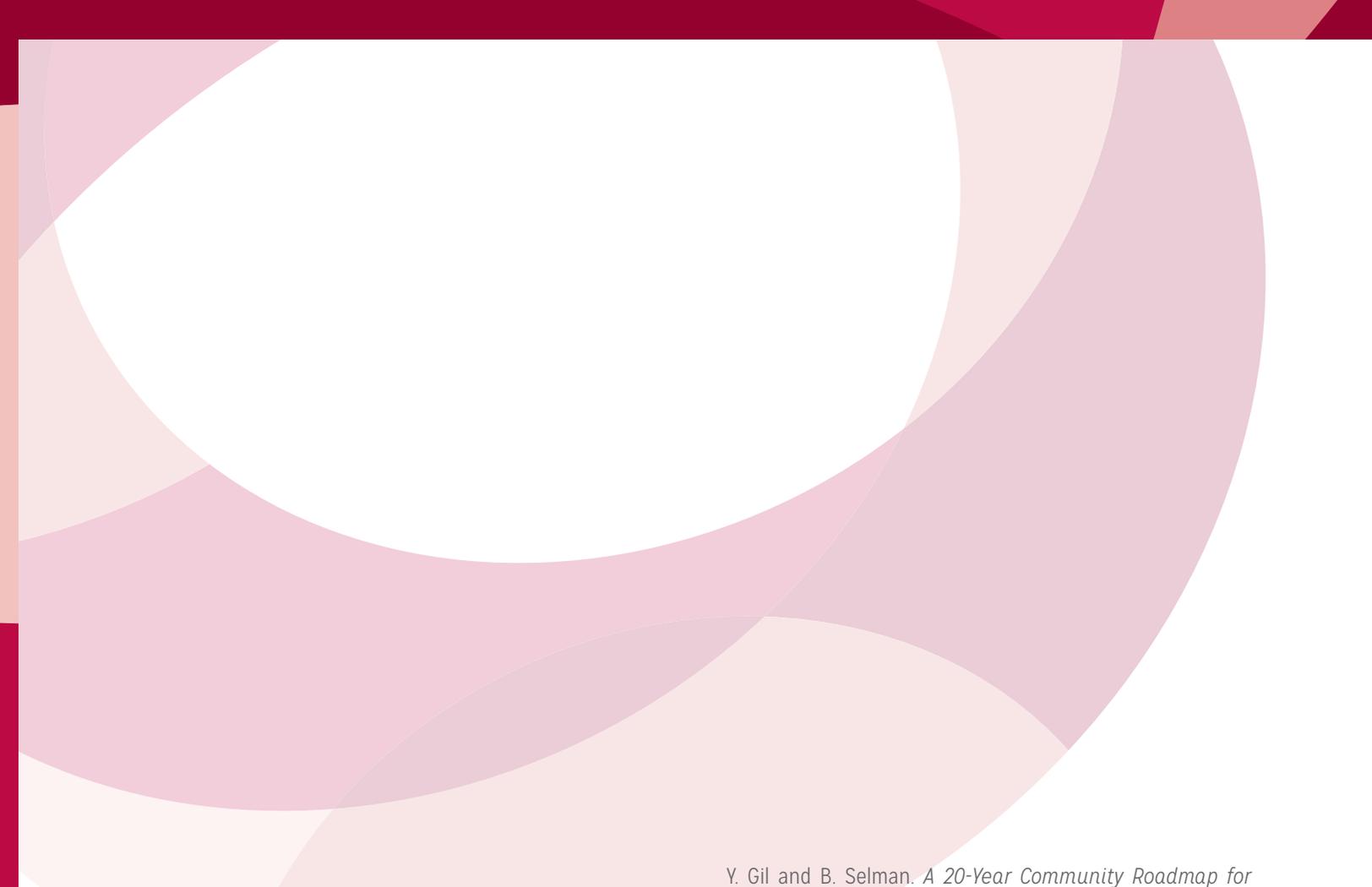

This material is based upon work supported by the National Science Foundation under Grants No. 1136993 and No. 1734706. Any opinions, findings, and conclusions or recommendations expressed in this material are those of the authors and do not necessarily reflect the views of the National Science Foundation.

# A 20-Year Community Roadmap for Artificial Intelligence Research in the US

**Roadmap Co-chairs:**

Yolanda Gil, AAAI President / University of Southern California
Bart Selman, AAAI President Elect / Cornell University

**Workshop Chairs:**

Tom Dietterich, Oregon State University
Ken Forbus, Northwestern University
Marie desJardins, Simmons University
Fei Fei Li, Stanford University
Kathy McKeown, Columbia University
Dan Weld, University of Washington

**Steering Committee:**

Elizabeth Bradley, CCC Vice Chair / University of Colorado Boulder
Ann Schwartz Drobnis, CCC Director
Mark D. Hill, CCC Chair / University of Wisconsin-Madison
Daniel Lopresti, Lehigh University
Maja Mataric, University of Southern California
David Parkes, Harvard University

August 2019

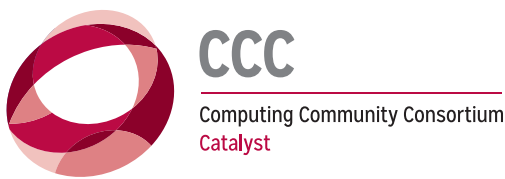
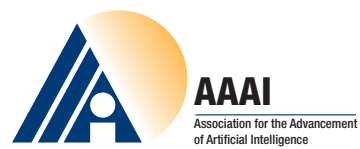

## Executive Summary

Decades of research in artificial intelligence (AI) have produced formidable technologies that are providing immense benefit to industry, government, and society. AI systems can now translate across multiple languages, identify objects in images and video, streamline manufacturing processes, and control cars. The deployment of AI systems has not only created a trillion-dollar industry that is projected to quadruple in three years, but has also exposed the need to make AI systems fair, explainable, trustworthy, and secure. Future AI systems will rightfully be expected to reason effectively about the world in which they (and people) operate, handling complex tasks and responsibilities effectively and ethically, engaging in meaningful communication, and improving their awareness through experience.

***Achieving the full potential of AI technologies poses research challenges that require a radical transformation of the AI research enterprise, facilitated by significant and sustained investment.*** These are the major recommendations of a recent community effort coordinated by the Computing Community Consortium and the Association for the Advancement of Artificial Intelligence to formulate a Roadmap for AI research and development over the next two decades.

## Societal Benefits of AI

AI systems have the potential for transformative impact across all sectors of society and for substantial innovation and economic growth. At the same time, there are many concerns about the security and vulnerability of systems with these capabilities, as well as about the future of work in such a world. The Roadmap articulates AI benefits in several specific areas: 1) boost health and quality of life, 2) provide lifelong education and training, 3) reinvent business innovation and competitiveness, 4) accelerate scientific discovery and technical innovation, 5) expand evidence-driven social opportunity and policy, and 6) transform national defense and security. The Roadmap includes detailed vignettes describing how AI innovations could impact individual lives, organizations, and society. Potential AI innovations include health monitors and advisors, mental and behavioral health coaches, enhanced education for remote students, effective natural disaster response, rapid materials discovery, accurate models of water resources, expeditious cross-disciplinary medical research, business innovation in personal devices, supply chain delay resolution, and resilient cyber-physical systems. All these innovations require a range of fundamental research advances in key areas of AI.

## Research Priorities to Realize Societal Benefits

Major research priorities that arise from motivating societal drivers include:

- **Integrated intelligence**, including developing foundational principles for combining modular AI capabilities and skills, approaches for contextualizing general capabilities to suit specific uses, creation of open shared repositories of machine-understandable world knowledge, and understanding human intelligence both to inspire novel AI approaches and to develop models of human cognition.

- **Meaningful interaction**, comprising techniques for productive collaboration in mixed teams of humans and machines, combining diverse communication modalities (verbal, visual, emotional) while respecting privacy, responsible and trustworthy behaviors that can be corrected directly by users, and fruitful online and real-world interaction among humans and AI systems.

- **Self-aware learning**, developing robust and trustworthy learning, quantifying uncertainty and durability, learning from small amounts of data and through instruction, incorporating prior knowledge into learning, developing causal and steerable models from numerical data and observations, and learning real-time behaviors for intentional sensing and acting.

## Challenges of the Current AI Landscape

Through the Roadmap activities, several critical challenges were identified. To begin, the field has matured beyond its initial academic focus on algorithms and theories and into a context of continuous data collection, social and interactive experimentation, and massive amounts of knowledge about a constantly changing world. Building from those foundations, the tech industry has compiled and leveraged massive resources—datasets, knowledge graphs, special-purpose computers, and large cadres of AI engineers—to propel powerful innovations. Tackling the research priorities above will require appropriate resources that can drive basic research of a more experimental nature. Without the right resources, academic AI research is limited—without answers to foundational questions, AI applications in industry will bring limited innovations. The constraints, incentives, and timelines are very different, too: Industry is largely driven by practical, near-term solutions, while academia is where many of the fundamental long-term questions are asked. Moreover, AI challenges span all areas of computer science and computer engineering, as well as cognitive science, psychology, biology, mathematics, public policy, ethics, education, and communication, to name just a few. The talent pool is another critical issue in the current AI ecosystem: The need for AI expertise far exceeds the supply and the gap will only continue to grow if not addressed. Many AI faculty have moved to industry to pursue new opportunities brought about by unique data and massive resources. US PhD graduates find attractive opportunities abroad as Asia and Europe are making multi-billion dollar investments in this area. Finally, there are many concerns about the security and vulnerability of AI systems, ensuring ethical uses of AI, and the future of work.

## Recommendations

*Surmounting these challenges will require a reinvention of the AI research enterprise to create a comprehensive national AI infrastructure and re-conceptualize AI workforce training.* To that end, the Roadmap offers the following specific recommendations:

**I – Create and Operate a National AI Infrastructure** to serve academia, industry, and government through four interlocking capabilities:

- **Open AI platforms and resources:** a vast interlinked distributed collection of "AI-ready" resources (such as curated high-quality datasets, software, knowledge repositories, testbeds for personal assistants and robotics environments) contributed by and available to the academic research community, as well as to industry and government.

- **Sustained community-driven AI challenges:** organized sequences of challenges that build on one another, posed by AI and domain experts to drive research in key areas, building upon—and adding to—the shared resources in the Open AI Platforms and Facilities.

- **National AI Research Centers:** multi-university centers with affiliated institutions, focused on pivotal areas of long-term AI research (e.g., integrated intelligence, trust, and responsibility), with decade-long funding to support on the order of 100 faculty, 200 AI engineers, 500 students, and necessary computing infrastructure. These centers would offer rich training for students at all levels. Visiting fellows from academia, industry, and government will enable cross-cutting research and technology transition.

- **Mission-Driven AI Laboratories:** living laboratories for AI development in targeted areas of great potential for societal impact. These would be "AI-ready" facilities, designed to allow AI researchers to access unique data and expertise, such as AI-ready hospitals, AI-ready homes, or AI-ready schools. They would work closely with the National AI Research Centers to provide requirements, facilitate applied research, and transition research results. These laboratories would be crucial for R&D, dissemination, and workforce training. They would have decade-long funding to support on the order of 50 permanent AI researchers, 50 visitors from AI Research Centers, 100-200 AI engineers and technicians, and 100 domain experts and staff.

**II — Re-conceptualize and Train an All-Encompassing AI Workforce,** building upon the National AI Infrastructure listed above to:

- **Develop AI Curricula at All Levels:** guidelines should be developed for curricula that encourage early and ongoing interest in and understanding of AI, beginning in K-12 and extending through graduate courses and professional programs.

- **Create Recruitment and Retention Programs for Advanced AI Degrees:** including grants for talented students to obtain advanced graduate degrees, retention programs for doctoral-level researchers, and additional resources to support and enfranchise AI teaching faculty.

- **Engage Underrepresented and Underprivileged Groups:** programs to bring the best talent into the AI research effort.

- **Incentivize Emerging Interdisciplinary AI Areas:** initiatives to encourage students and the research community to work in interdisciplinary AI studies—e.g., AI safety engineering, as well as analysis of the impact of AI on society—will ensure a workforce and a research ecosystem that understands the full context for AI solutions.

- **Highlight AI Ethics and Policy:** including the importance of the area of AI ethics and policy, and the imperative of incorporating ethics and related responsibility principles as central elements in the design and operation of AI systems.

- **Address AI and the Future of Work:** these challenges are at the intersection of AI with other disciplines such as economics, public policy, and education. It is important to teach students how to think through the ethical and social implications of their work.

- **Train Highly Skilled AI Engineers and Technicians:** support and build upon the National AI Infrastructure to grow the AI pipeline through community colleges, workforce retraining programs, certificate programs, and online degrees.

**III — Core Programs for basic AI Research are critical.** The new resources and initiatives described in this Roadmap cannot come at the expense of existing programs for funding AI research. These core programs—which provide well-established, broad-based support for research progress, for training young researchers, for integrating AI research and education, and for nucleating novel interdisciplinary collaborations—are critical complements to the broader initiatives described in this Roadmap, and they too will require expanded support.

All of this will require substantial, sustained federal investment over the course of the 20-year period covered by this Roadmap, but the outcomes will be transformative. The recommendations above are not only a scaffold for interdisciplinary, forward-looking R&D that will drive scientific and economic advances while taking into consideration issues around security, vulnerability, policy, and ethics. The recommendations in this Roadmap will also allow the retention of the best talent in fertile research settings, creating extensive human capital in this crucial technology area—another important benefit to society and the economy.

# A 20-Year Community Roadmap for Artificial Intelligence Research in the US

Yolanda Gil (USC) and Bart Selman (Cornell), co-chairs

Few of AI's challenges can be solved as piecemeal academic research projects

### ARTIFICIAL INTELLIGENCE (AI) LANDSCAPE
- Data-driven AI methods are highly effective but have important flaws
- Industry focuses largely on practical, near-term solutions using massive proprietary resources
- Academia asks many of the fundamental long-term questions that lay the foundations for AI

### ASPIRATIONS
- Reduced healthcare cost
- Universal personalized education
- Evidence-driven social opportunity
- Accelerated scientific discovery
- Unprecedented innovation for businesses
- National defense and security

**AI-driven capabilities**
- Behavioral health coaches
- High payoff experiments
- Opportunistic education
- Resolve supply chain delays
- At-home robot caregivers/helpers
- Effective natural disaster response
- Novel business processes
- Address food and water insecurity
- Resilient cyber-physical systems

### AI RESEARCH PRIORITIES

**Integrated Intelligence**
- Science of integrated intelligence
- Contextualized AI
- Open knowledge repositories
- Understanding human intelligence

**Meaningful Interaction**
- Collaboration
- Trust and responsibility
- Diversity of interaction channels
- Improving online interaction

**Self-Aware Learning**
- Robust and trustworthy learning
- Deeper learning for challenging tasks
- Integrating symbolic and numeric representations
- Learning in integrated AI/robotic systems

### CROSS-CUTTING ISSUES
- Security & vulnerability, ethics, resources (data, hardware, software, storage, people...)

## RECOMMENDATIONS FOR AI RESEARCH

**National AI Infrastructure**     **Core AI Programs**     **Workforce Training**

Interdisciplinary AI     National AI Research Centers     Open AI Platforms and Facilities     Mission-Driven AI Laboratories

Broadening AI curriculum     Community-driven challenges     Recruitment and training programs     Engaging underrepresented groups

# Table of Contents

# 1. Introduction

This Roadmap articulates the potential of artificial intelligence (AI) research advances to significantly improve our world and benefit our lives. It is the result of a community activity to formulate AI research priorities for the next 20 years, and make recommendations in support of their pursuit. This research agenda is motivated by a broad range of societal drivers aiming to improve health, education, science, innovation, justice, and security. The recommendations are supported by major findings resulting from the community discussions.

The rest of this chapter provides a brief background on AI and an overview of the community activities that led to this Roadmap document. Chapter 2 (page 12) gives an overview of major societal drivers, emphasizing the potential value of AI innovation in different sectors. Chapter 3 (page 16) details the outcomes of the three community workshops held on the core technical areas of the AI research Roadmap, including specific vignettes that describe AI systems for those societal drivers and articulates the research required that spans a wide range of areas of AI. The major findings and recommendations that resulted from reflecting on those efforts are detailed in Chapter 4 (page 81) and Chapter 5 (page 84).

## 1.1 What Is Artificial Intelligence?

In this report, we adopt the definition from the One Hundred Year Study of AI[1] that views AI as "a branch of computer science that studies the properties of intelligence by synthesizing intelligence." The bolding is ours:

> Notably, the characterization of intelligence as a spectrum grants no special status to the human brain. But to date *human intelligence* has no match in the biological and artificial worlds for sheer versatility, *with the abilities "to reason, achieve goals, understand and generate language, perceive and respond to sensory inputs, prove mathematical theorems, play challenging games, synthesize and summarize information, create art and music, and even write histories."*[2]

> This makes human intelligence a natural choice for benchmarking the progress of AI. It may even be proposed, as a rule of thumb, that any activity computers are able to perform and people once performed should be counted as an instance of intelligence. *But matching any human ability is only a sufficient condition, not a necessary one. There are already many systems that exceed human intelligence, at least in speed,* such as scheduling the daily arrivals and departures of thousands of flights in an airport.

> [...]

> AI can also be defined by what AI researchers do. *This report views AI primarily as a branch of computer science that studies the properties of intelligence by synthesizing intelligence.*[3] Though the advent of AI has depended on the rapid progress of hardware computing resources, the focus here on software reflects a trend in the AI community. More recently, though, progress in building hardware tailored for neural-network-based computing[4] has created a tighter coupling between hardware and software in advancing AI.

---

[1] "One Hundred Year Study on Artificial Intelligence (AI100)," Stanford University, accessed August 1, 2016, https://ai100.stanford.edu.

[2] Nils J. Nilsson, The Quest for Artificial Intelligence. New York: Cambridge University Press, 2009.

[3] Herbert A. Simon, "Artificial Intelligence: An Empirical Science," Artificial Intelligence 77, no. 2 (1995):95–127.

[4] Paul Merolla, John V. Arthur, Rodrigo Alvarez-Icaza, Andrew S. Cassidy, Jun Sawada, Filipp Akopyan, Bryan L. Jackson, Nabil Imam, Chen Guo, Yutaka Nakamura, Bernard Brezzo, Ivan Vo, Steven K. Esser, Rathinakumar Appuswamy, Brian Taba, Arnon Amir, Myron D. Flickner, William P. Risk, Rajit Manohar, and Dharmendra S. Modha, "A Million Spiking-Neuron Integrated Circuit with a Scalable Communication Network and Interface," Science, vol. 345, no. 6197 (August 8, 2014): 668-73. https://science.sciencemag.org/content/345/6197

*"Intelligence" remains a complex phenomenon whose varied aspects have attracted the attention of several different fields of study, including psychology, economics, neuroscience, biology, engineering, statistics, and linguistics. Naturally, the field of AI has benefited from the progress made by all of these allied fields.* For example, the artificial neural network, which has been at the heart of several AI-based solutions[5,6] was originally inspired by thoughts about the flow of information in biological neurons.[7]

AI research is sometimes focused on mimicking human intelligence. For example, one approach to learn from data is to design artificial neural networks that are composed of neuron-like units that adjust how they fire based on stimuli (the data). In many cases the approaches developed are engineered for performance and do not have any resemblance to human intelligence. For example, chess playing programs consider many more possible moves than humans are able to do.

Although AI has been a field of study for several decades, the term "AI" has become a colloquial term that is used very loosely.

Sometimes "AI" is equated with machine learning, and specifically with learning from large amounts of data in order to make predictions. Although learning from data is an important area of research in AI, there are many other learning techniques that are studied in AI, including learning from tutorial instruction, learning from demonstrations, learning from a single example, learning to work in teams, etc. And although learning is ubiquitous in intelligent behaviors, machine learning is one of many areas of study in AI, such as understanding and generating language, acting based on sensor inputs, representing knowledge in machine-readable structures, coordinating with others to work in teams, and reasoning about complex problems to generate solutions.

Sometimes "AI" is used to describe robots and other computer systems that have a physical dimension such as intelligent sensors. These AI systems can operate independently or be governed by humans. Many AI systems lack a physical dimension, such as conversational interfaces and intelligent search or recommendation systems.

Sometimes "AI" is used to describe very simple programs that give the illusion of intelligence, such as "game AIs" that govern synthetic players in computer games. In these cases, the meaning is simply "a computer program that generates behaviors that are more interesting than traditional ones."  A consequence of this is that AI is a moving target: what is considered AI today may not be considered AI tomorrow as more sophisticated programs are created.

"AI" is often used to refer to systems that exhibit any intelligent capability or aspect of intelligence. For example, any computer system that is able to learn from some data, or solve a problem, or generate language, is often called "the AI."

This document uses the term "AI system" to refer to a computer system that exhibits some intelligent behavior, and the surrounding text will be clear on what the intelligent behaviors are. It is important to note that just because a system includes some aspect of intelligence, it does not mean that it has general intelligent capabilities. In fact, there are no AI systems today that have general intelligent capabilities. This document lays out a Roadmap for AI research in the next 20 years, and even at that point AI systems will be far from having general intelligent capabilities.

## 1.2 AI Roadmap Process

As part of its continuing efforts to advance computing research, the Computing Community Consortium (CCC) reached out to the leadership of the Association for the Advancement of Artificial Intelligence (AAAI) to spearhead a 20-year Roadmap for AI.

The mission of the Computing Research Association's CCC is to catalyze the computing research community and enable the pursuit of innovative, high-impact research. CCC conducts activities that strengthen the research community, articulate compelling research visions, and align those visions with pressing national and global challenges. CCC communicates the importance of those visions to policymakers, government and industry stakeholders, the public, and the research community itself.

AAAI is a nonprofit scientific society devoted to advancing the scientific understanding of the mechanisms underlying thought and intelligent behavior and their embodiment in machines. AAAI also aims to increase public understanding of artificial intelligence, improve the teaching and training of AI practitioners, and provide guidance for research planners and funders concerning the importance and potential of current AI developments and future directions. AAAI has more than 200 elected fellows, and over 4,000 members.

Yolanda Gil (AAAI President; University of Southern California) and Bart Selman (AAAI President Elect; Cornell University) led the effort and designed a series of three community workshops:

◗ **Integrated Intelligence**, November 14-15, 2018, co-led by Marie desJardins (Simmons University) and Ken Forbus (Northwestern University).

◗ **Meaningful Interactions**, January 8-9, 2019, co-led by Kathy McKeown (Columbia University) and Dan Weld (University of Washington)

◗ **Self-Aware Learning**, January 17-18, 2019, co-led by Tom Dietterich (Oregon State University) and Fei-Fei Li (Stanford University)

Altogether, 91 researchers from 41 academic institutions and nine industrial partner organizations took part in this effort. Detailed participant lists appear in the Appendix.

Workshop participants discussed the potential benefits of AI for several societal drivers: reducing healthcare cost, accelerating scientific discovery, facilitating universal personalized education, innovation for business, and evidence-driven social opportunity. The workshop agendas combined full-group discussions and breakout sessions, some centered on the societal drivers and some centered on the technical challenges in the topical area. Participants discussed research priority areas and formulated 5-year, 10-year, and 15-year milestones toward 20-year stretch goals. Workshop participants also reflected on related topics such as the historical evolution of the field of AI, the multidisciplinarity of the AI research agenda, AI research infrastructure, the rapid growth of AI in industry labs, the interplay between academic and industry research, AI funding programs and agencies, and the public understanding and perception of AI technologies. Following the workshops, individual reports were drafted based on participant discussions at each event, then edited by the co-leaders and circulated to the workshop participants for comments. Findings and recommendations brought up in discussions throughout the workshops were also drafted.

A preliminary version of the Roadmap was presented to the community in a town hall meeting at the AAAI 2019 Conference on January 28, 2019, which was attended by more than 500 participants. A draft of the Executive Summary was posted on the CCC website on March 13, 2019, for preliminary comment. Numerous suggestions were collected during those months and incorporated into the Roadmap. The findings and recommendations were further refined.

A full draft of the report, including the findings, recommendations, individual workshop reports, and supporting materials, was distributed on May 13, 2019, for community comment.

## 1.3 Why Now?

Decades of research in AI have produced formidable technologies that are providing immense benefit to industry, government, and society. The bulk of AI research in industry at present is driven by advertising and product recommendation systems. The flood of recent developments in this area—catalyzed by the availability of massive datasets, vast knowledge graphs, powerful special-purpose computers, and large cadres of AI engineers—have created a trillion-dollar industry that is projected to quadruple in three years. While AI solutions have the potential for transformative impacts across all sectors of society and the economy, there are concerns about the security and vulnerability of these systems. As the resources that are driving the AI revolution continue to grow, the development and deployment of these technologies is poised not only to continue, but to accelerate. For all these reasons, the time is now to undertake a thoughtful and comprehensive envisioning of interdisciplinary, forward-looking R&D that will drive scientific and economic advances in AI while taking into consideration issues around security, vulnerability, policy, and ethics.

## 1.4 How to Read This Report

While the entire report is important for understanding and carrying out the goals of the Roadmap, we realize that it is long.  All of the sections in this report are written to be read independently, so readers can jump around and read individual sections of their choosing.  On your first read, it is suggested that you begin with the executive summary followed by the introduction, major findings, recommendations, and then conclusions.  On your next read, please go back and read the major societal drivers and core technical areas.

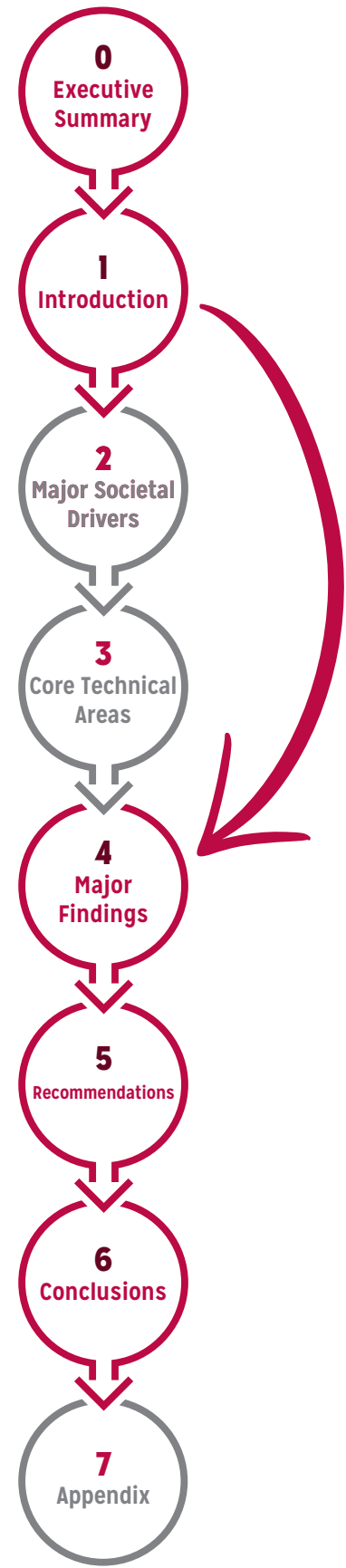

*Figure 1. How to read the report*

## 1.5 Reference Documents

We refer to the following documents for additional information about AI and its current context:

▸ **The state of the art in AI:** The "One Hundred Year Study on Artificial Intelligence"[8] provides a thorough review of the state of the art in AI research and applications, with contributions from researchers across all areas of AI and related fields. The report reviews the state of the art in a diversity of areas of AI research and their relevance to a range of applications with societal benefits of AI. The report provides citations and references to the published literature. This Roadmap document builds on and extends this study by providing a more in-depth motivation and description of areas for future research.

▸ **US government programs and priorities in AI:** The "National Artificial Intelligence R&D Strategic Plan"[9] was developed by the Networking and Information Technology Research and Development (NITRD) Program through its Task Force on Artificial Intelligence. The report reviews existing programs that fund AI research, identifies strategic priorities, and makes recommendations to guide federally funded AI research. This Roadmap document is consistent with the strategies proposed in the NITRD plan, and extends it with more specific recommendations and more elaborate descriptions of research priorities.

▸ **Related research Roadmaps:** "A Roadmap for US Robotics"[10] gives a thorough treatment of research priorities and applications in diverse areas of robotics and cyberphysical systems. This document is consistent with that report, and gives more extensive details on the cognitive capabilities needed in intelligent robots as well as expanding on many aspects of AI systems that inhabit the digital rather than the physical realm.

## 2. Major Societal Drivers for Future Artificial Intelligence Research

AI has permeated our lives and has become an engine for innovation across all sectors of society. Government investments can have a profound impact and transform society for the betterment of all citizens. Below we describe six societal areas that will be transformed by AI. These are meant to be examples of AI's potential impact to society, but in no way are they an exhaustive list. Detailed discussions for each of the societal drivers are included in the workshop reports through a series of vignettes that describe how a person or organization could interact with AI systems or would benefit from AI technologies, and discuss the capabilities that would be required and that motivate the AI research in this Roadmap. By looking at and working on the research challenges grounded in AI for social good, we will not only be changing society, but we can motivate K-12 students to study AI by presenting exciting problems to be worked on. These same societally relevant problems will help to engage women and other underrepresented populations in AI research, creating a more diverse workforce to better tackle the problems. In highlighting AI for social good, we will dispel misconceptions about AI while solving real-world, tangible problems.

### 2.1 Enhancing Healthcare and Quality of Life

Although the potential benefits for AI in healthcare have been demonstrated, this technology is largely untapped in clinical settings. AI applications can now diagnose some skin cancers more accurately than a board-certified dermatologist, and do so faster and more efficiently.¹¹ AI techniques already developed can examine high-risk spots on mammograms and provide advice on whether a biopsy is necessary, potentially reducing the number of biopsies by 30 percent.¹² In the near term, chronic health conditions like diabetes, cancer, and heart and neurological diseases are likely to benefit most from new applications of AI, according to a survey of healthcare professionals.¹³ In addition to AI's ability to aid in the medical diagnosis process, the deployment of AI applications in hospital workflows could further enhance care delivery. An analysis by Frost & Sullivan suggests such cross-cutting uses of AI "have the potential to improve patient outcomes by 30 to 40 percent, while cutting treatment costs by as much as 50 percent."[14] The impacts of AI enabled technologies over the longer term are likely to be even more profound.

---

[8] "One Hundred Year Study on Artificial Intelligence (AI100)," Stanford University, accessed August 1, 2016, https://ai100.stanford.edu.

[9] "The National Artificial Intelligence Research and Development Strategic Plan," National Science and Technology Council, October 2016, https://www.nitrd.gov/PUBS/national_ai_rd_strategic_plan.pdf.

[10] "A Roadmap for US Robotics: From Internet to Robotics, 2016 Edition," October 31, 2016, https://cra.org/ccc/wp-content/uploads/sites/2/2016/11/Roadmap3-final-rs-1.pdf

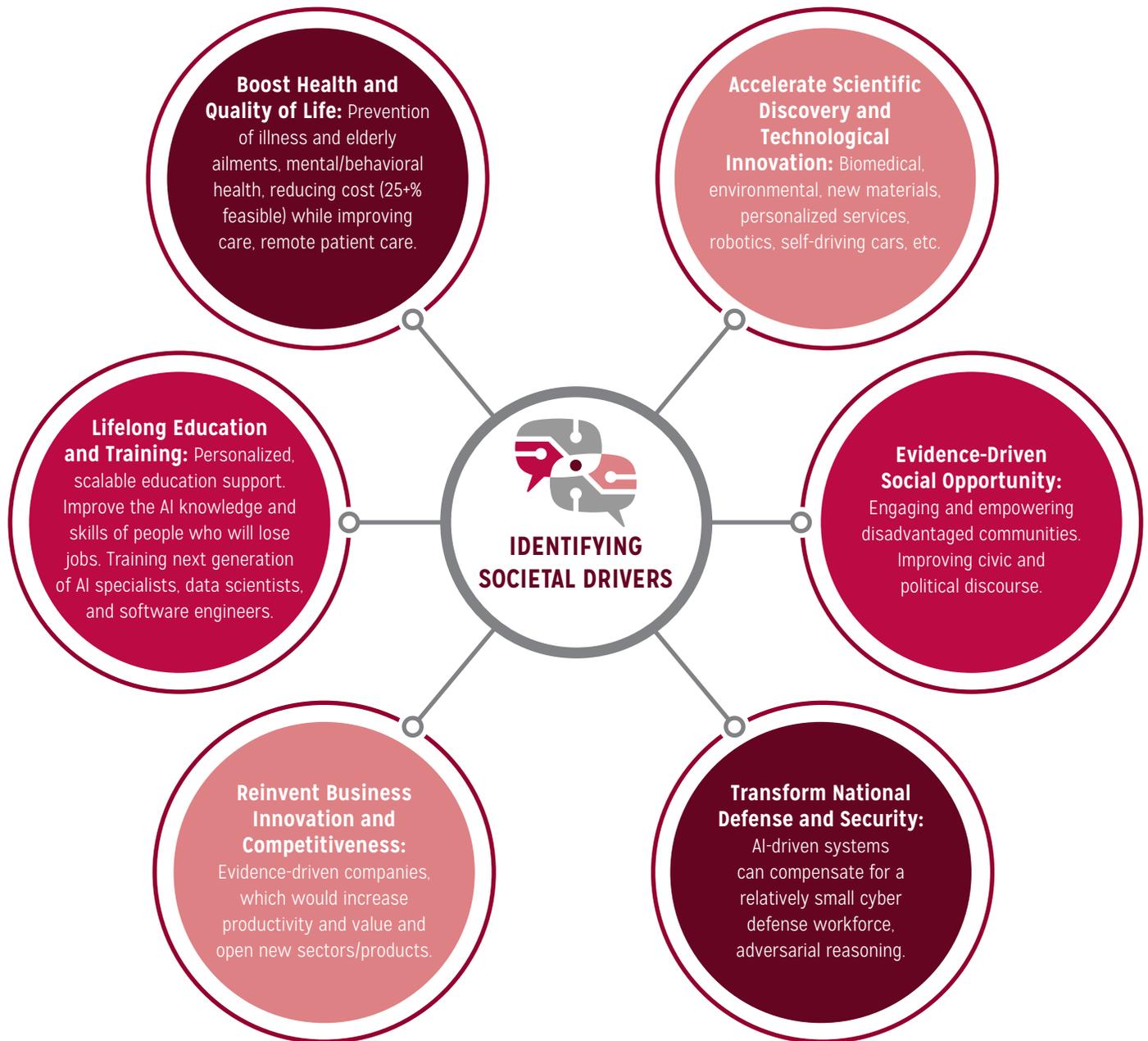

*Figure 2. Identifying Societal Drivers.*

---

11   Michael J. Rigby, "Ethical Dimensions of Using Artificial Intelligence in Health Care," AMA Journal of Ethics, vol. 21, no. 2 (Feb. 2019): 121.

12   Richard Harris, reporter, "Training A Computer To Read Mammograms As Well As A Doctor," National Public Radio, April 1, 2019. https://www.npr.org/sections/health-shots/2019/04/01/707675965/training-a-computer-to-read-mammograms-as-well-as-a-doctor and https://pubs.rsna.org/doi/10.1148/radiol.2017170549

13   Tom Sullivan, "3 charts show where artificial intelligence is making an impact in healthcare right now," Healthcare IT News, Dec. 21, 2018. https://www.healthcareitnews.com/news/3-charts-show-where-artificial-intelligence-making-impact-healthcare-right-now

## 2.2 Lifelong Education and Training

The US education system currently makes use of computing technologies, some of which are enhanced by AI, throughout the education environment, but there is still great room for improvement of these technologies, and even greater opportunity for full adoption to enhance our education system. In K-12 classrooms, students use adaptive reading and math software from a variety of vendors to receive personalized curricula tailored to their own progress. Content providers use machine learning to determine what material works well in each area. Teachers make use of feedback and scoring systems to help grade assignments, guard against plagiarism, and assess student progress. Augmented reality and virtual reality are becoming useful teaching and training tools, and AI systems that understand gesture and voice recognition are delivering more effective simulations. Assistive technologies powered by AI technologies are helping special-needs students recognize voices and text, thereby easing their communication. Even well outside the classroom, technology is changing how education is delivered in the US Student transportation services are increasingly optimized using AI, as is staff scheduling and substitute management. Grounds and facilities management, including smart building-management software and intelligent security products, is aided by AI. The market for AI technologies in education and training in the US, including higher-education, K-12 education, and corporate training, was $400 million in 2017, and is forecast to grow more than 45 percent per year to about $5.4 billion by 2024.[15] This Roadmap envisions integrated interactive AI systems that will expand education and career opportunities across broader segments of our society and level the playing field for students facing socio-economic, health, and disability challenges in our nation's schools.

## 2.3 Reinvent Business Innovation and Competitiveness

AI is helping drive US business innovation and competitiveness. Indeed, it is arguable that US industries' embrace of the AI technologies birthed from fundamental work in our research labs and institutions, notably machine learning, has provided a key competitive advantage to US companies in maintaining a leadership role in the global economy. By some estimates, AI contributed $2 trillion to global GDP in 2018 and is expected to be $15 trillion by 2030.[16] Business leaders worldwide see AI technologies as increasingly crucial to their competitiveness. When surveyed, executives at companies deploying AI cite its capability for generating new revenue, helping retain existing customers and acquire new ones, and providing a competitive differentiation with other companies in their sector as their top reasons for using the technology.[17] However, because of the limitations of the current state of the art, these deployments tend to be somewhat narrowly focused, applied to particular problems and constrained by the inability of current AI systems to integrate intelligence from a wide range of sources and contexts. The Roadmap describes a broader set of capabilities for business innovation and competitiveness, from developing broad ranging recommender systems to integrating robotic and AI systems into standard practices.

---

[14] Frost & Sullivan (press release), From $600 M to $6 Billion, Artificial Intelligence Systems Poised for Dramatic Market Expansion in Healthcare, Jan. 5, 2016. https://ww2.frost.com/news/press-releases/600-m-6-billion-artificial-intelligence-systems-poised-dramatic-market-expansion-healthcare

[15] "Artificial Intelligence (AI) in Education Market," Global Market Insights, Jan. 2018. https://www.gminsights.com/industry-analysis/artificial-intelligence-ai-in-education-market

[16] Frank Holmes, "AI Will Add $15 Trillion to the World Economy by 2030," Forbes, Feb. 25, 2019. https://www.forbes.com/sites/greatspeculations/2019/02/25/ai-will-add-15-trillion-to-the-world-economy-by-2030/#795251a71852

[17] Warren Knight, "Is Artificial Intelligence the Future of Business?" Business2Community, Nov. 9, 2018. https://www.business2community.com/business-innovation/is-artificial-intelligence-the-future-of-business-02137459

## 2.4 Accelerate Scientific Discovery and Technological Innovation

Some areas of modern science are benefiting from the use of AI technologies. The oceans of experimental and observational data produced by today's scientific infrastructure—e.g., space telescopes, super-colliders, sequencing equipment, medical imaging—are simply too vast for humanity to wade through unaided. Machine learning algorithms and other AI technologies help us make sense of the chaos by noting anomalies and detecting patterns that would otherwise go unnoticed.[18] Yet we are still only working in narrow models. The Roadmap envisions new hybrid modeling approaches that could be facilitated via advanced AI learning research. Advancements in machine learning techniques will enable techniques to process multimodal, multi-scale data, handle heterogeneity in space and time, and accurately quantify uncertainty in the results. The continued improvement of AI systems will impact the pillars of science, from data collection to experiment to discovery.

## 2.5 Social Justice and Policy

There is no greater opportunity to enhance quality of life than by increasing the equity, effectiveness, and efficiency of public services provided to citizens. The opportunity for AI to have an impact in the social policy and justice space is tremendous. McKinsey has recently done an analysis of 160 AI social-impact use cases and determined that existing capabilities could contribute to tackling cases across all 17 of the United Nations sustainable-development goals.[19] One of the goals is to take immediate and effective measures to eradicate forced labor, end modern slavery and human trafficking, and secure the prohibition and elimination of the worst forms of child labor, including recruitment and use of child soldiers, and by 2025 end child labor in all forms.[20] Methods exist today to identify trafficking or forced labor, and methods exist to help people once they are removed from their engagement in trafficking situations, but across the globe, the translation from identification to removal is often fraught with missed opportunities. The Roadmap envisions a future where meaningful interactions between people and systems will enable a complete path from forced laborer to survivor.

## 2.6 Transform National Defense and Security

Artificial Intelligence is already an essential part of cyber security, providing commercial and military cyber defenders with comprehensive security monitoring, threat detection, and actionable insight. At the same time, AI provides adversaries with unprecedented ways to understand a target's vulnerabilities and vector attacks accordingly. Spearphishing attacks are one highly effective means for an adversary to gain access to a target machine, but they are also labor-intensive in that they require research on the subject to determine what content would be most likely to motivate a click. A 2016 experiment demonstrated that AI-based spearphishing attacks on Twitter could garner almost the same response rate as a human-led attack could (34 percent vs. 38 percent), while targeting almost eight times as many victims in the same time period.[21] Defending networks from an intruder using compromised legitimate network credentials, perhaps gleaned by spearphishing, requires deep learning to analyze the user's behavior over a series of actions.[22] Future approaches to cybersecurity and defense will benefit from the powerful capabilities provided by AI systems. This report envisions how future AI systems can aid in responses to other types of threats as well, including natural disasters.

---

[18] Dan Falk, "How Artificial Intelligence Is Changing Science," Quanta Magazine, March 11, 2019. https://www.quantamagazine.org/how-artificial-intelligence-is-changing-science-20190311/

[19] Michael Chu, Martin Harrysson, James Manyika, Roger Roberts, Rita Chung, Peter Nel, and Ashley van Heteren, "Applying Artificial Intelligence for Social Good." McKinsey Global Institute discussion paper, November 2018. https://www.mckinsey.com/featured-insights/artificial-intelligence/applying-artificial-intelligence-for-social-good

[20] Sustainable Development Solutions Network, "Target 8.7." Indicators and a Monitoring Framework for Sustainable Development Goals: Launching a Data Revolution for the SDGs. May 15, 2015 http://indicators.report/targets/8-7/

[21] Artificial Intelligence for Cyber Security, The Workshops of the Thirty-First AAAI Conference on Artificial Intelligence: Technical Reports WS-17-01 — WS-17-15. Palo Alto, CA: The AAAI Press, 2017. https://aaai.org/Library/Workshops/ws17-04.php

## 3. Overview of Core Technical Areas of AI Research Roadmap

As discussed in section 1.2 the CCC, with the support of AAAI, held three community workshops in order to catalyze discussion and generate this research Roadmap. The three workshops were:

◗ **Integrated Intelligence**
◗ **Meaningful Interactions**
◗ **Self-Aware Learning**

The research priorities from the three areas are summarized in Figure 3 below and expanded upon in the following three sections of the report.

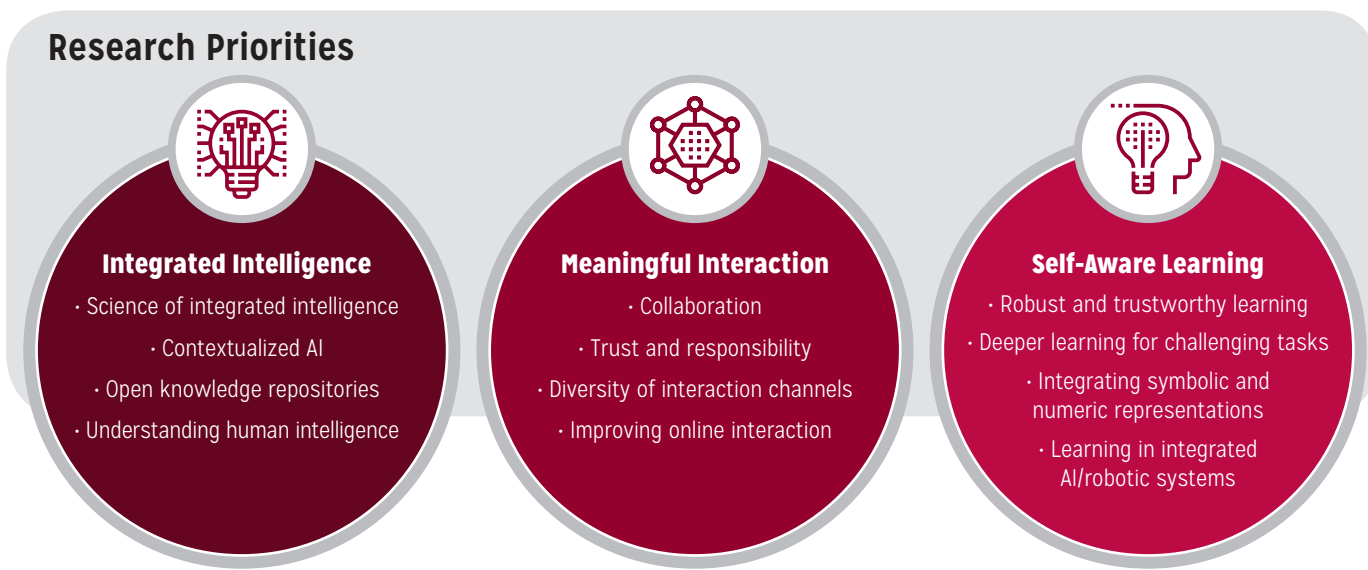

*Figure 3. Research Priorities.*

### 3.1 A Research Roadmap for Integrated Intelligence

#### 3.1.1 INTRODUCTION AND OVERVIEW

The development of integrated intelligent systems will require major research efforts, which we group into three major areas:

**1. Integration:** *Science of Integrated Intelligence* will explore how to create intelligent systems that have much broader capabilities than today's systems. Approaches include finding small sets of primitives out of which a broad range of capabilities can be constructed (like many of today's cognitive architectures) and composing independently developed AI capabilities (like many of today's deployed intelligent systems). Developing theoretical frameworks that provide analyses of performance and reliability will help us scale up and deploy more capable intelligent systems.

---

[22]   ibid

**2. Contextualization:** *Contextualized AI* refers to the ability to adapt general AI capabilities to particular individuals, organizations, or functional roles (e.g., assistant vs coach). This requires both the ability to easily incorporate additional idiosyncratic knowledge without undermining off-the-shelf capabilities, and the ability for the systems to maintain and extend themselves by continuous adaptation to their context and circumstances. This will enable the creation of lifelong personal assistants whose data belongs to their owner, not a third party, and are able to build relationships over time with the people that they work with.

**3. Knowledge:** *Open knowledge repositories* are needed to provide access to the vast amount of knowledge that is required to operate in the rich world we live in. The availability of open knowledge repositories will facilitate the development of a new generation of AI systems that can understand how the world works and behave accordingly.

This section begins with motivating vignettes for each of the societal drivers, highlighting the research required in each of these three areas, it then poses both stretch goals for the 2040 time frame and milestones to track progress along the way.

### 3.1.2 SOCIETAL DRIVERS FOR INTEGRATED INTELLIGENCE

The vignettes below illustrate with concrete example the impacts across human society that would be made possible by AI systems that have integrated intelligence in the next twenty years.

**ENHANCE HEALTH AND QUALITY OF LIFE**

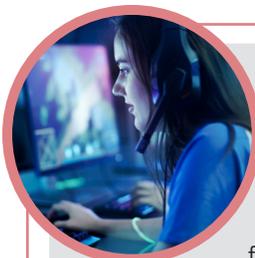

**Vignette 1**

Jane is a video game enthusiast and loves spicy food. She suffers from anxiety, has been under treatment for Type 1 diabetes since her early teens, and has a rare allergy to sesame seeds. Her health-focused personal assistant has been helping Jane manage her physical and mental health for years. It monitors Jane's vital signs and blood sugar, has access to her electronic medical records, and can draw on online health information from high-quality sources to generate recommendations and advice. It helps Jane manage her chronic illness, ensuring that the treatment is being administered correctly and has the intended effects. It stays up to date with the latest breakthroughs in diabetes treatment and reasons about how these might affect Jane. Jane works with the system via natural conversations, enabling her to express how she is feeling during high-stress situations that could affect her anxiety (e.g., a trip being planned) and to get immediate and continuous support and coaching with coping strategies. It was the system that helped Jane avoid a major hospitalization by identifying her rare allergy to sesame seeds early on, based on her reported symptoms and recent diet changes. It helps monitor her mental health, provides encouragement and lifestyle coaching, identifies signs of increased anxiety or depression, and contacts her physician (with Jane's permission) if these symptoms exceed established levels. When Jane sees a doctor, the system provides summaries of observed symptoms, highlighting known conditions and allergies. It double-checks and identifies potential drug interactions. The system also helps Jane track her overall health, choose health insurance programs that suit her needs, and file her claims. Jane is a proud data donor: she has given permission for certain categories of anonymized data in her medical files to be used by AI systems for medical research. In deciding to become a data donor, Jane received trustworthy advice from the system about the potential risks and societal benefits involved. In many ways, it acts like a caring, thoughtful, observant family member or close friend who is also a competent caregiver and medical professional.

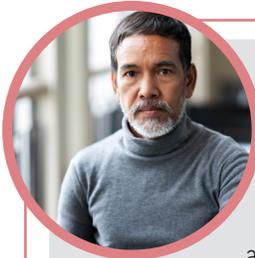

**Vignette 2**

Joshua is a 60-year-old veteran who was diagnosed with Alzheimer's one year ago. His health-focused personal assistant enables him to continue to live independently in his home. It provides cognitive support, helping him keep track of where he is and what he is doing, facilitating daily activities, alerts family members about changes in his patterns, and suggests social interactions (whether joining friends for a concert or arranging a virtual call on his brother's birthday). The system helps Joshua recognize people he knows, what past events they are referring to, and what their intentions may be. Joshua mostly communicates with the system via voice and gesture, with the system sometimes using augmented reality to overlay relevant information ("your daughter") or whispering through an earpiece ("remember to congratulate her on her promotion") as needed. The system can detect when Joshua seems down, and will suggest a walk in the park or start a conversation about planning a brunch for his friend's birthday. By Joshua's request, the system keeps his son and daughter apprised of his medication intake, and reminds him to call after important doctor checkups. It maintains an accurate model of Joshua's capabilities over time and a comprehensive view of Joshua's routine and lifestyle, so it can use daily behavioral data from thousands of other Alzheimer's patients in order to anticipate possible concerns before they arise. It can work with doctors, family, and other caregivers to help provide a safe and supportive environment for Joshua's independent living.

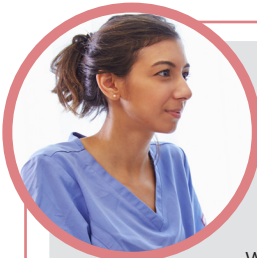

**Vignette 3**

It is a very busy night at the hospital. Nurse Parsa is in charge of monitoring 50 patients on the medical floor. She is able to manage such a large number of patients because in this hospital all patients are fully monitored—an AI system continuously tracks physiologic parameters to determine when a patient is at risk of deterioration and surfaces the information to Parsa. Further, instead of ordering labs every six hours, the system recognizes patients in need of additional tests and notifies the care team. Combined with advances in mobile imaging units, the AI also enables closer, deeper tracking of a patient's conditions via more frequent imaging. Normally, such extensive analysis would require a large support staff, but the AI can pre-process images nearly instantaneously to highlight areas of interest, enabling the current radiology and pathology teams to rapidly identify evolving conditions. This hospital experiences fewer diagnostic errors, fewer sudden events, and fewer unnecessary costs. The hospital also benefits from using real-time data to determine where staff should be sent—if a ward is experiencing busier than normal times, they're staffed accordingly. At 4:20 a.m., a patient in the ward next door is caught becoming hypotensive. Parsa is notified along with a message about the appropriate fluid dose, given the patient's history, helping Parsa remedy the situation quickly. Meanwhile a patient in Parsa's charge experiences sudden bleeding. The system immediately picks that up and asks for additional staff.

**Research Challenges.** These visions will require a number of breakthroughs in AI in the three areas mentioned above:

**1. Integration:** The complexity and criticality of the diverse capabilities involved in these scenarios will require well-engineered AI systems with assurances of their performance and predictability of their behaviors.

**2. Contextualization:** AI systems will need to understand how each individual lives in a different environment with particular health needs, lifestyle choices, household situations, and social settings. AI systems will need to adapt their initial knowledge over years or decades in order to fit the idiosyncratic lives of the individuals they serve, the people they know, and the ever-changing world around them.

**3. Knowledge:** Access to broad knowledge about routine activities and common pursuits, about expectations in social interactions, and generally how the world works will be crucial to developing such assistants. They also will require knowledge about human intelligence and capabilities, about the kinds of limitations that disease and aging pose on those capabilities, and what physical and emotional support and coping mechanisms provide effective assistance. Extensive, up-to-date, reliable sources of state-of-the-art medical research will make them more capable, along with advanced reasoning capabilities that can deploy this knowledge in support of their owners' health and well being.

## ACCELERATE SCIENTIFIC DISCOVERY

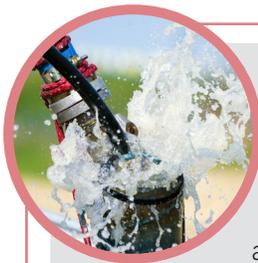

**Vignette 4**

The local businesses of a thriving tri-state area are concerned about the availability of water resources to support an anticipated 10-fold growth in local industry and an associated population increase. In addition, past experience with severe flooding suggests that thoughtful city planning and reservoir management can greatly mitigate risk and losses. The governors and mayors in the area request a study of predicted water needs, existing levels in aquifers and other water reserves, and possible management approaches for local reservoirs. They also ask for interventions and policies that would best support the anticipated growth. A group of local universities start a collaboration with state and local governments, key utility companies, representatives of different industry sectors, and public policy experts to study this problem. The universities bring to bear AI systems that, given the goals and scope of the study, locate additional relevant experts in weather and water modeling, agriculture and industrial economists, civil engineers, and potential data providers. Over the course of a few weeks, AI systems manage and track the collaboration across these different stakeholders, the identification and integration of relevant data, the creation of integrated simulations, and the development of predictive causal models that lead to a holistic understanding of the situation. On the science side, new hypotheses and models about the interacting aquifers and lakes result from the study, and the AI systems propose a fair credit scheme for publication authorship based on its tracking of contributions from each researcher. Local farmers are encouraged that the causes of flooding in certain areas are identified as resulting from an old policy that did not allow pumping in wells during the dry season, which opens up prospects for planting more profitable crops. On the policy side, the AI systems formulate possible interventions involving the development of infrastructure and policy to improve water availability, and generate explanations to different stakeholders to articulate why each intervention would be effective. Accordingly, a number of policies are discussed and enacted in the subsequent legislative period.

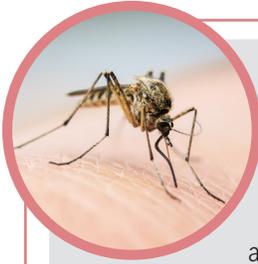

**Vignette 5**

In the blisteringly hot summer of 2031, a highly contagious mosquito-borne infectious disease breaks out in the US, rapidly spreading to major cities around the world. Medical centers, research universities, and government organizations start a collaborative program to investigate the disease and develop a cure and a vaccine. The teams in each organization are composed of AI systems as well as human scientists and clinicians. These AI systems start to track online news and social media to identify hospitals and medical professionals that can provide relevant data, and locate other seemingly unrelated data sources as indicators of how the general population is affected (e.g., through changes in work/school attendance or work absenteeism). In consultation with human scientists, AI systems design and execute experiments that efficiently coordinate dozens of robotic molecular biology labs to analyze samples and data gathered in numerous hospitals and field sites around the world—in real time. Before this event, biomedical literature has informed the AI system and characterized an enormous diversity of biomedical research data. Using this knowledge, they identify possible pathways where the virus might be interfering. After prioritizing the hypotheses, based on the known literature, they carry out targeted experiments in mice. Working with the results, scientists discover an interesting link to a rare neurological condition, and a treatment is developed. AI-accelerated discovery identifies the novel mechanism of viral action, and proposes an effective vaccine. Within days, the disease is under control and those affected are in remission.

**Research Challenges.** AI systems for scientific problem solving will have a significant impact in the future by supporting individuals, groups, and government agencies in scientific problem solving, decision making, research, and innovation. These systems will function as *problem-solving amplifiers*. Specific AI challenges involve:

**1. Integration:** AI systems for scientific discovery will need to seamlessly integrate human language processing, reasoning, planning, and decision making capabilities in order to read the literature and generate hypotheses about the data at hand. These capabilities need to be integrated with robotic sensors and actuators for designing and executing effective scientific experiments.

**2. Contextualization:** Although AI systems can have general knowledge and strategies for scientific discovery, each research problem requires identifying and incorporating specific information about the situation and updating it based on ongoing discoveries. In addition, in order to work effectively with diverse teams of scientists, AI systems will need to understand the expertise that each scientist brings to bear to the problem, and communicate information to each accordingly.

**3. Knowledge:** AI systems will need to automatically extract and integrate knowledge from the ever-expanding published literature. Supporting science (and science-policy) collaborations will also entail sophisticated knowledge of teamwork and human-computer interaction. Causal reasoning, a critical component of scientific thought, is still a nascent area of AI research.

**LIFELONG UNIVERSAL ACCESS TO COMPELLING EDUCATION AND TRAINING**

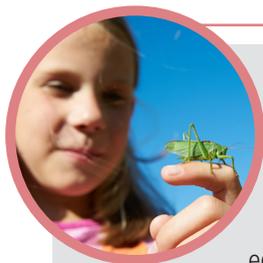

**Vignette 6**

Jody is a middle school student in rural Wyoming who has grown very interested in insects. Jody discusses her classroom lessons with her personal AI tutor, using it as a source of anytime/anywhere learning about insects in her farm and biology in general. The system draws on open educational resources to find interesting questions and topics to discuss, building on its extensive knowledge of what Jody knows in order to better challenge her in productive ways. When her family's crops are affected by an infestation, she sits with her parents and her tutor to read about what the pest could be. She asks the system detailed questions about different species, and narrows it down to three. She works with the system to learn more about the candidate species, then comes up with an experiment to determine which one it is, and discovers the culprit insect. When Jody wonders if her crop pest discovery would be a good start for a science fair project, the system helps her plan it and identify potential roadblocks, as well as ways to work around them. It suggests to her parents that they and she attend some local mentoring events, including a special museum exhibit and a talk by an astronaut. It also helps her find high school and college peers from her county studying biology. Ten years later, Jody is an accomplished veterinarian.

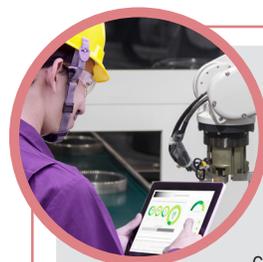

**Vignette 7**

By 2035, most factories are largely automated, but assembly-line workers are still employed in many industries to perform manual sorting, packaging, and piecework tasks. Recent advances in robotic hardware and control have produced haptic interfaces for augmented and virtual environments that can perform many of these tasks, at a considerably cheaper price. Wanting to retain the company's loyal and hardworking employees, the factory purchases AI training systems for management and quality control, as well as for other areas in which the factory is hiring. The AI training systems use virtual reality simulations that are automatically customized to the new factory processes and equipment, based on information provided by the company. They are customized to each employee, since each has a different background. Via this process, workers who would otherwise be laid off are matched with, and trained for, new positions within the factory that enable them to leverage their knowledge of the company's products and customer service, substantially reducing training costs compared to hiring new workers. The workers pool their knowledge to suggest improvements to the new production line as they work through the training, and help management implement changes. Some of their ideas lead to new product lines and partnership opportunities, resulting in new economic growth for the company.

**Research Challenges.** Creating systems that can be deployed across a wide range of education and training contexts to improve people's lives will require substantial advances in a number of areas of AI, especially:

**1. Integration:** In addition to traditional intelligence capabilities such as sensing, planning, and problem solving, AI training systems will need sophisticated skills such as collaboration, creativity, and critical thinking. These will be crucial in helping learners gain the knowledge and skills for a modern workforce that must generate non-routine creative solutions to challenging problems.

**2. Contextualization:** Since each person has a different background, topical interests, and ways of learning, it is essential that AI training systems accurately assess a learner's current knowledge state and design appropriate teaching strategies based on that state. AI training systems will need to optimize topics, support personalized individual and group projects, and provide social networks and resources that create unique and effective opportunities for individual students.

**3. Knowledge:** AI training systems will need extensive knowledge of traditional topics, including math, science, language arts, and humanities, as well as advanced technical topics such as business processes and machinery. In addition, they will require knowledge about how to connect those technical topics with the real world and the ability to generate rich scenarios and examples to effectively scaffold lessons.

## AMPLIFY BUSINESS INNOVATION AND COMPETITIVENESS

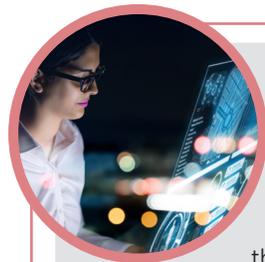

**Vignette 8**

In 2036, the new CEO of a small business and her management team use their organization's AI advisor systems to keep a close watch on international news that could affect their manufacturing operations. Their company uses a complex supply chain involving mostly overseas vendors. Before the new CEO came on board, the company's market value tumbled dramatically amid supply chain troubles and rumors of filing for Chapter 11 protection. The new CEO has improved the supply chain, but the company's market value has been rising only very slowly. When a news item arrives describing serious unrest in a region of Indonesia, it initially seemed irrelevant, since neither the company's direct nor indirect suppliers are based in that country. However, their AI advisory system drew upon broad knowledge of economics, politics, history, and geography to work out a troubling potential consequence: unrest in Riau suggested that the Indonesian government would divert security resources to control the unrest, reducing forces available through the Strait of Malacca, and likely leading to a rise in piracy in the coming months, which in turn could delay shipments from their suppliers. The CEO and her team worked with the AI advisory system to create a network of alternate suppliers and transportation routes and quickly put it in place, thereby minimizing all potential threats to their productivity.

**Research Challenges.**

**1. Integration:** A wide range of intelligent capabilities will need to be combined to attain this type of functionality. Deep language understanding and extensive reasoning will be needed to automatically process news stories and other information sources, including weeding out inaccuracies and deliberate misinformation.

**2. Contextualization:** AI advisory systems will need to develop detailed knowledge of their organizations, creating an institutional memory to inform future decisions that are suited to the particular ways in which the organization works in practice.

**3. Knowledge:** AI advisory systems for businesses will require massive knowledge that covers history, geography, politics, and economics, among many other areas. They will also need extensive knowledge about the structure and operations of organizations and their subsystems (e.g., supply chains). A key capability will be to make plausible inferences about the potential effects of current events on an organization. This will require accurate causal reasoning about consequences, including using historical precedents to help evaluate the plausibility of a predicted outcome and warn of potential pitfalls.

### 3.1.3 THE SCIENCE OF INTEGRATED INTELLIGENCE

Most AI research focuses on single techniques or families of techniques that share a common knowledge representation and are applied to isolated problems. However, there is increasing awareness that more advanced intelligent systems will require *multiple* forms of knowledge, reasoning, and learning, and will involve combinations of reactivity, deliberation, and reflection. These capabilities will be needed to build interactive systems, both cyber and physically embodied, that operate in uncertain environments and communicate with people. While we have ways to develop these individual capabilities, we lack an understanding or science of how to build systems that integrate them. A key open question is what is the best overall organizational approach: homogenous, non-modular systems; fixed, static modular organizations; dynamically modular systems that can be reconfigured over time; or some other variation. Beyond the overall organizational approach, there are also challenges in translation or co-existence of alternative knowledge representations, such as symbolic, statistical, and neural/distributed. Finally, we do not yet understand how to best design the overall control structure for integrated systems: that is, how an AI system can manage both sequential and parallel processes, and the extent to which control occurs in a top-down, goal-driven manner versus a bottom-up, data-driven manner. To achieve intelligent behavior, components will need to exchange rich information and coordinate their behaviors in a harmonious way, despite their expected interdependencies with other components.

To address these questions, we need a *Science of Integration*, focused in particular on integrated intelligence. A Science of Integration would define the space of possible organizations, possibly leveraging a formal language for specifying how systems are organized, similar to early work in computer architecture. The value of such a science is that it could support abstract and formal analysis, comparison, classification, and synthesis of integration approaches. It could provide a framework for both formal and empirical analysis of alternative methods. Such a framework would offer great benefits to researchers as a foundation for the rich systems required to perform the tasks of future AI.

Across the space of possible system organizations, AI systems will need to store and retrieve knowledge: that is, memory is fundamental to their function. In integrated systems, a unified shared memory enables individual components to share information and interoperate, as when knowledge of a play helps one understand how to interpret the actions of individual characters on stage. In turn, both memory and reasoning depend on the knowledge that they manipulate.

**Components of Intelligence**

Intelligent behaviors can be very sophisticated and complex. What are the main components and capabilities that produce these behaviors? A significant portion of AI research over the years has been inspired by the study of human intelligence, producing AI systems that include components of intelligence such as reasoning, problem solving, planning, learning, acting, reacting, understanding and generating language, collaborating, etc. At the same time, since human intelligence is not yet well understood, many AI researchers have focused on engineering new approaches that solve intelligent tasks in ways that had little connection with human intelligence, effectively creating new components of intelligence. This has created an enormous heterogeneity in both the design and functionality of these components. It is very challenging to compare and contrast the capabilities of a given component unless it can be characterized and compared in terms of the intelligent behaviors it supports.

AI systems should, ideally, be created to have minimum complexity while satisfying their design requirements. Associated research challenges include how to specify requirements for AI systems, how to map those requirements into intelligent components, and how to select appropriate architectures that can accommodate those components. Designing the effective and parsimonious AI systems of the future would be greatly facilitated if such modular approaches to integrated intelligence were possible.

Much is known in neuroscience about the structure of the brain at the anatomical and functional levels. The human brain combines specialized, modular structures with broad distributed connectivity. Brain modules do not generally correspond to the components of intelligence adopted by AI and often interact at a finer temporal grain scale. Resolving discrepancies between AI and neuroscience perspectives on the nature and interaction of the components of intelligence could bootstrap a more productive relation between the two fields.

**Stretch goals:** By 2040, AI systems will integrate a variety of intelligent components and capabilities for adaptive low-level and high-level reasoning in complex environments. Milestones along this path include—

**5 years:** AI systems can be created by identifying what major intelligent capabilities are needed, and then selecting and configuring appropriate off-the-shelf components.

**10 years:** AI systems will combine a broad spectrum of components and learning mechanisms.

**15 years:** AI systems will handle new and unexpected situations gracefully by resorting to first principles, analogies, causal reasoning, and other creative processes.

**Memory**
The success of future AI systems will depend on developing new approaches to memory. For example, future intelligent personalized assistants will have to retain and organize memories that are lifelong and life-wide. This presents challenges for computer scientists, behavioral researchers, educators, and ethicists. Lifelong systems will require the capability to organize, synthesize, and retrieve large quantities of information in meaningful ways. Life-wide systems will need to retrieve across contexts and to support a spectrum of tasks. Because some system tasks and task environments will be unknown when the systems are first fielded, these AI system will need to adapt, reorganizing their memories and developing new retrieval schemes.

Although many learning approaches can be seen as building up memories from their training, the associated storage is often narrowly tuned for specific tasks, storing only generalizations, rather than the examples themselves. Capturing and retaining rich episodes that augment examples with their context will enable one-shot analogical learning and will increase flexibility. It will also increase explainability, by enabling systems to point to intelligible factors underlying their decisions. To provide this functionality, memory technologies for future AI systems will need to recognize and retrieve near-miss examples, to retrieve relevant information across different domains, to generalize in appropriate ways, and to re-organize the contents of their memory as new information is encountered. They will also need to integrate this memory with other processes—for example, using stored experiences to provide expectations for an intelligent agent to guide its understanding of a new phenomenon.

Human memory is perhaps the most thoroughly studied system in cognitive science, with its capabilities and limitations the subject of systematic experimental studies. Human memory relies on a hierarchical system of working memory, episodic memory, and semantic memory that have distinct, complementary characteristics in terms of time span and generalization. Working memory provides limited, efficient storage of local task context; episodic memory stores the rich, specific multimodal episodes that provide the basis of our identity; while semantic memory abstracts and generalizes knowledge for broad, long-term use in a variety of contexts. Because of capacity and access limitations (not unlike those of, say, robotic systems) human memory has evolved mechanisms to efficiently store a lifetime of information in constantly changing, uncertain environments and heuristically access it in real time under severe task constraints.

Memories captured and used by future AI systems will range from those that are specific to a particular individual at a particular time, to population-wide event and cultural memories, to metadata about all levels of a massive structure of varied, interconnecting topics. Humans now make extensive use of external knowledge sources, in multimodal formats that include text, video, and images. In order for AI systems to achieve coverage on a par with—or beyond—humans, and to be able to understand the knowledge of humans with whom they interact, future memory systems will need to exploit the enormous digital knowledge sources that now exist.

As AI systems build up new knowledge from interactions with people and other sources, they will need to assess the reasonableness of, and set limits to, the knowledge that they capture and queries they serve. Both mined information (e.g., on social media), and the memories they reconstruct may be inaccurate. AI systems will need to manage conflicting information, both at storage and reconstruction time. For example, AI research assistants will need to ingest and understand vast amounts of scientific literature, including experimental results that may sometimes conflict. More generally, they will need to reason about their own knowledge, in order to "know what they know, or don't know," to provide information with confidence when warranted and appropriate caution when not, and to guide the search for new knowledge.

**Stretch goals:** By 2040, AI memory systems will provide lifelong and life-wide repositories of experiences and other information, with fully context-aware and task-relevant retrieval. They will integrate the experience of systems and information captured from many sources, including the experiences of people captured by a wide range of heterogeneous, vast external multimodal sources. These memory systems will be able to flexibly reorganize themselves and reconstruct information as needed to appropriately exploit knowledge available across different systems. They will reason about the limitations of their own processes and knowledge, enabling people and AI systems to understand those limitations and uncertainties. The security, privacy, and sharing needs for individuals, institutions, and groups will be embedded in the structure of these systems. Milestones along this path include—

**5 years:** AI memory systems will be able to integrate and retrieve knowledge across tasks, knowledge types, and levels of abstraction, with active inference and tracking of uncertainties associated with observed and inferred information.

**10 years:** AI memory systems will flexibly assemble information from multiple sources, assess the quality of the knowledge from each source, and identify gaps.

**15 years:** AI memory systems will be easily reconfigured for new tasks and will integrate/assess multiple vast information sources, maintaining a clear distinction between bedrock memories, reconstructed plausible memories, and imaginings for the future.

**Metareasoning and Reflection**

Today's high-performance AI systems require hundreds to thousands of human engineers to build, maintain, tune, and extend them. To achieve the goal of creating lifelong personal AIs, we need to learn how to make AI systems that do self-maintenance. This will require them to build and use models of themselves and data about their performance. Already, some cognitive architectures accumulate statistical metadata about their own knowledge for the purposes of estimating its utility and so better deploy it in new problems. More advanced metaknowledge will also include the ability to prioritize goals and estimate the timelines to achieve them and the likelihood of success. More accurate metaknowledge will enable AI systems to balance their workloads better, including trading off effectively between responding to their users and investing in learning to do their jobs better.

**Stretch goals:** By 2040, metaknowledge and self-knowledge models will be robust enough to scaffold AI systems that maintain themselves over years of operation, learning continually and adapting to new challenges with little to no intervention by human designers. Milestones along this path include—

**5 years:** The need for support staff in maintaining and extending AI systems will be significantly reduced due to the ability for the systems themselves to take on more of the support burden. At least 20 percent of the maintenance and extensions will occur via user interaction with the system, rather than through redesign and reimplementation by an AI engineer.

**10 years:** Support staff will be required only for periodic maintenance, with autonomous operation and updates otherwise, including balancing activities for self-improvement with effective collaboration. At least 80 percent of maintenance and extensions will occur via routine user interaction with the system.

**15 years:** No routine manual inspections and checkups of AI systems will be needed; instead, support staff will be used on demand, when requested (infrequently) by either the system or the people working with it.

### 3.1.4 CONTEXTUALIZED AI

Contextualization refers to the adaptation of general intelligent capabilities for a particular individual, organization, situation, or purpose. This includes contextual knowledge about the world, historical context of the circumstances of that particular entity, and its circumstances as it interacts with other entities and with the world. Contextualization is necessary in order to develop AI systems that assist people in their daily lives at work and home, for leisure and for learning. It is also necessary for the development of AI systems that are tailored to the culture and processes of specific organizations. Contextualization will allow AI systems to be *individual*, in that they use data and yet remain the property of their owner, and also be *personalized*, in that they prioritize the interests of the people interacting with them rather than third parties (e.g., their manufacturers). Contextualization will also allow AI systems to identify what sort of language is most familiar to their owner, which feedback improves a person's well-being, and which approaches are most engaging. It will also support effective human-system collaboration, automated support of high-quality teamwork and rapid execution, and reasoning about social context and the social consequences of actions.

**Customization of General Capabilities**

AI systems with a broad range of intelligent capabilities will need to be personalized to an individual, tailored to an organization or group, redirected to particular purposes, and adapted to changes in the environment where they operate. To accomplish this, off-the-shelf, general AI components will need to be transformed in appropriate ways to operate in their particular contexts. Today, many AI systems can be customized, but only in very narrow ways. For example, current personal assistants (e.g., Alexa) can learn a few preferences based on data from user interactions. However, they cannot take instructions about how a user would *like* them to behave, or assist a user differently in different contexts (their role as a parent at home, or a manager at work, or a gym member), or understand when the same preference applies in different contexts—or not.

The need for AI systems to adapt to their context has many dimensions, best exemplified by personal assistants. To be effective, these will require: 1) *lifelong* adaptation, supporting a user for the long haul through different stages of life; 2) *life-wide* adaptation, supporting all aspects of life at work and at home, for leisure and for learning; and 3) *continuity*, determining which customizations are relevant and incorporating them into new AI systems that a user may adopt later on.

**Stretch goals:** By 2040, AI systems that start with general abilities and over time can adapt effectively to specific users, organizations, or purposes. These adaptations will continue over time as their environment changes, will be coordinated across tasks and activities, and will be transferable to new AI systems. Milestones along this path include—

**5 years:** AI systems will be able to extend their initially given knowledge and behaviors to fit into a particular environment.

**10 years:** AI systems will adapt their initial behaviors over long periods of time to reflect the changing situations and environments around them.

**15 years:** AI systems will be able to use any acquired preferences and customizations and adapt them for new tasks and goals.

**Social Cognition**
Social cognition seeks to understand interactions between individuals and how they understand each other. This has been a topic of study in multiple areas of cognitive science (e.g., psychology, linguistics, anthropology, neuroscience). To create AI systems that work effectively with people as collaborators, the systems must themselves be capable social beings, able to participate in social interactions. This will require significant advances in the understanding of social cognition from a computational perspective. The rest of this section outlines some relevant key areas.

To begin, AI systems will need to align with human values and norms to ensure that they behave ethically. They will need to take into account potential risks, benefits, harms, and costs. In order to do this, AI systems will have to incorporate complex ethical and commonsense reasoning capabilities that are needed to reliably and flexibly exhibit ethical behavior in a wide variety of interaction and decision making situations. They will potentially also be able to encourage humans toward ethical behaviors and even encourage respectful behaviors and decision making.

Finally, AI systems will need to build models of others as independent, intelligent beings. This will require having a model of their knowledge, their capabilities, their goals, and their awareness of the situation at hand. It will also require an awareness of how another human or AI system might, in turn, be modeling others.

**Stretch goals:** By 2040, AI systems will be capable of reasoning about social context, with behaviors appropriate to social events, awareness of people's emotional states and reactions, and acknowledgment of shared and differing goals and reactions. Milestones along this path include—

**5 years:** AI systems will take social norms and contextual information into consideration when deciding how to pursue their goals.

**10 years:** AI systems will effectively reason about how to behave when faced with conflicting social norms and goals.

**15 years:** AI systems will handle situations where the motivations and goals of others interfere with accomplishing tasks, responding in appropriate ways: by changing their own tasks or negotiating with others.

### 3.1.5 OPEN KNOWLEDGE REPOSITORIES

AI researchers have long acknowledged the vast amount of knowledge about the world that humans acquire continuously from birth. This kind of knowledge, which is currently very difficult to incorporate into AI systems, includes understanding how people behave, commonsense knowledge about the physical world, and encyclopedic knowledge about objects, people, and other entities in the world.

Open knowledge repositories with massive amounts of world knowledge could fuel the next wave of knowledge-powered AI innovations, with transformative effects ranging from innovations in scientific research to the commercial sector. These knowledge resources could enable new research in machine learning systems that take human background knowledge into account, natural language systems that link words and sentences to meaningful descriptions, and robotics devices that operate with effective context and world knowledge. These open knowledge repositories could be extended to include domains of societal interest such as science, engineering, finance, and education. Through these systems, scientific research could exploit a vast trove of knowledge to augment data and support interdisciplinary research.

Industry is already benefiting from rich knowledge repositories. As technology companies push the envelope in markets such as search, question answering, entertainment, and advertising, they have tapped into the power afforded by massive amounts of knowledge about the world to make their systems ever more capable. This knowledge is typically captured as a graph containing entities of interest (e.g., products, organizations, notable people) interlinked in diverse dimensions and highly structured to enable abstraction and generalization. Currently, these knowledge bases focus on information particular to ecommerce and web search (e.g., information about products and geography), and do not include other kinds of knowledge about the world that are important for intelligent systems more broadly. Most are not openly available to academia, government, or smaller businesses.

Open knowledge repositories have the potential to impact science, education, and business if we mount an open effort to develop and expand shared resources. For example, already open knowledge repositories are ubiquitous in biomedical research. They represent entities of interest such as genes, proteins, organisms, diseases, and many other entities referenced and reasoned about in biomedical research. These knowledge repositories are created and maintained by different communities. They have enabled major advances in terms of data integration, pathway discovery, and machine reading. As powerful as they are, they would be even more useful if they could be linked to representations of everyday knowledge, to better understand the context of texts, and to expressive representations of biomedical theories and the evidential links between theories and experiments.

Many government agencies have been investing in well-scoped areas to create specialized knowledge networks, but fusing these small islands requires enormous effort at the touching points where key integration and innovation projects reside. Open knowledge repositories would provide a semantic infrastructure that would build upon and significantly enlarge these existing limited capabilities. Key steps in achieving this vision include expanding representations for expressive knowledge and developing advanced reasoning and inference methods to operate over these representations.

**Heterogeneous Knowledge**
Knowledge bases express beliefs about the world, in machine-understandable form, to support multiple uses. For example, billion-fact knowledge bases are used to recognize entities in web search engines, and thereby improve precision. (Every time you see a sidebar box in a web search, or a question answered directly, you are seeing their knowledge bases in use.) Knowledge bases are also used in reasoning systems, natural language understanding, and interactive task learning systems. Some are built by hand and some are built by machine reading. Most open and proprietary resources currently focus on facts about specific entities, such as the population of a town or which actors starred in what movies. These facts are stated via relations that link two or more entities (e.g., a country is linked to the city that is its capital). These kinds of facts are the easiest type of knowledge to extract from texts and databases, and they constitute a major part of human knowledge. But people know much more than just facts, and AI systems need to as well. This section outlines several key types of more expressive knowledge that need to be explored to create integrated intelligent systems. In all cases there has been some progress, but much more research is needed.

An important kind of knowledge are cultural norms, which include etiquette, conventions, protocols, and moral norms, such as prohibitions against murder. Including information about cultural norms is crucial for AI systems to understand not only how to use things but also what is required, permitted, and forbidden. In people, norms are often tacit, which can lead to serious cross-cultural misunderstandings. In AI systems, such knowledge needs to be contextualized, so that the norms of many cultures can be expressed, which will support perspective-taking by AI systems. Another important type of knowledge is information about what things look like, sound like, and even taste and smell like, which helps ground AI systems in the everyday world that we all experience. Understanding the significance of particular types of clothing worn by people, for example, is important in order for machines to understand the world around them.

Decades of research in cognitive science indicate that causality is a key component of human conceptual structure from infancy, and that it is multifaceted. In medicine and biology, causality is tightly tied to statistics, with careful argumentation from trials needed to disentangle causality from correlation. To assist or automate such reasoning will require representing these ways of establishing causality. In commonsense reasoning about the physical and social world, qualitative representations have been developed to provide frameworks for causal knowledge, but these have not yet been applied at scale. In engineering problem solving research, qualitative models have also been used to capture aspects of professional reasoning in several domains, but, again have not yet been explored at scale. Causality in social cognition requires models of human psychology and theory-of-mind reasoning that can represent the beliefs, desires, and intentions of other agents.

Another arena where causality is crucial is in understanding processes, plans, and actions. While AI planning techniques are already richly developed, the focus has mostly been on actions taken by an agent or set of agents, rather than agents continually interacting in a rich world where knowledge is required to operate. AI systems that understand how a business works well enough to participate in evaluating courses of action in response to a crisis, for example, would require deeper understanding of processes, plans, and actions. There is already research on gathering and sharing how-to knowledge for robots, which may suggest a cooperative model for knowledge capture that could be used more broadly. Reasoning about how things work occurs in everyday life, as in trying to figure out what is wrong with a car. Being able to express alternative hypotheses about what is wrong, compute the implications of these hypotheses, and make observations to settle the questions raised are essential aspects of troubleshooting.

**Stretch goals:** By 2040, our scientific understanding of causal models will be broad and deep enough to provide formalisms that can be used across diverse application areas, including education, medicine, eldercare, and design. A repository of causal models that covers both everyday phenomena and expert performance will enable AI systems to act as collaborators in these domains. Milestones along this path include—

**5 years:** Causal models at the level of elementary school science knowledge will be used for linking evidence and open questions in science and design, and for supporting advanced decision making.

**10 years:** Causal models at the level of high school science knowledge in several scientific domains (e.g., biomedicine) will enable AI systems to read and analyze scientific papers, and will support decision making in eldercare support and design.

**15 years:** Causal models for multiple professional domains will support AI systems working as collaborators with human users.

**Diversified Use and Reasoning at Scale**

Reasoning puts knowledge to work in answering questions, solving problems, and troubleshooting. Broadly speaking, reasoning combines existing statements to create new ones. Multiple types of human reasoning have been captured to varying degrees in machine reasoning systems. Rarely is one form of reasoning sufficient for a complex task. For example, working out the geopolitical implications of a policy change can involve reasoning about economics, geography, and human motivations. In addition to progress on specific types of reasoning, we must better understand how to make them interoperate flexibly, so that their strengths and weaknesses complement each other to provide better performance and more accurate conclusions in wider ranges of circumstances.

Everyday cultural knowledge is estimated to require in the range of $10^9$ facts, which is orders of magnitude larger than the knowledge bases that most reasoning systems have been used with to date. One of the most difficult challenges will be scaling up reasoning services to operate at that level. Part of the solution is likely to include reformulating open tasks into a set of more tightly constrained problems. For example, the industrial use of satisfiability solvers and model checkers in design indicates that, even when reasoning methods have exponential complexity, they can still be successfully used on practical problems. But the reasoning of model formulation today mostly resides in the minds of the people using those tools. Another tool for scale that people use is reasoning by analogy. By retrieving relevant experiences, long inference chains can be avoided. Statistical reasoning can also help, with metaknowledge providing estimates of utility of approaches and accuracy of conclusions, to enable systems to focus on the most productive reasoning paths. Today's methods of integrating statistical and relational knowledge vary in degree of formal integration; to date, none of the more formal methods scale to real-world problems.

While people do reason deductively sometimes, evidence from cognitive science indicates that non-deductive reasoning plays a major role in human cognition. One important form of non-deductive reasoning is abduction, which is reasoning to the best explanation. For example, if the windshield of a car is wet, it could be because it is raining, or because its path took it into an area where sprinklers were in use. Abductive reasoning is heavily used in scientific reasoning and troubleshooting, but has been difficult to scale up. Another form of non-deductive reasoning is induction, which makes general conjectures from specific instances. For example, if you see that several cars have speedometers, then you might conjecture that all cars have speedometers. There has been research on a variety of inductive algorithms in machine learning. When the data consists of entities represented by a well defined set of individual attributes (which can be viewed as a simple special case of ground facts), these systems can operate successfully at industrial scale. Induction methods that work with more relational representations (i.e., statements connecting multiple entities) have been explored in limited contexts in inductive logic programming and other areas. Progress in this area is especially important, since relational representations are required for many forms of human knowledge (e.g., plans, explanations, and theories).

Analogy is a third form of non-deductive reasoning that handles relational representations. Analogy, including the particular set of techniques known as *case-based reasoning*, involves reasoning from one example to another. For example, if trucks are like cars and cars have engines, then one might conjecture that trucks have engines, too. Analogical inferences can be used for explanation as well as prediction. Moreover, *analogical generalization*, where multiple examples are combined by analogy to yield probabilistic structural representations, provides a form of induction. For example, if someone observes many examples of vehicles on the road, they start to build up models of different kinds of vehicles (e.g., cars and trucks). Evidence from cognitive science suggests that people rely heavily on analogy in their everyday reasoning, and heavy reliance on analogy may help explain why human learning is so much more data-efficient than today's machine learning techniques. While there are models of matching, retrieval, and generalization that have been used to explain aspects of human reasoning and learning, and applied in several medium-scale domains, today's accounts of analogical reasoning and learning have yet to be tested at scale—something that open knowledge repositories will enable.

**Stretch goals:** By 2040, we will have a deep scientific understanding of the integration issues involved in different forms of reasoning, and AI systems will be able to combine them to achieve desired accuracy/performance tradeoffs. A suite of standard, open-source methods for integrating reasoning, using off-the-shelf deductive and non-deductive systems, will be used to rapidly build new AI systems. Milestones along this path include—

**5 years:** Robust multimodal reasoning will be possible across million-, and billion-fact knowledge bases to carry out decision-support and design tasks.

**10 years:** Inductive and analogical techniques will incrementally build up and maintain models of tens of thousands of concepts, from hundreds of examples per concept.

**15 years**: AI systems will be able to understand long complex sequences of events with many actors (e.g., a movie) through analogy and abduction.

**Knowledge Capture and Dissemination**
Human intelligence is fueled by knowledge accumulated over time by individuals and societies. Similarly, AI systems need to be fueled by knowledge repositories that store the information required to learn or reason about the world. While some of the human knowledge has already been documented and stored in distributed raw or semi-structured documents, there is also knowledge where the people themselves are the main medium of preservation. For example, a major goal for AI has been the accumulation of commonsense knowledge repositories, but to date commonsense knowledge is still largely a human construct. Human knowledge continues to grow due to scientific advances, events (e.g., an election), creativity (e.g., new songs or buildings), discoveries (e.g., a new archeological site), etc.

Several online knowledge sources exist that are quite comprehensive, though never complete. They are semi-structured in nature, or follow certain templates. Examples of such repositories include crowdsourced encyclopedias such as Wikipedia, DBPedia, or Wiktionary, or vertical websites that include specialized collections such as "How To" step-by-step instructions, recipe websites, course offerings for colleges, travel websites, and others. These sources are not directly machine-understandable, that is, in a format that AI systems can directly use. Much of the research work done to extract knowledge and represent it in machine-understandable formats, though, has been focused around using patterns to extract specific types of information.

Important information is often structured in tables or item lists, though the entries and items may still need some processing. For example, a table with basketball players for a team may include their names, but there may be two players named "Maria Smith." A variety of approaches have been developed for information extraction that take such semi-structured sources and generate knowledge bases that are machine-understandable. An important future challenge is knowledge curation and maintenance over time. Several of these sources are highly dynamic (e.g., Wikipedia), with periodic new contributions and changes. New approaches that can track such temporal changes and update the knowledge repositories to be consistent with the changes and capture the nuances the validity of each piece of knowledge (e.g., if a football player leaves the Miami Dolphins in 2019 then they were a Dolphins player in 2018 but not in 2020) are needed.

The vast majority of human knowledge is stored in raw documents, in the form of text, images, videos, sounds, and other kinds of formats that have no particular structure. Although effective approaches have been developed to extract target categories of entities (e.g., colors or animals), more research is needed to extract other kinds of entities and relations, and to extract knowledge from other media, including images, videos, sounds, and tables.

Recent years have also witnessed a significant increase in the use of crowdsourcing to create knowledge repositories for AI systems. Crowdsourcing and information extraction are often used jointly, where crowdsourcing can fuel automated text extraction, and automated text extraction output can be corrected through crowdsourcing. The quality of the results obtained through crowdsourcing can be improved using a variety of techniques (e.g., by avoiding spam contributions), and knowledge collection pipelines can be used divide the task into smaller working subtasks and subsequently effectively combine their outputs.

Important areas of future research include evaluating the quality and trustworthiness of knowledge repositories, improving them, resolving inconsistencies in the knowledge sources or in the knowledge base, updating the knowledge over time, and reasoning over such knowledge to answer questions.

**Stretch goals:** By 2040, we will have the ability to create knowledge repositories on target domains or topics as needed by AI systems. Given a target domain, we will be able to identify the sources needed to create such knowledge bases, and extract entities and the relations between them. We will also have algorithms that will validate and maintain these knowledge repositories. We will also have the ability to extract significant amounts of commonsense knowledge from text, images, videos, and other unstructured formats. Milestones along this path include—

**5 years:** Automatic creation of knowledge bases from raw data, covering different media (text, images, tables, etc.). A suite of methods for creating effective crowdsourcing pipelines for knowledge base construction, sometimes in close interaction with algorithms.

**10 years:** Knowledge repositories that continuously grow through extraction from sources. Large commonsense knowledge repositories. Algorithms that can effectively validate and maintain very large knowledge repositories.

**15 years:** AI systems that can understand their own needs in terms of knowledge, and can identify the required sources and construct the knowledge bases they need. Methods for effective reasoning over knowledge bases.

### Knowledge Integration and Refinement

The days of a single human, or a team of humans, having a complete understanding of an entire large-scale knowledge base are already over. Given that today's large-scale knowledge bases are already at $10^9$ facts to cover a reasonable subset of common cultural knowledge, it would not be surprising for extending coverage to include multiple domains of professional knowledge and multimodal knowledge to lead to a total size of $10^{10}$ facts. Moreover, most of the new knowledge will be gleaned by AI systems learning, either by reading, watching, or interacting with people. This makes it inevitable that inconsistencies and issues can creep in. Thus, we need to develop reasoning processes that help with the integration of large amounts of knowledge.

Knowledge refinement processes will need to integrate feedback from systems that use knowledge from the repositories, to gather the data needed to diagnose and repair knowledge structures. One difficulty is that there will be a variety of such systems, with different levels of engineering quality, whose properties need to be considered during diagnosis. Moreover, the existence of bad actors, which may try to poison the system by providing misleading reports, also needs to be taken into account.

**Stretch goals:** By 2040, a scientific understanding of reasoning processes will support knowledge base curation**,** maintaining accuracy even in the face of low-quality and adversarial inputs. A suite of open-source reasoning processes will enable open knowledge repositories to grow past $10^9$ facts. Milestones along this path include—

**5 years:** Tools and techniques for knowledge integration will work with expert human curators to build out everyday and professional knowledge in multiple practical domains.

**10 years:** Tools and techniques for knowledge integration will be semi-autonomous, relying on human experts only when needed or requested, and will be capable of handling adversarial inputs.

**15 years:** Tools and techniques for knowledge integration will be sufficiently reliable that humans will only be called in a few times per year, by the knowledge service itself or by its users.

### 3.1.6 UNDERSTANDING HUMAN INTELLIGENCE

Human intelligence is studied not just by AI researchers but also in many other disciplines, such as psychology, linguistics, neuroscience, philosophy, and anthropology. As in AI, these disciplines have mostly focused on particular phenomena, rather than developing integrated accounts of cognition. Cognitive science was formed with the idea that the computational ideas that AI was using to study intelligence might become a language that could also be used to understand minds. The research involved in integrated intelligence that is discussed above will provides a unique opportunity to deepen that connection, using the study of artificial intelligence to inform our theories of natural intelligence.

**AI Inspired by Human Intelligence**
Work on cognitively and neutrally inspired AI has to date produced two of the most dramatic successes in the history of artificial intelligence: rule-based expert systems that capture high-level articulable patterns and relationships, and neural-network-based deep learning systems that capture low-level non-articulable patterns and relationships. As we turn our sights toward the future, a critical challenge is to understand how humans effectively combine both high-level articulable and low-level non-articulable capabilities to leverage the strengths of each while offsetting the weaknesses of the other, yielding flexible, effective, transparent, and efficient integrated reasoning.

A second challenge is to expand the range of machine learning capabilities to span the myriad of ways in which humans learn from heterogeneous inputs and experiences in order to improve the scope, accuracy, and speed of both individual learning abilities and of the overall system.

A third key research challenge is learning from (and ultimately replicating) how humans reason beyond their local conceptual contexts, which allows them to exhibit critical global thought processes that can be characterized as out-of-the-box creativity, analogy, and the distal transfer of learned knowledge.

A fourth key topic concerns social cognition (as mentioned above), which will enable AI systems to model other intelligent systems—both natural and artificial—so that they can produce individual and group behaviors that are more appropriate and effective than thinking in isolation. These modeling capabilities will need to incorporate adversarial reasoning: As AI systems become ever more broadly deployed and have ever greater impact on human lives, people will inevitably attempt to manipulate them for their benefit, for example by submitting unrepresentative data to machine learning algorithms. AI systems will thus need to be able to reason strategically in order to be robust in the face of such manipulation.

Finally, metacognition is perhaps the least well understood human ability. Part of that capacity is likely due to the flexibility and heterogeneity of knowledge representation in various brain modules, especially the capacity to combine symbolic and statistical information and access it in flexible, introspective ways. Areas most directly involved in metareasoning and reflection include structures of the anterior prefrontal cortex, whose recent development is most characteristic of human evolution, responsible for building, managing, maintaining, and reasoning about goals.

Solutions to no one of these challenges will be sufficient to provide the most comprehensive forms of integration found in the study of human intelligence: Creating this level of artificial intelligence will require cognitive architectures that embody hypotheses concerning more comprehensive combinations of necessary and/or sufficient capabilities. Lessons from such architectures can potentially inform solutions to the individual challenges, while also highlighting possible paths for integration across these areas.

**Stretch goals:** By 2040, AI systems will exhibit effective human-like integrated adaptive low-level and high-level thinking in complex social environments. Milestones along this path include—

**Milestones**
**5 years:** AI systems employ cognitive models that incorporate the full strengths of high-level articulable reasoners, planners, and problem solvers with low-level non-articulable inference networks and learners.

**10 years:** AI systems use cognitive models that combine a broad spectrum of high-level and low-level learning mechanisms.

**15 years:** AI systems based on cognitive models that incorporate strong models of the cognitive and social aspects of people and other agents, individually and in groups.

### AI to Understand Human Intelligence
A wide range of capabilities in human intelligence have been studied in AI, including symbolic and probabilistic processing, reinforcement learning, and artificial neural networks. Much of this work was originally inspired by models of brain structures and processing. AI has, in turn, led to important insights about the nature of human intelligence. This virtuous cycle of reciprocal benefits that highlights the ideal of what is possible with bidirectional influences.

An area in which this synergy has been particularly notable is the study of learning algorithms. Reinforcement learning has become one of the most important machine learning algorithms in AI. In the brain, it has long been associated with subcortical structures such as the basal ganglia that control the procedural aspect of our behavior, from internal operations that route and request information across brain areas to external actions, including active perception and motor actions. Detailed work has mapped specific aspects of reinforcement learning algorithms onto brain mechanisms such as neurotransmitters. Hebbian learning, initially identified as a principle for local learning at the synaptic level, has led to AI algorithms such as backpropagation that have been central to deep learning. New principles of Hebbian learning, such as spike-timing-dependent plasticity, hold the promise of further advances in learning algorithms that are more robust and self-regulating. More broadly, the study of how those processes operate in concert within the complex structure of the human brain can shed light on how to integrate a variety of learning algorithms in complex AI systems.

**Stretch goals:** By 2040, broad-coverage integrated models of mind and brain will be applied in concert to enable intelligent reasoning spanning time scales from one millisecond to one month. Milestones along this path include—

**Milestones**
**5 years:** AI systems could be designed to study psychological models of complex intelligent phenomena that are based on combinations of symbolic processing and artificial neural networks.

**10 years:** Integrated architectures are the standard vehicle for modeling the results of complex psychological experiments.

**15 years:** Progress in AI on neural networks and integrated architectures yields major advances in neural/brain modeling.

**Towards Unifying Theories of Natural and Artificial Intelligence**
Ultimately, a comprehensive theory of integrated intelligence is desirable that spans all possible forms of intelligence, whether natural or artificial. Such a grand challenge will likely take much more than 20 years to complete, but a start can be made within this time frame by focusing on more modest goals. Recently, a new community has begun to coalesce around the goal of designing human-like cognitive architectures that model both human and artificial intelligence. Among other things, such a common architectural framework could provide a useful intermediary in evaluating and comparing cognitive architectures and create a pathway for unifying models of natural and artificial intelligence.

An essential aspect of those architectures that has increasingly guided their development is the ability to validate them using neural imaging data. Neural imaging data, assembled in large databases, such as the Human Connectome Project covering substantial subject populations performing a diverse range of tasks and using a number of imaging techniques, provides constraints regarding both the structural and functional organization of brain modules as well as the details of knowledge representation within those modules. This new source of data expands on the wealth of existing behavioral data accumulated over more than a century of research, provides converging evidence for and against proposed architectures, and holds the promise to considerably speed up convergence to a consensus theory of natural and artificial intelligence.

**Stretch goals:** By 2040, a common cognitive architectural model will yield deep understanding across at least one full arc of cognition, from perception to behavior, for a complex task in a real environment. Milestones along this path include—

**Milestones**
**5 years:** The full space of existing cognitive architectures (i.e., integrated models of human-like intelligence) is mapped onto a single common model of cognition.

**10 years:** Strong connections are demonstrated between AI architectures and cognitive models that can be mapped at the level of major brain regions, their functional connectivity and mechanisms, and their communication patterns.

**15 years:** Shared implemented models of cognition are in wide use by both the AI and computational cognitive science communities.

## 3.2 A Research Roadmap for Meaningful Interaction

### 3.2.1 INTRODUCTION AND OVERVIEW
Research on AI for Interaction has resulted in significant advances over the last 20 years. In fact, many of these advances have seen their way into commercial products. In the four focus areas of AI for Interaction, we have seen successes but also major limitations:

◗ AI systems that act as personal assistants have seen wide-scale success; These systems can interact using spoken language for short, single turn commands and questions, but they are not able to carry on intentional dialog that builds on context over time.

◗ AI systems can use information from both speech and video when processing streams such as broadcast news to build an interpretation, but they are not able to fuse information from other sources such as smart home sensors.

◗ AI systems are already able to detect factual claims in text and to distinguish fact from opinion but they cannot reliably determine when claims are false.

◗ AI systems have seen widespread deployment and trust for a number of practical applications where the system is very accurate or the stakes are very low, as in the case of music recommender systems, but they are not yet trusted in high-stake domains such as medical diagnosis.

Many future AI systems will work closely with human beings, supporting their work and automating routine tasks. In these deployments, the ability to interact naturally with people will be of unparalleled importance. Developing AI systems that have this capacity for seamless interactivity will require major research efforts, spanning 20 years, in four interrelated areas:

**1. Integrating Diverse Interaction Channels:** AI systems can integrate information from images and their text captions, or from video and speech when processing related streams such as broadcast news. Combining different channels of interaction, such as speech, hand gestures, and facial expressions, provides a natural, effective, and engaging way for people to communicate with AI systems. Jointly considering multiple input modalities also increases the robustness and accuracy of AI systems, providing unique opportunities that cannot be accomplished by considering single modalities. Today's AI systems perform well with one or two modalities. We have seen tremendous advances in speech-based commercial products such as home personal assistants and smartphones. As we move into the future, however, AI systems will need to integrate information from many other modalities, including the many sensors that are embedded in our everyday environments, thus leveraging the advances that have taken place in the Internet of Things. AI systems must also be able to adapt their interactions to the diversity of human ability, enabling their use by people with disabilities, by people with non-standard dialects and by non-English speakers. AI systems must be able to take context into account and handle situations when information is missing from a modality. Multiple modalities also pose a particular challenge for preserving privacy, as much sensitive information is revealed during interactions, including images of faces, recordings of voices, and views of the environment in which people live. AI systems also need to be capable of integrating new, more efficient communication channels, as well as directing and training users in how to best interact with the systems on these new channels.

**2. Enabling Collaborative Interaction:** Interactions with today's AI systems are often stilted and awkward, lacking the elegance and ease of typical interactions between people. For example, today's AI systems rarely understand the context underlying the user's desires and intentions, the potentially diverging objectives of different users, nor the emotional states and reactions of humans with whom they are interacting. In contrast, when humans interact, they intuitively grasp why someone is asking a question, they seek to find common ground when faced with differing opinions within a team, and they reciprocate in their emotional expressions, offering understanding and encouragement when their teammate seems despondent. Building AI systems that interact with humans as fluently as people work together will require solving challenges ranging from explaining their reasoning and modeling humans' mental states to understanding social norms and supporting complex teamwork.

**3. Supporting Interactions Between People:** With the increased presence of social media and other online discussion venues, increasingly more human social interactions take place online. AI can help in facilitating these interactions and in mining online data for understanding the emerging social behaviors of online communication. Understanding what people expect from online media, and how they want to use it, can help in developing AI systems that address those needs. We categorize future research in this area into three subsections: deliberation, collaborative creation, and social-tie formation. In the area of deliberation, AI systems are needed that can distinguish fact from falsehood and fact from opinion. AI systems that can detect bots will also be critical moving forward. Online deliberation data can help researchers understand how opinions form and how influence spreads. In the area of collaborative online content creation, Wikipedia already uses AI bots to detect vandalism and enforce standards. Future research should extend this work to new settings such as creative arts or collaborative software engineering. Here there is a role for AI systems to support collaboration: identifying inconsistencies, assumptions, and terminological differences among collaborators. Finally, social-media analysis is an active area of research, with commercial efforts focusing on identifying phenomena such as hate speech, albeit with the human in the loop. We foresee increasingly sophisticated human-machine hybrid technologies in this space, which may ultimately enable new forms of human-human interaction, and even support emergent social structures.

**4. Making AI Systems Trustworthy:** Today's AI systems often use methods that offer high performance and accuracy, but are not able to appropriately describe or justify their behaviors when asked. Generating appropriate descriptions of their capabilities and explanations of the resulting behaviors will require research to develop new approaches that align the machine's internal representations with human-understandable explanations. Based on their understanding of an AI system, users should be able to assess and improve any undesirable behaviors when appropriate. AI systems should also have provisions to avoid undesirable persuasion and manipulation. Finally, appropriate mechanisms need to be developed to enable AI systems to act responsibly and ethically, to convey to users the acceptance of responsibility, and to engender and uphold people's trust.

The report's next section contains motivating vignettes for each of the societal drivers, highlighting the research required in each of these four areas. We then discuss the challenge areas in more detail, posing both stretch goals for the 2040 time frame and milestones to track progress along the way.

### 3.2.2 SOCIETAL DRIVERS FOR MEANINGFUL INTERACTION WITH AI SYSTEMS

The vignettes below illustrate the impacts across human society that the proposed research could enable by the 2040 time frame.

**ENHANCE HEALTH AND QUALITY OF LIFE**

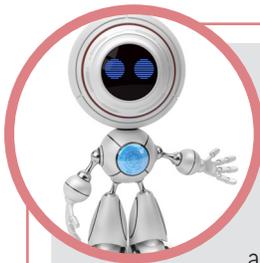

**Vignette 9**

Andrei is a precocious 4-year-old who has had difficulty making friends in the neighborhood and complains of interrupted sleep. Their home AI system, embodied as a robot called Nishi and responsible for managing many household duties, has performed routine screenings for ADHD, ASD, and dyslexia based on Andrei's behavior, play patterns, and sleeping patterns (inferred from a wearable activity tracker and in-home monitors). Since Andrei's behavior included traits of autism spectrum disorder, his parents authorized remote screening by a specialist, Dr. Marie, which led to an in-person visit. During this exam, an automated note-taking system records their conversation and highlights specific phrases that indicate medically relevant information and changes from previous checkups. It also notified Dr. Marie of a potential drug interaction correlated with poor sleep, reported in a research study published the previous month.

Andrei now receives an hour of personalized social skills training each day from Nishi, using prescribed therapy games to help Andrei and his parents relate to each other. Nishi also summarizes Andrei's progress to his parents and the doctors, providing personalized views to each physician, highlighting the medical details each one might deem important. The physicians confer to configure personalized updates for the therapy robot.

**Research Challenges.** Realizing this vision of improved health care will require a number of breakthroughs in the four challenge areas:

**1. Integrating Diverse Interaction Channels:** Whether conversing with the elderly, a child with emotional challenges or a patient from any walk of life, AI systems must be able to handle a wide diversity of human speech. As we develop systems that use vision, speech, and gesture, we will need a new research ecosystem that makes it possible to create and collect datasets appropriate for these different healthcare settings and that represent all modalities, as well as experimental platforms that enable testing new approaches that were computationally expensive in just one modality. To infer qualities of social interaction and to monitor behavioral health problems such as sleep, interactive systems need to develop new modalities (e.g., facial expressions) and fuse together modalities that have previously not been explored together (e.g., vision, sleep monitors, and speech). With the advent of systems that interact with the elderly to catch cognitive problems and that evaluate a child's ability to socialize, we will need to develop new methods to safeguard their privacy as well as the privacy of those around them.

**2. Enabling Collaborative Interaction:** To interact with a child like Andrei in a way that is understanding and caring, research in AI must enable systems to carry on a more natural interaction than currently possible and must be better capable of recognizing human emotions and motivations. AI systems must safeguard individuals who are not able to adequately care for themselves, acting reliably and ethically when carrying out critical tasks. To enable appropriate social experiences for Andrei, the AI system must be able to quantify and evaluate Andrei's experiences and progress.

**3. Supporting Interactions Between People:** One of the robot's goals is to connect Andrei with his neighborhood community, building stronger social ties. Nishi watches in the background as Andrei plays with his friend, but monitors their interaction and may suggest a new activity when discord occurs. The AI system also facilitates collaboration between Andrei's pediatrician and an ADHD specialist, providing different summaries of Andrei's health records corresponding to the key aspects that each specialist requires. Similarly, the AI system tracks recommendations given by different physicians to ensure consistent care and to follow-up on treatment plans.

**4. Making AI Systems Trustworthy:** To fully take advantage of what interactive AI systems can offer, patients must come to trust the AI systems assisting them. This can only happen if AI systems can explain their actions and suggestions, and understand how to seek the trust of their users.

## ACCELERATE SCIENTIFIC DISCOVERY

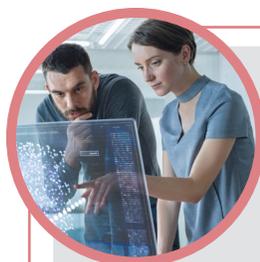

**Vignette 10**

Charlie and her distributed team of top-notch materials scientists have a goal of identifying a material with certain properties, obeying various constraints that will render the material practical and cost-efficient.

Charlie manages the project using an AI system that is connected to robotic lab equipment. The AI system makes it easy for Charlie to articulate research goals, hypotheses, and experiment designs. Most experiments are performed automatically with results recorded for full reproducibility. In addition, the AI system tracks inventory and orders replacement materials, tracking the resulting orders. The AI system can mine the literature on materials, gathering information about past experiments and reporting the results in a way that is consistent with Charlie's existing database of past experiments and results. Charlie directs the AI system to find possible patterns in the data and works with it to conjecture causal relationships between factors and outcomes, rather than simply correlative

> **Vignette 10 Continued**
>
> relationships. The AI system also performs a type of optimal experimental design to offer options to the team, carries out the next experiment, given constraints and directives from the scientists. The AI system estimates both an experiment's cost and its possible benefit in the context of the latest published results, providing clear reasons for its uncertainty estimates and accepting redirection and priorities from Charlie's team.
>
> When discussing their project, Charlie's team uses a mixture of online platforms, from text messages and video to immersive augmented and virtual reality. These platforms are moderated by another AI system that supports discussions. This AI system is designed to be an active participant in the conversation that helps synthesize their different contributions into a consensus on the team's goal. The modeled goal is constantly updated and provided in an interpretable way that is consistent with the scientific language of the community. Charlie notices that an important constraint regarding the desired properties of the material is missing from this goal model and she directly updates it. Thanks to dynamic visualizations and an intuitive control interface, her teammate notices that the AI system is using a weak causal relationship and suggests to the system to ignore it. Together the team decides on the next course of action and the resulting experiment proves successful. The IMS drafts many sections of the paper submission, including experimental methods and related work, and the final paper is a landmark.

**Research Challenges.** Achieving this vision requires a number of breakthroughs in the four challenge areas:

**1. Integrating Diverse Interaction Channels:** The AI system renders 3D models of the proposed materials, sometimes printing them out and testing their material properties directly and comparing the properties to those predicted by the simulation. The AI system also facilitates virtual meetings, so that the scientists can interact fluently despite being in different locations. Effective use of AR and VR for visualization and collaboration requires coordination of voice, gestural, and other inputs, along with inferring and acting on user intentionality. The AI system must understand natural input and convert user actions between different media (flat screens and video or immersive 3D AR and VR).

**2. Enabling Collaborative Interaction:** The AI system must be able to use knowledge in this role. In synthesizing contributions, identifying related work, and dividing work among the team, the system must take context into account, communicate naturally, model the team members' mental states, learn to adapt to expectations, and support complex teamwork. In its role of helping the team, the AI system must be able to explain its decisions and to gain the trust of team members.

**3. Supporting Interactions Between People:** As an active participant in the team members' conversations, the AI system helps build consensus from possibly diverging observations shared by all team members, and helps design metrics based on its knowledge of the most recent metrics developed by that scientific community. The AI system is tasked with using interactive structures to produce documents (e.g., a task list or a consensus on team goals).

**4. Making AI Systems Trustworthy:** The AI system must be able to explain its rationale to the team members (e.g., why it suggested certain experiments) and it must make its level of uncertainty clear in a way that team members can truly understand. Given the critical outcomes of experimentation, the system must act reliably. Science requires a high standard of ethics and must adhere to IRB regulations and thus, the system must be able to address ethical concerns.

**LIFELONG UNIVERSAL ACCESS TO COMPELLING EDUCATION AND TRAINING**

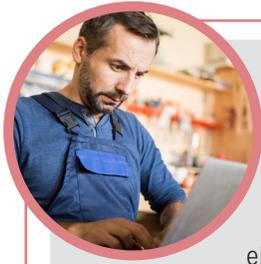

**Vignette 11**

Joe is a worker who was laid off in a company restructuring. He wants to retrain, but needs income in order to support his family and cannot afford to embark on full-time education. A free AI system helps him plan for career change—what is a feasible job he could take that would either build the skills he needs along the way or would pay the bills while giving flexibility to study and advance his career?

To explore his short- and long-term career opportunities, Joe navigates to an interactive AI system and describes his skills and interests. The system visualizes a number of possible career paths for him, including both short- and long-term steps he can pursue to make progress on those paths. The system is sensitive to his individual experiences and goals and builds models of what he wants from a new job. For each path, the system determines missing skills and good first steps to take in achieving that career. In some paths, Joe may enroll in online classes where human instructors supplemented with AI assistants that can provide personal tutoring.

Where a human teacher leads the instruction, the AI system tracks student engagement and attention; prompts the teacher to engage with each student as necessary; assists the instructor in giving feedback to students, even on open-ended work; gives feedback to the teacher on each student's progress, mastery, and challenges; and intervenes with students to increase student retention in the classes. It observes communication between students and the teacher to understand when they agree and disagree. Students are matched together efficiently to collaborate on projects according to their interests, skills, and collaboration style.

Where students must practice material on their own, Joe may use virtual reality to develop some skills. Joe uses an online simulator to practice interacting with simulated customers, and AI helps provide explainable feedback on how he is doing and how he can improve.

The interactive system helps match Joe to open jobs that fit his skills and interests, matching constraints he may have. Joe works with the AI system to identify additional challenges of his new job where he could use further training. The AI system allows Joe to easily and continuously learn and practice a variety of new skills, even including those that require teamwork with other employees.

**Research Challenges.** Achieving this vision requires a number of breakthroughs in the four challenge areas:

**1. Integrating Diverse Interaction Channels:** A diversity of students may use the AI system, and their speech must be understood in order to identify their learning goals and evaluate their progress. Extensive datasets that include video, speech, gestures, and other modalities are required for developing AI systems that give on-the-job guidance in real-world situations and that accurately allow students to practice what they learn using virtual reality or augmented reality environments. User models are necessary for understanding and evaluating the interactions students have during their training and providing feedback. Those user models may involve students' movements, vision, speech, facial expressions, etc.

**2. Enabling Collaborative Interaction:** Common sense is required for the system to recommend career paths (e.g., Joe cannot simply become a manager in a field where he has no experience; he must take an entry-level job in that field) and for on-the-job training in which students must interact with the world or with others. An understanding of context is required for the system to recommend career paths (in this case, students' background and prior interactions with the system), and for understanding students' social and emotional backgrounds when giving teachers feedback on their mastery and challenges. Natural interaction is required when the system interacts with the user to discuss career paths and the users' constraints, and during training when students have questions or must interact with others. Modeling students' mental states is required to model their learning: the system must know what students figure out and do not figure out. Modeling students' emotional states is required to understand their engagement and progress in their classes and training, in career change settings, and when the system is attempting to predict when students may withdraw (so that it can monitor and intervene to increase retention). These models may need to be built with limited data about the students. Personalization and adapting to expectations is required to enable personalized content generation and curriculum generation (in order to select the best materials and best way to convey them for each student). An understanding of human experience is required for the AI systems described to understand users' desires in career paths and their engagement with or enjoyment of training or studying. If multiple students are involved (e.g., in a group project, or a team job), the AI system may need to facilitate complex teamwork among them.

**3. Supporting Interactions Between People:** AI systems that facilitate constructive online collaboration may be required when multiple students interact to complete work (e.g., a group project). Gathering and linking social media data to real-world experiences may be required when giving users advice about potential career paths or developing practice problems.

**4. Making AI Systems Trustworthy:** The AI systems described must be trustworthy and explainable so that students learn effectively, in classes or on the job: If Joe is practicing patient interaction, the simulator must be able to explain why his bedside manner is or is not ideal. The feedback that teachers are given on students' engagement and progress must also be explainable for effective teaching. Since the AI systems described may be helping people make personally important decisions about their life paths, or guiding them through critical jobs, ethically designed systems that protect against and do not perform malicious manipulation are necessary.

**AMPLIFY BUSINESS INNOVATION AND COMPETITIVENESS**

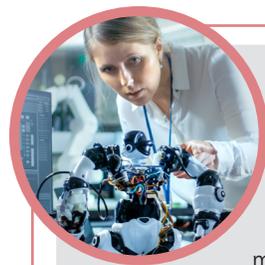

**Vignette 12**

Hollis runs a small online business, where she sells customized personal devices and customized robots, which she designs and builds on demand. Some objects are aesthetic, such as integrating light and motion sensors with embedded LED lighting to add responsiveness to jewelry; others are more functional, such as customized wristbands that integrate her designs with medical sensors and small displays.

An interactive AI systems allows Hollis to rapidly develop specialized products for her customers, enabling new business opportunities. Hollis can focus on her creative skills, while the AI system monitors her storefront and manages the manufacturing and logistics of her business. Her online storefront shows existing designs that can be customized with different electronics: The AI system ensures requested customization will work (taking into account size, layout, power and other constraints), and notifies Hollis only when a customer requests something that requires adjusting the base designs. The AI system uses smart analytics to inform Hollis about product trends by integrating data from her storefront with opinion data mined from relevant blogs as well as knowledge about what sorts of

> **Vignette 12 Continued**
>
> electronics are popular and available from the suppliers in her supply chain. Hollis uses this information to explore new designs, using another AI service to select appropriate focus groups and collect feedback.
>
> Hollis creates the company's production designs in 3D using direct manipulation in an augmented reality display. The AI system enforces practical constraints in real time and can be taught new guidelines and skills during operation using a combination of demonstration (e.g., showing the system how to do it) and verbal commands (e.g., "Repeat those actions on all sides of the object.")
>
> Hollis's products are made on demand by a set of manufacturers around the world, each with different materials and electronics chosen through automated negotiations that suggest the right supplier for the part. The AI system considers price and build times, but also checks the reputations of the manufacturer and of the equipment they use. The AI system can also be directed to summarize other factors, such as expected shipping times at different times of year, social issues such as political upheaval and working conditions, and so on. The AI system presents its final suggestions to Hollis using an automatically synthesized graphical rendition that highlights the pros and cons of each choice. After asking clarification questions, Hollis makes the final decisions with confidence, knowing that the AI system continuously monitors pricing and availability conditions, alerting Hollis only when there are significant changes to consider. Occasionally, Hollis needs to communicate directly with a supplier; in these situations, the AI system translates between languages during real-time videoconferencing, and advises Hollis during negotiation and conflict resolution.

**Research Challenges.** Achieving this vision requires a number of breakthroughs in the four challenge areas:

**1. Integrating Diverse Interaction Channels:** Hollis's AI system finds video blogs depicting users interacting with her products, extracts facial expressions and analyzes acoustic emotional cues to deduce user sentiment. When Hollis uses the design program, the AI system relies on several sensors to augment her creative experience, including haptic feedback and touching sensors. The AI system provides the ability to migrate from augmented to virtual reality, allowing Hollis to see, touch, feel, and wear new prototypes or alternative designs, which the AI system can create in short time. The AI system also supports natural speech interaction between Hollis and suppliers around the world, breaking language barriers. The system handles accent variations from specific locations and analyzes nonverbal behaviors for their meaning when interacting with people from other cultures, avoiding misunderstandings.

**2. Enabling Collaborative Interaction:** The AI system should have an understanding of the sort of work Hollis is engaged in at any moment (based on watching and learning, but also leveraging her calendar, and so forth), so that it interrupts her at appropriate times. Different people have different preferences (e.g., when Hollis considers different suppliers), which need to be learned by AI systems. AI systems must also support natural cross-cultural communication and interaction (e.g., when responding to customers and negotiating with suppliers).

**3. Supporting Interactions Between People:** In order to recommend suppliers, the AI system must estimate reliability from a variety of online traces, including noisy or fraudulent reviews. In order to help predict trends and preferences for Hollis's products, the AI system must link social media signals to real-world processes.

**4. Making AI Systems Trustworthy:** In order for Hollis to trust the AI system, it must be able to explain its reasoning, answer questions about which information it is taking into account and quantify its uncertainty. The AI system also must understand that it must treat customers and suppliers without misleading them and use customer data ethically. All these skills require significant progress over current methods.

**SOCIAL JUSTICE AND POLICY**

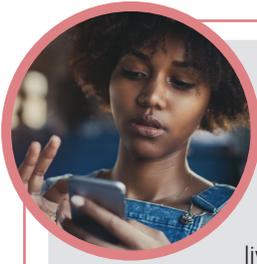

**Vignette 13**

Susan comes home to an eviction notice on her door, which gives her 24 hours to remove all of her possessions from her apartment. She has had problems with her landlord and feels that the eviction is unjustified and possibly illegal. But her immediate concern is to look for another place to live, so she asks her AI assistant to search for short-term housing in her city. The AI system returns a set of results, but in a side channel (visual or auditory) asks if something is wrong with her current apartment, since it knows that her lease runs for another eight months. She tells the system that she was evicted.

One possible path for the AI system is to provide legal support. The system queries relevant tenant rights law for her city and state. The result is a set of legal documents, which are long and difficult for non-experts to read. The AI system therefore identifies critical parts of each document, specifically relating to eviction. Susan can highlight parts that she doesn't understand, and ask for clarification; this can be done either by generating explanations of an appropriate level of technicality and dialect for Susan, or it can identify snippets of relevant documents. The AI system offers to connect with her with a set of legal aid resources in the morning. The AI system also provides Susan with a set of just-in-time social connections to individuals who have been in a similar situation and are willing to share their experiences. This group can include individuals who have fought an eviction (successfully or otherwise), individuals who have used short-term housing in the same area, others who have used free legal aid services, and other current and former tenants of the same landlord. Keeping in mind that these individuals may not have shared these experiences widely, the AI system needs to decide when and under what conditions to share what information to support Susan's interaction with the others. In addition to person-to-person connections, the AI system can aggregate social media content from these individuals related to their experiences, which will create a set of resources to help Susan deal with her situation both practically and emotionally.

**Research Challenges.** Achieving this vision requires breakthroughs in all four technical areas.

**1. Integrating Diverse Interaction Channels:** Susan might interact with the system partially via speech (necessitating accurate speech recognition for a diverse range of speakers and across many contexts), and the system may make additional inferences about the environment and context via video input. Conversely, it can offer information to Susan across multiple modalities, including visual and audio display when appropriate in context. Multimodal integration will support more advanced capabilities, such as a combination of speech and gesture to identify salient features of the environment, such as lack of upkeep of the apartment.

**2. Enabling Collaborative Interaction:** The AI system works collaboratively with Susan to quickly identify solutions to her situation. There are important contextual factors, and the system must recognize the priority of this situation with respect to other plans and goals that may have been previously indicated by Susan. There is also an affective component to the interaction: Eviction is likely to provoke stress, which may affect the user's decisions as well as her interactions with the system. The system should account for this and adapt its behavior accordingly.

**3. People Interacting Online:** Several of the system's goals require capabilities relating to online interaction. The AI system must identify factual claims about the legal situation and verify them against external sources. This requires not only finding trusted resources that support or reject existing factual claims, but ensuring that those resources are germane to the specific situation at hand: for example, that the laws and regulations pertain to the correct jurisdiction and type of housing. The system should also be able to distinguish legal consensus from theories or strategies that only some community members believe will work. In proposing just-in-time social connections, the system should optimize for constructive and civil discourse. The scenario also pertains to collaborative creation: The system may help non-experts to assemble a set of guidelines for dealing with eviction, while soliciting targeted advice. The system could identify and help to bridge linguistic gaps between experts and non-experts, including multilingual support.

**4. Making AI Systems Trustworthy:** Like a trusted human advisor, we expect the system to act entirely on behalf of the user, without the possibility of external manipulation. This includes blocking the acquisition of data about the situation that might impact the user negatively. In this case, because eviction has potential long-term social consequences, the system should not leak information about the situation without explicit user approval. The scenario is relatively high-stakes and time-sensitive, so the AI system's advice should be appropriately scoped and have appropriate justifications. In particular, the advice should be grounded in original sources (laws and regulations), and the connection between these sources and the advice should be made clear. This may necessitate reasoning about linguistic transformations that preserve meaning while helping readers with variable levels of legal experience, formal education, and reading ability. Any risks should be accurately explained, and system uncertainty must be communicated to the user.

### 3.2.3 TECHNICAL CHALLENGES FOR MEANINGFUL INTERACTION WITH AI SYSTEMS

We group the technical challenges into four high-level capabilities: integrating diverse Interaction channels, enabling collaborative interaction, supporting better interactions between people, and making AI systems trustworthy. While people routinely use two or more modalities (e.g., voice, gestures and facial expressions) when communicating with each other, today's AI systems cannot do this in a robust and flexible manner. Similarly, today's AI systems aren't collaborative; They poorly model the context underlying a human's user's desires and intentions, the human's emotional state, and the potentially diverging objectives of different users. Increasingly, humans are using technology to mediate interactions with other humans, whether through social media or telepresence; There are many opportunities for AI systems to facilitate these interactions if we can solve key technical challenges. Today's AI systems perform well in low-stakes domains but are often unreliable in high-stakes environments such as medical diagnosis or judicial support; to realize the full potential benefits of AI, we need to make these systems more trustworthy and easier to control. We elaborate these challenges below.

**Integrating Diverse Interaction Channels**

Multi-channel inputs and outputs are everywhere, where wearable sensors can be interconnected with other computing devices, leveraging the advances in the Internet of Things. Day-to-day environments are replete with multimodal sensors in everyday consumer items: For example, doorbells now have video cameras, in-home personal voice assistants contain microphones, and thermostats are equipped with proximity sensors. People wear suites of sensors, such as heart rate monitors on their smart watches and accelerometers on their smartphones, which provide continuous, high resolution, and real-time multimodal data. Information of all kinds is being presented multimodally, as well. For example, smartphone apps use both vibration and text to send messages to users. Online communication, even from traditionally text-based organizations such as news agencies, is increasingly in the form of video and audio in addition to written media. The use of multiple channels for inputs and outputs is increasing as data bandwidth becomes cheaper and sensors become more integrated into common products. People and companies are already using multimodal data to derive insights about sleep, physical activity, traffic, and other information.

In addition, multimodal interaction provides a natural, effective, and engaging way to communicate with AI systems. Jointly considering diverse interaction channels increases the robustness and accuracy of AI systems, providing unique opportunities that cannot be accomplished by considering single modalities. Information incorrectly inferred from one channel can be corrected by another modality, leveraging the complementary nature of multimodal data. For example, compared to a decade ago, voice-based systems have improved in the ability to recognize user input accurately. Improvement in these technologies has led to more mainstream use of some channels (e.g., voice) in systems designed for leisure and entertainment but also work and activities of daily living. The advances in automatic speech recognition and the success of speech-based commercial products that are now ubiquitous in our lives (e.g., watches, smartphones, home personal assistants) have created a paradigm shift in multi-channel interaction. These devices have several heterogeneous sensors that are promoting applications relying on multimodal interactions.

The advances in multimodal sensors has also lead to better models. Over the last years, we have seen emerging algorithmic development for multi-channel processing with machine learning architectures that create representations that can be shared between modalities. Current research efforts have often focused on merging a few modalities, with prominent combinations of audio plus video (e.g., audio-visual automatic speech recognition and audio-visual emotion recognition), text plus images (visual question answering), and text plus video (semantic labeling, image/video retrieval).

At the same time, multimodal technology may be inaccessible to many people. For example, voice activated devices do not work well for people with speech impediments, visual output cannot be easily parsed by a screen reader for the blind, and speech output is unusable by people who are deaf. The fusion of more than two channels is also a challenge, especially when the data is incomplete (e.g., missing facial features due to occlusions).

In the remainder of this section, we identify major challenges for interaction using multiple modalities.

**Handling Diversity of Human Ability and Context**

Despite progress, many input modalities do not account for diverse human abilities limiting access to some who desire to use these systems or may benefit from them. Voice-based systems still find it challenging to recognize input from users with non-standard speech (e.g., speech impediments, deaf speech, different dialects) limiting access for some users. The ability to handle languages other than English is also critical in today's world, even enabling the possibility of translating between different languages, each spoken by a different conversation participant. Furthermore, some modalities have become commonplace (visual, speech, audio), while others such as haptics, proprioception, and other sensory inputs are underexplored. Multimodal systems will also need to take into account the context in which the signals are created to reach correct conclusions. Context includes the particular environment, interaction partners, prior knowledge, motivation, preferences, and a host of other information that dictates how the data should be interpreted. In the future, single modalities must continue to improve their capabilities for handling diverse human abilities. In parallel, multimodal systems must also take into consideration ways to leverage context as well as other underexplored modalities as opportunities for increasing access and improving interactions.

**Stretch goals:** By 2040, natural communication with virtual and physically embodied AI systems will be achieved through multiple modalities in both task-specific and open domains. Milestones along this path include—

**Milestones**

**5 years:** Multimodal systems will combine multisensory data (e.g., speech and a perceived visual environment) to infer communicative goals and to capture shared experience in constrained environments, communicating shared perceptual experience in language (i.e., generating language descriptions based on first-person view and grounding language to objects using well-trained computer vision models). Human-machine dialog will explicitly model common ground and shared intentionality with respect to the dynamic physical environment in specific domains.

**10 years:** Multimodal systems will capture long-term past joint experiences, using persistent memory and constructed models of past interactions.

**15 years:** Multimodal communicative actions will be generated for achieving shared communication goals. This will include capabilities for easily extending and integrating additional communication channels and task-specific languages aimed at improving human-AI communication efficiency.

**Explainable, Interpretable, and Data-Limited Multimodal Interfaces**
Transparency in AI systems has always been a concern among some in the AI community, beginning with early work on explanation generation for expert systems. Due to the pervasiveness of AI systems in recent years, there has again been increased interest in developing explainable, transparent, and trustworthy AI systems that also respect user privacy needs. Often, multimodal systems are developed to collect as much data as possible about a person using different modalities to improve accuracy or to supplement the limitations of one or more modalities. Some systems may use visual input as an alternative modality to supplement the limitations of speech as an input. Therefore, a system could collect quite a bit of data about an end user through the use of different modalities. It may be difficult, however, for end users to fully understand what data is collected, how it is collected, or how it is being used to support their interactions with the AI system. Some systems that collect data about users attempt to notify users about the data collected about them. However, the choice to opt out of having certain data collected often prohibits use of the system (i.e., if the user does not allow the system to use the data it desires, the user cannot use the system). There is a need to identify methods for creating multimodal systems that allow users to easily understand the data being collected about them, and how that data is used. Systems must also provide enough flexibility to allow users to customize the manner and time in which data about them is collected as well as the kind and amount of data that is shared in a way that aligns with their preferences. AI systems must consider options for end users to customize their interactions to provide access without having to abandon the system completely.

**Stretch goals:** By 2040, customizable multimodal systems that allow end users to choose how much privacy they retain while allowing for different levels of access to the system. Milestones along this path include—

**Milestones**
**5 years:** Methods to explain what data about a person is collected and necessary for interaction.

**10 years:** Auditing algorithms to efficiently determine how channels of streaming data (or the lack thereof) affects AI prediction capabilities, while ensuring privacy.

**15 years**: Methods for ensuring privacy for systems that use multiple modalities as input, including speech, vision, and other sensor input.

**Plug and Play Multimodal Sensor Fusion**
Current solutions for multimodal interactions are not flexible, mainly because: 1) they are often restricted to a small number of interaction channels, 2) they cannot handle missing information, and 3) they assume that the target sensors are predefined, where new (possibly unforeseen) modalities cannot be easily integrated. Yet new input devices are coming: haptic interfaces, brain implants, and more. An important technical problem is increasing the flexibility of multi-channel interaction systems to handle such new and diverse modalities. We envision plug and play sensor fusion approaches that can address these challenges.

Existing fusion algorithms often consider only two or three modalities. With new advances in multimodal sensors, it is important to design machine learning solutions that can robustly and effectively combine heterogeneous data coming from multiple sensors. When adding new channels, current solutions to fuse sensors exponentially increase the model parameter space, which leads to models that either are undertrained due to limited data, or cannot capture the true relationships across modalities. New multimodal fusing solutions should properly scale as more modalities are added. They should also provide capabilities to synchronize data streams, regardless of their sampling rates.

An important technical challenge is to design fusion algorithms that are robust even when data from some modalities are incomplete or absent. There are several scenarios that can lead to missing or incomplete data from modalities. From an accessibility perspective, individuals with physical impairment may not be able to operate or use certain modalities. The users may also decline the use of a sensor to protect their privacy. The placement of the sensors and the environment may temporarily lead to missing information from the users (e.g., occlusion, out of field of vision, and acoustic noise). When this happen, it is important that the next generation of AI systems be able to leverage the remaining modalities to complete their task. This technical challenge requires more effective leverage of the complementary nature of heterogeneous data coming from different sensors and design strategies to train and test models with partial information.

Current multi-channel fusion approaches assume that the sensors are predefined. We envision models that are modality agnostic, so they can work with whatever channels are available. Advances in this area will allow systems to scale to different modalities based on user preferences. The fusion formulation should enable replacement of current sensors with better modalities over time by creating models that easily integrate with new sensors. We envision data agnostic modules that can be interchangeably implemented with different sensors. By defining data modules, these systems can also scale by adding new, possibly unforeseen modalities.

**Stretch goals:** By 2040, plug and play multimodal sensor fusion systems will adapt to individual users, seamlessly handle the addition of new modalities as the developed, and scale to different modalities based on user preferences. Milestones along this path include—

**Milestones**
**5 years:** Multimodal system can learn from realistic datasets, exercising multimodal capabilities in the context of a range of real-world tasks.

**10 years**: Voice-based interfaces will recognize inputs from user with diverse speech patterns, such as deaf speakers, strong accents, and rare dialects.

**15 years**: Systems using three or more modalities, fusing input from voice, vision, and sensors allow for natural interaction.

**Privacy Preservation for Multimodal Data**
Multimodal interaction algorithms leverage sensors such as cameras and microphones to capture inputs like speech and gesture, but the data captured by these sensors can also capture and reveal sensitive information and the user, their environment, and bystanders. Systems should be designed such that the potentially sensitive data collected from these sensors is not shared further than it needs to be, and neither retained nor used for secondary purposes without permission. Some data is needed locally for real-time interaction (e.g., watching someone's hands to detect gestures such as pointing) and can be discarded immediately after use. Other data needs to be used to train personal models to support long-term interaction, and some might be used to train or refine more global models and algorithms. When possible, sensitive data could be processed in the platform (e.g., sensor or web browser) where it is collected, which would then return processed results (e.g., removing parts of video that are not close to interaction areas, or removing faces of bystanders) to applications. Systems that run in the cloud should separate user-identifiable data from anonymized data and try to use algorithms or techniques that can operate with as little access to sensitive data as possible.

**Stretch goals:** By 2040, multimodal interfaces will respect user privacy. Milestones along this path include—

**Milestones**
**5 years:** Privacy preserved for systems using voice and vision as input.

**10 years**: Multimodal interfaces will work across different levels of privacy and fidelity.

**15 years**: AI systems will ensure privacy preservation through differential privacy, running only in trusted code, or running models and fusion locally or at the edge.

**Collaborative Interaction**

In the last decade, we have witnessed a broad adoption of virtual personal assistants such as Apple's Siri, Google's Assistant and Amazon's Alexa. These systems aim to help users to accomplish simple tasks, such as checking the weather, looking up directions, and playing music, through multi-turn interactions. Furthermore, they aim to detect and answer a user's factual questions. However, they are still far from being as efficient as human assistants. Furthermore, these systems are designed domain-by-domain (for example, weather or music), resulting in good coverage of few commonly used domains, but gaps for the much larger set of less common topics. Although there are a few examples of use of context (for example, previous turns of a conversation, identity of the user), these behaviors are laboriously engineered by hand and the broader context is not considered. Except a few example cases, these systems do not integrate common sense (for example, "later today" can mean 4-5 p.m. in the context of business meetings, but 7-8 p.m. in the context of dinner discussions). In general, today's AI systems use special purpose data structures rather than general representations of meaning; such representations have been studied and challenges highlighted, but we don't yet have practical and effective means for representing the content of utterances. In addition, today's AI systems are largely unable to converse with two or more people in a group.

There has also been significant research undertaken in the space of representing, detecting, and understanding human emotions in fields such as computer vision, speech, and language processing. Different representations of emotions have been proposed, including discrete categories (e.g., joy or fear) and the continuum space of valence and activation. The area of sentiment and opinion analysis has also received significant attention from both academia and industry. We have seen significant advances in some areas, for instance systems are now available that can reliably detect the sentiment (positive or negative polarity) of text for certain domains, such as product or movie reviews; we have deployed computer vision tools that can detect smiling or frowning; we have speech models to sense the presence of anger. Progress has also been made in the use of multiple modalities for emotion and sentiment detection, and the fusion of different modalities is an active area of exploration. There is, however still much to be done to understand how to best represent emotions and how to detect emotions of different granularities and in different contexts. Moreover, the research area of emotion generation is just starting, and significant advances will have to be made to develop systems that can produce emotions and empathetic behaviors.

In order to make our AI systems be better partners, they must be collaborative, which requires the following technical advances.

**Enabling Natural Interaction**

Many of today's interactions with artificial intelligence technologies are limited, stilted, or awkward; they frequently lack the richness, elegance, and ease of typical interactions between people. As AI technologies become more commonplace in our daily lives, there is a need for these technologies to allow for more natural interactions that match the expectations of the human users. Much of the work in this area in the past has focused on the recognition of subtle perceptual cues (such as facial expressions or gaze direction) or on the production of fluid, naturalistic behavior (such as human-sounding speech or smooth, coherent gestures). While most of our systems are currently designed for short, transactional interactions, we must push toward building systems that support long-term interactions, remembering the content and context of previous interactions and shaping current interactions based on this knowledge, adjusting the expectations and behavior of the system as experience grows, and working to maintain long-term stability of behavior and personality. While most of our current systems are reactive (e.g., providing answers to posed questions or responding when tasked) AI systems of the future must be proactive in their activity by initiating conversations with relevant information, asking questions, and actively maintaining trust and positive social relationships.

Finally, in order to enable natural and beneficial interaction and collaboration with people, AI systems should be able to change their behavior based on people's expectations or state. AI systems should be able to provide personalized experiences that provide high levels of improvements on people's success and experiences. Personalization and adaptation capabilities are enabled through an understanding of context and by modeling of users, including their mental models (both described below).

**Stretch goals:** By 2040, AI Systems that can have extended meaningful, personalized conversation taking advantage of the context of the conversation. Milestones along this path include—

**Milestones**

**5 years:** AI systems that can have an extended dialog over a single topic.

**10 years:** AI systems that can have extended, personalized dialogs over a single topic, using context.

**15 years:** AI systems that can shift topics and can return to previous dialogs, keeping track of conversational state and conversational participant's intentions.

### Alignment with Human Values and Social Norms

AI systems must incorporate the social norms, values, and context that are hallmarks of human interaction. Systems must begin to understand properties about social relationships (e.g., knowing that I am more likely to trust my spouse than a stranger), about objects and ownership (e.g., knowing that it might be acceptable to take and use my pen but not my coffee mug), and about social roles and responsibilities (e.g., knowing that the answer to the question "what are you doing?" should be answered differently when asked by my employer than by my co-worker).

Alignment of AI systems with human values and norms is necessary to ensure that they behave ethically and in our interests. Human society will need to enact guidelines, policies, and regulations that can address issues raised by the use of AI systems, such as ethical standards that regulate conduct. These guidelines must take into account the impact of the actions in the context of the particular use of a given AI system, including potential risks, benefits, harms, and costs, and will identify the responsibilities of decision makers and the rights of humans. Currently, AI systems do not incorporate the complex ethical and commonsense reasoning capabilities that are needed to reliably and flexibly exhibit ethical behavior in a wide variety of interaction and decision making situations. In contrast, future AI systems will potentially be capable of encouraging humans toward ethically acceptable and obligatory behavior.

**Stretch goals:** By 2040, AI systems will constantly and reliably reason about the ethical implications and social norm adherence of actions they engage in or observe. They will plan and adjust their own behavior accordingly, and will act to prevent others, both human and AI systems, from engaging in unethical behavior. In situations in which the ethical implications are complex and do not lend themselves to simple ethical acceptability verdicts, AI systems will engage in thoughtful conversations about these complexities. Milestones along this path include—

**Milestones**

**5 years:** AI systems will take ethical requirements and contextual clues into consideration when deciding how to pursue their goals.

**10 years:** AI systems will effectively reason about how to behave when faced with conflicting social norms and ethical acceptability tradeoffs.

**15 years:** AI systems will reason broadly about the ethical implications of their actions and the actions of others, perhaps more efficiently than humans.

### Modeling and Communicating Mental States

Humans are capable of creating a mental model of intentions, beliefs, desires, and expectations of others. Understanding what someone else is thinking makes it easier to interact and collaborate with them. As AI systems become more widely accessible, it is important for them to accurately model mental states of their users and enable frictionless interactions.

Current, widely used AI systems have only very limited modeling a user's mental states. They mainly focus on determining which of a small pre-defined set of possible intentions the user might have. Furthermore, these systems aim to track user's intentions throughout conversations, the user specifies more information and sometimes change their mind. However, even in a simple interaction, the possible number of mental states is much larger than what these systems can model. Furthermore, as a user does not necessarily know about the set of intentions that were pre-defined by the AI system builders, they often try to interact with these systems with different intentions. The mismatch between what builders include in these systems and a user expects to result from inefficient interactions ends up limiting these interactions to a few simple use cases and prevents the application of these AI systems to many scenarios and domains. While the work on modeling and tracking users' intentions should continue to improve in these limited scenarios, future AI systems need to extend to the scale and complexity of real-world intentions.

To support collaboration, AI systems need to be able to reason and generate persuasive arguments that are not only relevant to the current situational context, but are also tailored to individual users based on shared personal experience. Argumentation and negotiation methods could be learned through extracting and learning from text.

**Stretch goals:** By 2040, AI systems will work collaboratively with multiple human and AI-based teammates using verbal and nonverbal communications, engage in mixed-initiative interactions, have shared mental models, be able to predict human actions and emotions, and perform collaborative planning and negotiation in support of shared goals. Milestones along this path include—

**Milestones**
**5 years:** Persuasive arguments will be generated based on shared experiences and goals.

**10 years:** Human-machine teams will regularly tackle hard problems collaboratively, using defined measures of engagement and perseverance in human-machine teams. AI systems will communicate with human teammates in task-specific contexts, using verbal and nonverbal modalities to establish shared mental models.

**15 years:** AI systems will exhibit persistent memory, enabling them to serve as long-term teammates. They will take both reactive and proactive actions to complement their human teammates' rational and emotional actions, and will communicate in natural language dialogs to plan, re-plan, and negotiate alternatives to support the team's goals.

**Modeling and Communicating Emotions**
Building systems that can represent, detect, and understand human emotions is a major challenge. Emotions must be considered in context—including social context, spatial and temporal context, and the context provided by the background of those participating in an interaction. Despite multiple proposed representations of emotions to date, there is still no agreement on what are the ideal representations that would fully cover the multiple facets of human emotion. The detection and tracking of emotions is also challenging, and it will require significant advances in data construction, unimodal and multimodal algorithms, and temporal models. The understanding of human emotions will assume reliable context representations, and the ability of the algorithms to account for the variability brought by different contexts.

People often reciprocate in their emotional expressions. Empathetic communication forms the ties of our society—both in personal settings and in more formal settings (e.g., patient-doctor relations, student-instructor interaction). Our expectation is that the people we communicate with will understand and eventually respond to our emotions: if we are sad, we expect understanding and encouragement; if we are happy, we expect similarly joyful reactions. Yet, most current AI systems lack the ability to exhibit emotions, and even more lack the ability to create empathetic reactions. Thus, significant research will need to be devoted to emotion generation, coupled with methods for emotion understanding. Since human emotions arise in the context of experiences that a person is undergoing, research on emotion must connect with that research for modeling context and common sense. A key technical challenge is how to model human experiences in a way that the AI system can assess its quality. This is difficult because experiences are extremely varied, nuanced, and involve subjective judgments. To deal with the subjective aspects of experiences, we need to develop methods for modeling and reasoning with subjective data, whereas in contrast, previous work has focused almost exclusively on objective data. Ultimately, we hope to produce empathetic AI systems that act in a manner that is emotionally congruent with the users they interact with.

In addition to modeling a human's mental state, an AI system must ensure that its behavior is explicable and comprehensible to the humans with whom it is interacting. One way of ensuring explicability is for the AI system to explicitly consider the mental models of humans when planning its own behavior. In some cases, the system may decide that adapting to people's expectations is either infeasible or poses undue burden (cost) on the AI system or the team. For example, a robot in a collaborative search and rescue scenario might find that it is unable to rendezvous with the human partner in the agreed location because of collapsed pathways and debris. In such cases, the AI system should be able to provide an explanation, which involves communicating to the human the need to change the rendezvous point. These explanations need to be tailored to the mental models of the humans.

**Stretch goals:** By 2040, AI systems will reason about the emotional components of encountered situations and about the emotional effects of their actions and the actions of others involved in a wide range of situations, both real and fictional (e.g., stories, films, worst-case scenarios). Milestones along this path include—

**Milestones**
**5 years:** AI systems will reason about how their actions emotionally affect people with whom they interact.

**10 years:** AI systems will predict human emotions based on observing an individual at work, play, or during training, for as little as 30 seconds. When reading a short story, AI systems will construct models of the emotions of the characters in the story and how they change as a consequence of the story's events.

**15 years:** AI systems will reason about complex affect-aware interactions, e.g., anticipating how a person might be impacted by a team interaction, personal history, or particular context.

**Supporting Interactions Between People**
Human social interactions are increasingly conducted through online media and networks. Artificial intelligence can play an important role in facilitating these interactions and in mining online data for insights about social phenomena. While artificial intelligence already has a strong presence in online interaction, there are a number of possibilities for future research. These research possibilities can be structured by the purpose of the online interaction, which can include 1) deliberation, 2) collaborative creation, and 3) social-tie formation.

In deliberative online communication, existing state-of-the-art techniques use supervised machine learning to identify factual claims in text and distinguish statements of fact from opinions. However, like all learning-based approaches to natural language, there is still a question of domain specificity, as well as the development of technology for "low-resource" languages that lack labeled data. Online deliberations also provide valuable data for social scientists who are interested in understanding the dynamics of opinion formation, and we see a trend toward interdisciplinary research that leverages such data. However, online data poses unique challenges, and methodological advances are required to ensure the validity of the resulting inferences.

Artificial intelligence already plays a limited role in collaborative creation. Wikipedia, the online encyclopedia, uses AI bots to prevent vandalism and ensure that content meets existing standards. More advanced capabilities are at the level of research prototypes: for example, recommending collaborators and summarizing the existing literature in an area of scientific research.

Social media analysis is an active area of research, with substantial progress toward commercialization. For example, AI is already widely used to detect objectionable content such as hate speech. However, deployed solutions for content filtering in online media are typically human-machine hybrids, with AI used only as a preliminary filter, reducing the workload of human moderators. Questions remain about whether online platforms and communities can reach consensus about definitions of objectionable content; this is a baseline requirement for the deployment of any artificially intelligent system. AI is also playing an increasingly active role in content generation and manipulation: for example, through predictive text entry and machine translation. We foresee increasingly sophisticated human-machine hybrid technologies in this space, which may ultimately enable new forms of human-human interaction and emergent social structures. In order to realize this vision, we need progress in all three high-level directions. We also foresee more sophisticated machine-machine interactions with AI systems that represent people and that can help to promote new human-human and human-machine modes of interaction, be this with new or strengthened social ties (including coordinated activities), new possibilities for discourse, and new opportunities collaborative creation. Machine-machine interaction is also, in and of itself, an important component of AI interactivity and it is important to continue to develop theory, algorithms, and formalisms for multi-agent systems so that ecosystems of interacting AI systems will achieve intended outcomes and represent successful extensions (and reflections) of human society.

Another active area of current research involves linking online social communication to external phenomena, such as economics, politics, and crisis events. However, existing approaches are typically "one-off" technical solutions that are customized for the event and data sources of interest. We still lack a set of robust, general-purpose methods for jointly reasoning about these very different types of data.

**Improving Deliberative Interactions Between People**
Individuals are increasingly using online platforms to participate in deliberative discussions. This trend provides new opportunities as well as new challenges for new AI systems to utilize and address. Below, we identify and elaborate on four technical challenges.

**Identifying factual claims**
The Web and social media have democratized access to information. However, the decentralized nature of these platforms, the lack of a filtering mechanism, and the impossible challenge of inferring credibility of individuals and information online also enable the widespread diffusion of misinformation and disinformation, threatening to the health of online conversations. Given the growing pace with which misinformation and disinformation spread online and the societal implications of this trend, it is crucial to build AI systems that detect and combat these phenomena. Such systems need to be scalable, accurate, and fair. One important building block here is the detection and validation of factual claims in online data. For instance, given a text (e.g., a post on social media), what are the claims of facts (e.g., "the moon is made of cheese"), and can we check these facts against external resources (e.g., Wikipedia, textbooks, research articles, knowledge bases, etc.)?

Current research in claim detection often relies on fact databases and conventional statistical methods. These solutions have limited accuracy and generalizability—due to the limited scope of fact databases. Future research in claim detection should broaden the application domains, work across different communication modes, and be interactive and responsive. The claim validation step is currently at its infancy. We expect the next 10 years to lead to more accurate and adaptive AI systems. Such systems should have a strong fairness emphasis (e.g., how are the false positives of a false-claim classifier distributed across individuals from different communities?). Furthermore, current systems focus on batch claim validation. Yet in cases of misinformation, timely action is crucial. This requires investment in real-time detection methods.

**Understanding and designing metrics**
Current research on the health of online deliberation has a strong emphasis on civility. There are systems (e.g., the Google Perspective API) that accurately model civility of text data across different domains (e.g., Wikipedia, news commenting systems). However, there are various other qualities beyond civility—identified by social science research and grounded in the principles of deliberation—that are currently unexplored by AI systems. Some of these qualities concern how participants treat one another over the course of the discussion (e.g., acknowledging others' contributions, feelings, or beliefs). Other qualities concern the ways that participants engage with the topic (e.g., providing evidence for held beliefs). We expect future research in AI to model such dimensions to provide a richer understanding of conversation health. Furthermore, current research mostly focuses on modeling content at the individual content level (e.g., a single tweet, a Reddit post). We expect future research to move beyond this building block and accurately model users, discussions, and communities. Such granularity of modeling will enable better intervention systems.

The diversity of online discussion spaces introduces yet another challenge. Markers of conversation health revered by one community can be irrelevant for another. Communities might have different significance they assign to each conversation quality (e.g., Do we value civility or diversity of thoughts more?). We expect future AI systems to learn these community priorities and goals through passive and active interactions with community members (which will map to revealed and stated preferences) and label content accordingly. In addition, the interpretation of conversation qualities might differ across communities (e.g., What counts as civil in the "change my view" and "parenting" subreddits may be quite different.). Current AI systems heavily rely on crowd-workers to label content according to the codebooks constructed by researchers with limited community input. Accordingly, the models do not reflect the values of the communities they are meant to model/guide. We would like future AI systems to engage the community both at the codebook construction and data labeling to identify community-level deliberation health measures.

It is important to note that online communities do not exist in a bubble. What should an AI system do in the case of a community that aims to share and spread health misinformation? Identifying when a local community-level measure of conversation quality is to be preferred over the more traditional definition put forth by social scientists will be yet another important challenge for responsible AI.

### Identifying consensus on facts and goals

Given the transcript of an online multi-party discussion (or set of discussions), can we identify the facts that are universally agreed upon? For example, a transcript of a conversation between economists might reveal strong agreement on the rational choice model of decision making. Current human-computer interaction research leverages human intelligence to perform discussion summarization (e.g., Wikipedia discussions, slack channels). Future AI+HCI+network science collaborative solutions can use text and network features to scale these solutions up.

Similarly, given the transcript of an online multi-party discussion (or set of discussions), can we identify the goals that are shared by the participants/community members? Identifying community goals is a more challenging task, as the revealed and stated preferences might diverge. Furthermore, individual participants may not be cognizant of their own goals. Future AI systems that integrate user modeling with text (stated goal) summarization techniques can provide better models. In the long term, AI could help communities elicit shared goals (e.g., a mission statement for a subreddit).

### Affecting and facilitating online conversations

Future AI systems should go beyond simply identifying and understanding. The aforementioned measures and models should feed into systems that facilitate healthier, higher quality, and more fruitful conversations online. For instance, future AI systems can help online users improve the quality of conversation spaces they participate in by helping them craft messages that 1) directly contribute to quality and 2) indirectly inspire others to behave similarly. A message-crafting AI system might pull parts of the conversation thread that need attention (e.g., an unanswered question) or pull relevant external information from credible sources to enrich conversations. In order to convince individuals to follow algorithmic message-crafting suggestions, such systems should model users accurately and identify the optimal messaging and timing for these algorithmic suggestions. Furthermore, in order to indirectly inspire others, such systems should accurately model relationships and contextual cues—predicting the likely impact of a given message on the quality metrics for subsequent conversations. We believe the next 20 years will bring similar applications for facilitating factually correct conversations, helping communities identify community-specific health metrics, as well as facilitating community goal-setting processes.

### Supporting Complex Teamwork

Some scenarios of collaboration move beyond short-term interactions with atomic tasks. People are often involved in collaborations that span a long time horizons and require team members to carry on complex activities while supporting each other's work. AI systems becoming effective partners within teamwork opens up new challenges in modeling of context and people, and requires algorithmic advances in planning and multi-agent reasoning. Early work in models of collaboration and teamwork relied on constructs such as joint intentions or intention toward teammates' success. These important breakthroughs illustrated that teamwork is more than the sum of individuals doing their tasks, but these early models lack the ability to handle uncertainties and costs present in more complex real-world settings. Subsequent models offer a rich representational framework, but face problems due to computational complexity. Future research must combine these expressive representations with explicit models of intention, while yielding computationally practical algorithms. Furthermore, we must address other crucial aspects in effective human-machine teamwork, such as delegation of autonomy, modeling the capabilities of others, authority, responsibility, and rights.

**Collaboration and Interaction in Social Networks**

Social interactions are crucial for addressing many societally beneficial interactions, such as spreading information about HIV prevention in an at-risk population of homeless youth in a city. Facilitating this type of operation is a problem of collaborative interaction on a massive scale, which raises new challenges that bridge theory and practice. For example, consider the challenge of spreading information about health. In AI, this problem is recognized as one of influence maximization, i.e., given a social network graph, recruit a set of nodes in the graph (opinion leaders) who would most effectively spread this information. Decades of work have yielded important models of influence spread with theoretically attractive properties and algorithms with formal guarantees. However, it is still unclear whether these models of influence spread are reflective of how influence actually spreads in the real world. We need real-world studies to validate these models in physical (as opposed to virtual or electronic) social networks; the work should be interdisciplinary, combining AI methods with efforts in social work, sociology, and other related areas.

**Stretch goals:** By 2040, AI systems can monitor human discussions and automatically identifying consensus on facts and goals, as well as collaborative workflows and emergent plans. Milestones along this path include—

**Milestones**

**5 years:** Defining metrics for measuring various dimensions of the quality of online discussions and constructing methods for automatically estimating these metrics.

**10 years**: AI systems that can reliably identify factual and fraudulent claims from a conversational transcript, making use of background material available on the Web.

**15 years:** AI systems that can understand domain-specific conversational contexts, where it is also necessary to model communities (in terms of communication patterns, goals, and processes).

**Collaborative Creation of New Resources**

*Organizational structures and collaborative creation:* Online interactions are a vital part of collaborative creation of many different types of artifacts. While there have been studies on both the artifacts themselves (e.g., Wikipedia articles or open-source software), and the process of their creation (e.g., talk pages or pull requests), there is a significant opportunity in better connecting the social structures that result in collaborative creations. Clearly, the creations themselves are of significant interest and value (Wikipedia articles, open-source code, collaboratively crafted papers, legal documents, etc.). However, the process by which they were created is equally important. The process here includes the various online interactions of the participants as mediated by their organization (the social network, corporate structure, etc.). The relationship between the interactions and the created artifacts are both important in understanding the creative process but also in building AI-driven interventions that lead to better creation.

The collaborative process includes workflows that allow individuals and groups to manage and coordinate the creative goal. Individuals and groups coordinate to decide on how the work should be split, what each sub-component will look like (in form and function), and how the pieces will come together as a whole. Mining this data represents a unique challenge today as both the artifacts themselves and the interactive traces are varied in structure and form (text, images, video, code, votes, change logs, discussion forums, social networks, community tags, etc.). A key AI challenge is in extraction and the modeling of the creations by people, their online interactions, and the connections between interaction and the creative process. Doing so effectively requires improving natural language understanding and text mining for everything from Wikipedia text, collaboratively created scientific text, discussions around code (e.g., Stack Overflow), the code itself (e.g., Github). However, an increase in multimedia content such as images, videos, and fonts (e.g., Behance) and audio (e.g., Soundcloud) presents new modeling challenges. Modeling the interactions will also require addressing new challenges. Existing interactive media (e.g., email) are being supplanted by innovative social platforms that must be modeled in different ways. More advanced extraction tasks may include identifying causal structures between the process itself and the creative product.

A consequence of better models of the interactive process is better AI-driven interventions to support a community or individuals in their creative goals. By determining which workflows result in higher quality outcomes, a challenge will be to build interventions that can appropriately provide support. For example, an AI system that is aware of both the goal (e.g., the design of a video) and the way the community suggests changes can intervene in a group discussion to suggest task breakdowns, bring in related clips from search engines, or produce alternative video sequences as they are discussed.

**Synthesis and Context Bridging**
Retrieving information as needed or anticipating needs is a valuable contribution to the creative process. However, a more significant benefit is enhancing this participation to contribute novel synthesis and to bridge context. For example, an AI-system that can model what software engineers are discussing and what they have previously built may be constructed to provide alternative architectures for a new module. Such intervention requires a significant understanding of the goals of the community, where they are in the process, and the ability to bridge this model to other knowledge bases. The opportunity is an interactive system that can more actively participate in the creative process. Rather than simply retrieving relevant information, this information can be synthesized and translated into an appropriate form. The "query" for such a system is the ongoing interactions of the participants, a model of their goals, and the state of their creations (e.g., the text, software, lab notebooks, etc.). The research challenges include improvements to text modeling, goal and knowledge modeling, integrating causal models, understanding the network structures of the interactions and how the organization makes decisions. Further interactive challenges require creating the appropriate kind of query response, a process that requires synthesis, summarization, and translation across domains.

An example of this type of system could be one that supports scientists in generating hypotheses. For example, given a single research paper or a set of research papers, a task might be to mine collections of research articles to identify relevant prior work that might have been missed and then to provide a summarized synthesis of this information. A specific challenge for this is in bridging the vocabulary differences across communities (e.g., all the different names for factor analysis). A more ambitious intervention (20-year time frame) would be for the system to understand the discussions around the hypotheses, model the hidden assumptions of the researchers and then to identify additional hypotheses that have not been tested. These may naturally follow from existing results or tacitly underpin prior work. This requires not only analyzing the text of the documents but extracting the hypothesized causal models and reasoning about them.

**Stretch goals:** By 2040, AI systems model the hidden assumptions of participants and identify additional information (e.g., hypotheses) that have not been discussed, using information from external resources as needed. AI systems engage human collaborators by suggesting possibly fruitful activities and task decompositions. Milestones along this path include—

**Milestones**
**5 years:** Improved reputation systems for gauging the quality of human contributors and the effectiveness of team interactions.

**10 years**: Goal-sensitive monitoring of human team activity that learns which workflows are most productive for a given team and objective task.

**15 years:** Interactive systems that participate in the creative process while modeling the participants, their goals, their interactions, and the state of their "creations."

**Building Stronger Social Ties**

*Linking social media to real-world social data:* The increasing prevalence of social interaction in online media has enabled a new wave of social scientific research that aims at linking real-world events to their human causes and consequences. For example, metadata can be used to link messages on Twitter to real-world events such as natural disasters and civil unrest. The relevant messages can then be analyzed using natural language processing, computer vision, and social network analysis. This application of artificial intelligence produces real-time situational awareness of rapidly unfolding crises; and in the long term, it can offer new insights about the causes and consequences of events of interest. However, a key challenge for such research is making valid and reliable inferences based on online data. Online data is typically incomplete and non-representative; it may be confounded by existing algorithms such as content recommenders; and it can be sensitive with respect to the privacy of the individuals who are included in the data. Similarly, event logs may also be incomplete: for example, we usually have data only for crimes that were reported, rather than all crimes that were committed. Working with this data — and linking it to real-world events — therefore requires new methodological principles. We see an opportunity for the development of more sophisticated techniques for reasoning about causal structures and potential confounds. This research will require new bridges between the predictive technologies that are typically favored in computer science and the explanatory methods that are typical of the social sciences. Interdisciplinary teams of AI researchers and social scientists may be best positioned to make such advances. Future research in this space will also depend critically on the availability of data. One possibility is for public-private partnerships that might enable researchers to access data held by social media corporations, while preserving the privacy guarantees that these platforms offer their users. There is also a role for increasing accessibility to government data, so that social media can be linked against records of real-world trends and events.

*Social Assistants:* In a world of digital, multimodal communication with ubiquitous connectivity, smartphones, collaboration platforms, and online social networks, an increasing challenge is to manage the vast amount of information that is available and find ways to hold meaningful conversations. There is a role for AI in managing this communication and enabling healthy interactions, whether among our friends and acquaintances, in public spaces, or while at work. Today, we have simple email assistants that suggest that you sleep on an angry message before sending it, recommend possible responses to an email, and remind you when you should respond to an email. But these suggestions have limited personalization and don't grow to understand the human's social relationships in any deep way. Future research in this space could develop AI with the capability to more fully understand social context, to identify which kinds of communication, with whom, when, and on what topic will be especially valuable, and to automatically handle as much communication as possible. This can be done in a personalized way by AI that is responsive to our goals and our desire for privacy, is cognizant of our social and business relationships, and is considerate of our current context so as to know when to interrupt and when not, as well as what can be automatically handled and where interaction is required. AI can also play a role in helping us to identify opportunities for new kinds of interactions. A future capability is for personalized AI systems to better understand the preferences, goals, and current context of individuals (as described in Section 3.2) so that AI systems can interact with each and identify new opportunities. Examples of this include: suggesting a conversation with a stranger on a particular topic, identifying a goal that an ad hoc team of individuals are motivated to achieve (along with an outline of a plan for how to achieve it), or even enabling new forms of democracy by identifying representative groups of people who are well informed and willing to respond to a poll about an important issue that is affecting their city or state.

**Stretch goals:** By 2040, AI assistants will facilitate healthier, higher quality, and more fruitful conversations online and better processes for creative teamwork, as well as work to enable people to identify and then collaborate in achieving shared goals. Milestones along this path include—

**Milestones**
**5 years:** Learning methods for modeling human interaction, predicting emotional reactions, and tracking long-term happiness of users.

**10 years**: AI systems will track simultaneous news and sensor feeds, linking real-world events to their human causes and consequences.

**15 years:** AI systems can identify and automate valuable communication opportunities in the context of deep models of human relationships, social context, and privacy concerns

### Making AI Systems Trustworthy

AI systems have already seen wide deployment, adoption, and trust for a number of practical problems. They have been particularly successful when the AI system is very accurate, the problem has low stakes, or when there is an easy way for the human to supervise and correct the system's mistakes. For example, music streaming services allow people to steer recommendations via likes and dislikes, spell checkers and auto-completion systems display multiple alternatives to allow people to select correct suggestions, and personal voice assistants typically ask for confirmation of spoken commands when they are uncertain.

### Transparency and Explainability

If we cannot construct a system that can explain its behavior to users, a set of complicated questions come into play. If we are designing an inscrutable AI system, how do we transparently incorporate values into the design? How do we formally define and capture the different types of explanation and interpretation depending on the setting (explanation for recourse, explanation for understanding, explanation for persuasion, and so on)? How would we capture the subjective nature of many notions of trust, for example the fact that multiple stakeholders in a system might have different expectations from a system? These challenges are fundamentally socio-technical, in that they have to recognize the ambiguous and often messy human element at the core of any AI system. But this is a challenge that is in the spirit of AI and interaction, especially when viewed in conjunction with other ideas such as modeling human mental states and collaborations.

Explanation will be key to developing mechanisms to build trust between AI systems and their human interlocutors. While developing these methods is important, there are broader design concerns that also need to be addressed. These include: how to transparently incorporate human values into the design of AI systems, breaking down their inscrutability barriers; how to formally define and capture the different types of explanation and interpretation in ways that are tailored to a given setting (e.g., explanation for recourse, explanation for understanding, explanation for persuasion, etc.), and how to capture the subjective nature of many notions of trust, for example, the fact that multiple stakeholders in a system might have different expectations.

These are fundamentally socio-technical challenges, and effective solutions must recognize and handle the ambiguous human elements involved in interactions.

**Stretch goals:** By 2040, AI systems will be able to reason about trust in both one-on-one interactions and in teams, including how a person's trust might be impacted by a team interaction, personal history, or specific context. The operations of AI systems will be well enough understood by their human partners to build trust in the system over time, even as the system learns and evolves over months, years, and decades. Milestones along this path include—

### Milestones
**5 years:** AI systems will generate concise, human-understandable explanations for machine learning algorithms, summarize specific reasons for independent actions, and explain recommendations.

**10 years:** AI systems will analyze their decisions according to confidence, potential, uncertainty, and risks, and will compare alternatives along these dimensions with detail or at a high level of abstraction. Trust levels in AI systems will be measurably increased based on these explanations.

**15 years:** Human partners will be able to understand a new AI system based on explanations of its operations provided at a level that an untrained person can understand.

### User Control of AI System Behaviors

Users should be able to accurately assess and modify AI system behaviors when appropriate. Ultimately, AI systems are intended to improve human lives and it is therefore imperative we consider the human factors involved in correctly understanding their capabilities and limitations in order to prevent undesirable outcomes. Engendering appropriate levels of trust requires considering the psychology of people basing their actions or decisions on the outputs and explanations of AI systems. For example, research has shown that the mere length of an explanation can have a significant impact on whether or not a person believes an AI system. Moreover, because AI systems learn and therefore evolve over time, it is important to ensure people remain vigilant in vetting AI behavior throughout continued use: An AI system that worked in a specific situation in the past may not work the same way after it is updated.

It is thus critically important to develop approaches that AI system creators can use to convey model uncertainty, help people explore alternative interpretations, and enable people to intervene when AI systems inevitably make mistakes. Further, it is important to realize that the accountability for the effects of the AI systems resides with the developers. Policies should be enacted that will require research oversight through institutional review boards and industry equivalents. Doing so requires consideration of several factors, including characteristics of the target scenario and risk level. Furthermore, while confidence intervals may be effective for skilled end users interacting with AI systems through graphical interfaces, they may be inappropriate for dialog-based settings or time-pressured situations. Similarly, while AI behaviors can be easily adjusted in low-risk scenarios (e.g., email clients that automatically sort or prioritize emails based on importance often support correcting individual misclassifications), it is vital that we make advances in how to best support understanding and intervention in increasingly prevalent high-risk scenarios as well.

**Stretch goals:** By 2040, AI systems can be used in high-risk scenarios where users trust the system to convey uncertainty and allowing them to intervene as appropriate. Milestones along this path include—

**Milestones**
**5 years:** AI systems can be corrected by users when exhibiting undesirable behaviors.

**10 years**: AI systems can convey uncertainty of their knowledge and enable users to intervene accordingly.

**15 years**: AI systems communicate to users significant changes in their behaviors as they evolve over time.

**Preventing Undesirable Manipulation**
People are socially manipulative and persuasive as part of normal social interaction. We influence and are influenced by others' opinions, behaviors, emotions, values, desires, and more. We tend to be more influenced by those we trust and with whom we share a positive relationship. We have also long created technologies, artifacts, and media with the power and intent to persuade, influence behavior, and evoke emotion, such as movies, literature, music, TV, news, advertisements, social media, and more. Socially interactive, personified AI technologies can be designed to use multiple methods of persuasion: leveraging the interpersonal persuasiveness of a known and trusted other along with the technology-enabled persuasiveness of interactive media. To date, narrow socially persuasive AI systems have been designed and explored in research contexts for benevolent purposes, such as acting as health coaches, intelligent tutors, or eldercare companions to promote health, social connection, learning, empathy, and more. But just as social manipulation can be used to benefit people, it can also be used to exploit or harm. This raises important ethical considerations in the design of socially persuasive AI. For instance, a personal health coach robot that provides social support to help users change their behavior to be healthier is a welcome innovation. However, it would be undesirable if that same robot was operated by a company that exploited the robot's persuasive abilities to try to get users to sign up for expensive meal plans that are not needed. Whose interests does the AI serve and how would one really know? How can we mitigate unintended negative consequences or prevent exploitation by those who do not have our best interests at heart? There exist ethical and moral considerations in the design of socially persuasive AI, such as how to design and ensure fairness, beneficence, transparency, accountability, explainability, respect for human dignity and autonomy, promoting justice, etc.. We need to develop methods, technologies, and practices for responsible and trustworthy socially persuasive AI that will enable people to design, monitor, characterize, verify, and improve their behavior over time to benefit people and society.

**Stretch goals:** By 2040, users control and trust their AI systems to use persuasion when ethically appropriate, and to alert and protect them from intentional manipulation by others (humans or machines). Milestones along this path include—

**Milestones**

**5 years:** AI systems follow ethical guidelines when knowingly attempting to persuade users.

**10 years**: AI systems understand and notify a user when others (humans or machines) are acting persuasively against the preferences of the user.

**15 years**: AI systems detect and control malicious persuasion or intentional manipulation by others (humans or machines).

### 3.2.4 CONCLUSIONS

There is a long way to go before AI systems can interact with people in truly natural ways and can support interaction between people online. Building AI systems that are transparent and that can explain their rationales to end users and developers alike will help enable the development of systems that are unbiased and can be trusted, for it is through explanation that biases can be discovered. In 20 years, if research is properly supported, we will see AI systems that can communicate naturally and collaboratively with a diverse range of users regardless of language, dialect, or ability. We will see AI systems that can communicate using multiple modalities understanding how gesture, speech, images, and sensors complement each other. We will see AI systems that can support the human creative process, suggesting hypotheses and enabling better collaboration. We will see unbiased, transparent AI systems performing high-stakes tasks that humans can trust.

## 3.3 A Research Roadmap for Self-Aware Learning

### 3.3.1 INTRODUCTION AND OVERVIEW

The field of machine learning (ML) seeks to provide learning and adaptation capabilities for AI systems. While classical computing based on programming relies on human ingenuity to anticipate the wide range of conditions in which a system may find itself, AI systems must learn autonomously from a variety of sources including labeled training data, tutorial demonstration, instruction manuals and scientific articles, and through interactions with the physical and social world. They must also be able to adapt in order to compensate for changes in the environment and context, changes in the goals of the users, and changes in software, sensors, and computer and robotic hardware. All AI systems are limited by their sensors, their training experiences, and the vocabulary and representations in which their knowledge of the world is expressed; nonetheless, they need to behave robustly in the face of these limitations. This requires a deep understanding of their own uncertainty and a mandate to confront difficult decisions with caution. This is especially important in the face of adversarial attacks that seek to find and exploit the weaknesses and blind spots of all computer systems. Avoiding unintended biases that may be inherent in the way a system has been trained is another important consideration.

While machine learning methods transform the way we build and maintain AI systems, they do not eliminate the need for software engineering. Programming is still required, but at a higher level, where a human expert selects the training materials, designs the representation (i.e., vocabulary and structure) in which the learned knowledge will be captured, and specifies the measures of success. It is often the case that very large quantities of labeled data are needed for training and testing purposes, introducing significant labor and expense. Another critical task of AI designers is to check the correctness of what has been learned. This can be a real challenge, as it requires the AI system to be able to visualize and explain the learned knowledge. Ideally, an AI system would be aware of its own capabilities—it should be able to analyze what it has learned, characterize its boundaries and limitations, and proactively seek opportunities to improve its performance through further learning.

This section of the Roadmap focuses on machine learning and is organized into four areas: learning expressive representations, creating trustworthy systems, building durable systems, and integrating AI and robotic systems.

**1. Learning Expressive Representations.** Most machine learning today works by discovering statistical correlations between attributes of the input (e.g., a group of pixels in an image) and the value of some target variable (e.g., the kind of object in the image). While this produces excellent results on some tasks, experience reveals that those approaches are very brittle. Slight changes in image size, lighting, and so on rapidly degrade the accuracy of the system. Similarly, while machine learning has improved the quality of machine translation (e.g., from English to Spanish), the results often show that the computer has only a very shallow understanding of the meaning of the translated sentences.

A major challenge for the next 20 years is to develop machine learning methods that can extract, capture, and use knowledge that goes beyond these kinds of surface correlations. To accomplish this, researchers must discover the right representations for capturing more sophisticated knowledge, as well as methods for learning it. For example, scientists and engineers think about the world using mechanistic models that satisfy general principles such as conservation of mass and energy. These models are typically causal, which means that they can predict the effects of actions taken in the world and, importantly, that they can explain what would have happened if different actions had been taken in the past. Two important research directions in AI are how to effectively learn causal models and how to integrate existing mechanistic models into machine learning algorithms, yielding more robust systems while reducing the need for brute-force training and large quantities of labeled data.

Many AI systems represent knowledge in the form of symbolic statements, similar to equations in algebra. These symbolic representations support flexible reasoning, and they have been applied with great success in areas as diverse as manufacturing, transportation, and space flight. Deep neural networks offer an alternative approach in which symbols are replaced by numbers. For example, the word "duck" might be encoded as a list of 300 numbers. Since similar words have similar encodings, deep neural networks have proven very useful in natural language processing tasks such as translation. However, the meaning of the individual numbers may not bear any resemblance to how humans reason about the noun "duck"—that ducks are members of the class of birds and the smaller class of waterfowl, etc. Rather, the internal variables in a neural net are abstract quantities extracted by the network as it learns from the examples that it is given. There is important traction to be gained from symbolic knowledge, though, and a critical research challenge is to develop ways to combine numeric and symbolic representations to obtain the benefits of both; that is, systems that are both highly accurate and produce decisions that are more easily interpretable by humans.

**2. Trustworthy Learning.** Today's AI systems perform acceptably in low-stakes domains, but that level of performance can be unacceptable in high-stakes environments such as medical diagnosis, robotics, loan approvals, and criminal justice. To be trusted with such decisions, our AI systems must be aware of their own limitations and be able to communicate those limitations to their human programmers and users so that we know when to trust these systems. Each AI system should have a *competence model* that describes the conditions under which it produces accurate and correct behavior. Such a competence model should take into account shortcomings in the training data, mismatches between the training context and the performance context, potential failures in the representation of the learned knowledge, and possible errors in the learning algorithm itself. AI systems should also be able to explain their reasoning and the basis for their predictions and actions. Machine learning algorithms can find regularities that are not known to their users; explaining those can help scientists form hypotheses and design experiments to advance scientific knowledge. Explanations are also crucial for the software engineers who must debug AI systems. They are also extremely important in situations such as criminal justice and loan approvals where multiple stakeholders have a right to contest the conclusions of the system. Some machine learning algorithms produce highly accurate and interpretable models that can be easily inspected and understood. In many other cases, though, machine learning algorithms produce solutions that humans find unintelligible. Going forward, it will be important to assure that future methods produce human-interpretable results. It will be equally important to develop techniques that can be applied to the vast array of legacy machine learning methods that have already been deployed, in order to assess whether they are fair and trustworthy.

**3. Durable ML Systems.** The knowledge acquired through machine learning is only valid as long as the regularities discovered in the training data hold true in the real world. However, the world is continually changing, and we need AI systems that can work for significant stretches of time without manual re-engineering. This will require machine learning methods that can detect and track trends and other changes in the data and adapt to those changes when possible (and appropriate). Our AI systems also need to remember the past, because there are often cyclical patterns in data over time. A medical diagnosis system that learns to diagnose flu one winter should retrieve that learned knowledge the next winter, when the flu season begins again, and determine how to sensibly incorporate that knowledge into the current environment. Existing machine learning systems lack this form of long-term memory and the ability to effectively merge new and old knowledge.

People are able to detect novel situations (objects, events, etc.) and learn new knowledge from just a few examples. We need so-called "one-shot" or "few-shot" learning algorithms that can do the same; this is likely to require the kinds of expressive representations discussed above. People can also learn skills in one setting (e.g., driving on the left in Britain) and transfer them to substantially different situations (e.g., driving on the right in the US). Such "transfer learning" is not perfect, but it is much faster than having to learn to drive all over again from scratch.

Of course, it is not always possible to avoid the need for retraining. Hence, we need methods that can detect when the current learned model is no longer applicable and flag the system for wholesale re-engineering and careful re-testing.

**4. Integrating AI and Robotic Systems.** Robotics has made tremendous advances in the past two decades. Autonomous driving in both urban and off-road environments has entered a stage of rapid commercialization, attracting immense investments of capital and personnel from both private venture capital and large public technology and automobile companies. Robotic manipulation in structured or semi-structured environments is fairly well understood and has been widely adopted in industry situations that allow for carefully co-engineered workspaces that protect humans in proximity, such as assembly lines for manufacturing.

Still, there are many challenges facing robotics in the next two decades. Current robotic systems lack a holistic understanding of their own behavior, especially when AI technologies are introduced. Moreover, today's robots are not able to work in unstructured home/hospital environments in assisting and caregiving roles, where they must interact safely and effectively with people and where they must deal with non-rigid objects such as cloth and cables. Interaction with humans—both when humans are training the robots and when the robots are assisting or caring for the humans—requires much deeper knowledge about both the physical and social worlds. Acquiring and using this knowledge will require learning from data coming from multiple modalities such as speech, prosody, eye gaze, and body language.

In recent years, there has been impressive progress in both software and hardware for robotics. Sensors such as cameras, radars, and lidars have become smaller, faster, cheaper, and more accurate under a wider envelope of operating environments. Frameworks such as the open-source Robot Operating System (ROS) have grown in scope and quality and enjoy wide adoption and support in both industry and academia. As a major step toward more flexible, adaptable robot systems, there is both a strong need and a powerful opportunity to integrate advances in AI and machine learning to create intelligent software middleware layers that integrate sensing, reasoning, and acting in ways that are tuned to the current needs of the robot and the environment. More work is also needed to establish standards for testing, security, deployment, and monitoring of intelligent robotic systems.

### 3.3.2 SOCIETAL DRIVERS FOR EXPRESSIVE, ROBUST, AND DURABLE LEARNING

This section presents motivating vignettes for five of the societal drivers identified in the overall AI Roadmap. These highlight the research breakthroughs required in each of the four areas listed above that will be necessary to build AI systems that could successfully realize these vignettes.

**ENHANCE HEALTH AND QUALITY OF LIFE**

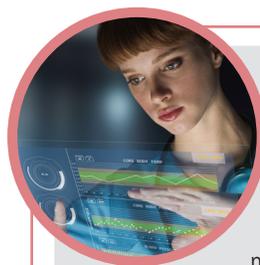

**Vignette 14**

In the spring of 2040, Sue founds a startup pharmaceutical company with the goal of creating a drug for controlling childhood asthma. If she had tried to develop a drug back in 2020, she would have needed to first identify a target receptor to inhibit. Then she would have needed to synthesize and test many different candidate compounds to find a good inhibitor. Following that, she would need to spend millions of dollars for phase I and phase II trials—trials that most drug compounds fail. But thanks to 20 years of investment in AI, medicine, biology, and chemistry, the process of drug design has been revolutionized. A consortium of research hospitals has standardized their health records and deployed a robust data-collection infrastructure that includes advanced wearable sensors as well as a carefully deployed sensor network over the city. This has produced an extensive set of data that capture fine-grained information about health and daily living, including disease and quality of life outcomes. Concurrently, machine learning analysis of the tens of thousands of experiments performed by biologists has produced high-fidelity simulators for basic metabolism and many other biological processes. By linking these simulations and medical records, AI algorithms can now identify causal hypotheses for many disease processes, including asthma. Sue's company uses these resources to identify a small number of candidate target receptors. The company focuses its effort on creating and validating an in vitro system for measuring binding affinity to those receptors. It then licenses large combinatorial libraries of small molecules (the result of federal investments in organic chemistry), and screens them against these receptors. Machine learning strategies developed in cooperation with chemists permits rapid optimization of the shape and structure of the candidate drug molecules. These are then processed by the metabolism simulators to predict required dosages and identify potential side effects. The handful of candidate drug molecules that pass these tests then undergo clinical trials. Health data from the research hospital consortium—maintained with algorithmic assurances of privacy and anonymity—make it easy to recruit participants for these trials. The candidate drugs survive all phases of the trials because of the strong data and simulations that have guided the design process. The availability of these open data and simulation resources dramatically lowers the cost of bringing drugs to market, greatly reduces side effects, and boosts overall health and quality of life.

**Research Challenges:** Realizing this vision of improved health care will require a number of breakthroughs in machine learning:

**1. Learning Expressive Representations:** To analyze the scientific literature and extract causal hypotheses, machine learning for natural language processing needs to go beyond surface regularities of words to a deep understanding of the organ systems and biological and chemical processes described in each paper, the scientific claims made by the paper, and the evidence supporting those claims. It must integrate this information into a growing knowledge base that carefully quantifies the uncertainty arising both from the experiments described in the papers and from the AI system's interpretation of those papers.

To relate the data collected from patients to the underlying biology, machine learning methods must form and evaluate causal models. These will lie at the heart of the simulators of metabolism and other biological processes.

Finally, to optimize the shape and structure of candidate drug molecules, machine learning models must be integrated with physical models of molecule conformation and dynamics.

**2. Trustworthy Learning:** To carry out clinical trials, Sue's company, the hospital consortium, and the patients (as well as their physicians) must all trust the results of the machine learning analysis. Proper quantification of all sources of uncertainty must be achieved, and all stakeholders (including the developers of the AI system) must be able to obtain useful explanations for why certain drug compounds were chosen and others rejected, why certain doses were chosen, and why side effects are predicted to be minimal. Privacy must be protected in appropriate ways throughout this process.

**3. Durable ML Systems:** The laboratory procedures in the published literature are under continuous evolution and improvement, as are the sensors and medical tests recording information about the patients. The AI systems analyzing the literature and patient health data must manage these changes as well as handling the extreme heterogeneity of the patient populations in different hospitals. As new biological discoveries are made, previous simulations must be re-executed to check whether existing drugs may be causing unobserved side effects or whether existing drugs may have new positive applications.

**4. Integrating AI and Robotic Systems:** The use of wearable sensors and the deployment of sensor networks over the city introduces many challenges. Sensors may not be fully reliable and may fail, so the data collected may not be consistent. As data is collected, sensor systems may learn a model of which sensors fail and how. The systems must discover which frequency of collection is most appropriate for the type of data being collected. For example, on rainy days the pollution may be reduced and data can be collected at longer time intervals, while on hot days the network should activate all sensors and collect data more often.

### REINVENT BUSINESS INNOVATION AND COMPETITIVENESS

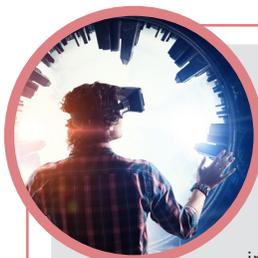

**Vignette 15**

Juan is an artist who wants to turn his passion for personalized augmented reality board games into a business. To design such games, Juan uses an intelligent design framework that plays the proposed game with him, perceives his behavior and enjoyment level, and suggests ways to improve the game's rules and layout. Once he is happy with his design, Juan uploads it to a prototype manufacturing site where the local manufacturer quickly teaches a robot how to build and package the game. The robot employs multimodal human interaction to learn from its trainer. Even though the game components are unique to each user, including a wide range of potential materials, the robot can now deploy robust manipulation skills to assemble the novel game components into customized packaging. The next morning, game testers receive initial prototypes and give it rave reviews, and demand for the game explodes. The trained robot shares what it learned through building and packaging Juan's game to hundreds of robots at other manufacturing sites. Because each robot is slightly different, the learned models have to be adapted to work with their particular hardware and software setup. The new robots immediately begin making games to meet the demand, and Juan is able to rapidly expand his sales.

**Research Challenges:** This vision requires research advances in machine learning along the four areas mentioned:

**1. Learning Expressive Representations:** Juan's game design framework understands his desire to build a fun game. It is able to learn from multiple modalities, including video cameras, audio microphones, and wearable physiological sensors, as well as from spoken statements that Juan makes. This allows it to understand Juan's intent and emotional response at a deeper level, which in turn allows it to propose improvements to the game.

**2. Durable ML Systems.** The game design framework is able to transfer what it has learned from interacting with Juan to learning from other customers. Similarly, the manufacturing robots can transfer their knowledge to other companies, where the robots are slightly different and are manufacturing similar, but not identical, game pieces.

**3. Integrating AI and Robotic Systems.** The prototype manufacturer is able to teach the robot how to build Juan's game using direct interaction, including verbal communication and visual demonstration. The robot assistant is adept at handling a wide variety of physical materials, both rigid and non-rigid, to produce games that are highly customized.

### ACCELERATE SCIENTIFIC DISCOVERY AND TECHNOLOGY INNOVATION

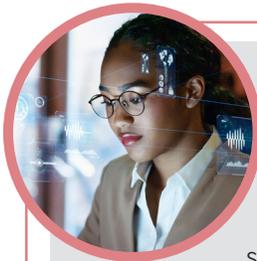

**Vignette 16**

Aishwarya is a climate scientist trying to make predictions of future climate at the local and regional scale. It is essential that such predictions correctly quantify uncertainty. She chooses a climate model that is based on mathematical models of atmospheric physics, solar radiation, and land surface-atmosphere interactions. Unfortunately, running the model at the required fine level of detail is not possible due to the computational cost and the lack of sufficient observation data. Fortunately, recent advances in ML research have produced new physics-aware ML approaches that learn from data while incorporating knowledge about the underlying physics. These approaches run efficiently and produce models at a much finer level of detail than the original climate models. This makes it easy to run multiple models efficiently, which in turn allows Aishwarya to provide clear uncertainty bounds for the resulting predictions.

The results of these models are then used by Jia, who works at FEMA. Using machine learning methods, she combines climate predictions under different policy scenarios (no change versus reduced carbon emissions, etc.) to identify regions that are most vulnerable to extreme weather events such as hurricanes, floods, droughts, heat waves, and forest fires. With the aid of these causal models, she can plan appropriate responses. For example, her physics-aware ML model produces inundation maps in response to extreme meteorological events (hurricane, heavy rain) to identify areas of flooding, which is fed into an AI system for smart cities to perform evacuation planning, emergency relief operations, and planning for long-term interventions (e.g., building a sea wall to ward off storm surge).

Thanks to the 20 years of research investment in AI, physics-aware machine learning techniques are available that can process multimodal, multi-scale data and also handle heterogeneity in space and time, as well as quantify uncertainty in the results. The combination of physics-based climate and hydrological models with machine learned components allows Jia to produce more accurate predictions than would be possible with pure physics-based or pure machine learned models alone. This hybrid approach also generalizes better to novel scenarios, identifying new threats that could result in injury or death. In 2035, these models are applied to revise flood maps, saving many lives in the floods caused by hurricane Thor in North Carolina.

**Research Challenges:** This vignette illustrates the productive integration of mathematical models based on first principles with massive quantities of real-world data. Each approach complements and extends the other and helps address the inherent limitations of each. The methods apply hard-earned physics knowledge but also detect when it no longer applies and must be adapted and refined to cover new situations.

**1. Learning Expressive Representations.** Mathematical models based on physical laws, such as conservation of matter and energy, are causal and generalizable to unseen scenarios. In contrast, the statistical patterns extracted by traditional machine learning algorithms from the available observations often violate these physical principles (especially when extrapolating beyond the training data). On the other hand, the mathematical models are not perfect. They are necessarily approximations of reality, which introduces bias. In addition, they often contain a large number of parameters whose values must be estimated with the help of data. If data are sparse, the performance of these general mathematical models can be further degraded. While machine learning models have the potential to ameliorate such biases, the challenge is to ensure that they respect the required physical principles and are generalizable to unseen scenarios.

**2. Durable ML Systems.** Building Aishwarya's model requires combining sparse data measured at different locations around the planet, measured at different spatial and temporal scales using different instruments and protocols, and stored in different formats. Much of the knowledge brought to bear is adapted from different circumstances, from past natural disasters that occurred in other locations but with similar features. One-shot and few-shot learning are applied to anticipate and address previously unseen situations that could have tragic consequences.

**3. Trustworthy Learning.** Aishwarya seeks to properly account for all sources of uncertainty, including those that result from transferring knowledge from some parts of the planet to others and from one model to another. In this context, machine learning methods are needed that can assess uncertainty in transfer learning. Jai and FEMA will be making life-and-death decisions based in part on the uncertainties in Aishwarya's predictions and in part on uncertainties in the hydrological models.

### SOCIAL JUSTICE AND POLICY

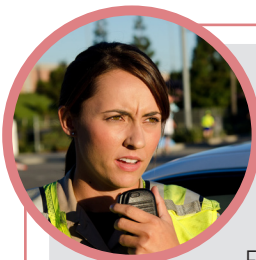

**Vignette 17**

During a training procedure in a virtual reality facility, Officer Smith and the more experienced Officer Rodriguez are dispatched to a house to handle a situation in which a husband and wife are having a very loud dispute with threats of violence that caused the neighbors to become concerned. Each one tells Officer Smith a story that conflicts with the other. They are very manipulative, using emotions, misrepresentations, and guilt to manipulate each other and the officers. Smith's challenge is to say the right things to de-escalate the situation and resolve the dispute. She also needs to say things in the right way: the virtual characters are perceiving her body position and body language, as well as her eye gaze and facial expressions. If she is either too assertive or not assertive enough (which will be indicated by both what she says and how she says it), things will go badly. In this case, Smith becomes angry and moves too abruptly; the husband pulls a gun and Officer Rodriguez shoots in reaction. During the after-action review, Officer Smith walks through the scenario, and an AI coach presents things she did badly and encourages her to think about how she could have done better. Officer Smith is also able to "interview" the virtual husband and wife to find out what they were thinking and how they perceived her statements and actions.

**Research Challenges:** The use of AI in advanced training involving life-or-death decisions and split-second reactions requires accurate modeling all of the visible and audible cues that humans depend on as well as hidden motivations, coupled with the ability to engage in a collaborative dialog on what happened after-the-fact.

**1. Learning Expressive Representations.** To create the virtual characters, machine learning and computer vision must be trained on data from actors in order to build models of the motivation and personality of each character enough to capture how those are revealed through their speech, emotions, and body language. This requires deep understanding of the meaning of their spoken language as well as prosody, facial expression, and physical interactions. The learned simulation models must be able to synthesize responses to all of the new behaviors that a trainee police officer might exhibit.

**2. Trustworthy Learning.** The virtual characters must be able to explain (in the after-action review) what they were thinking and why, which requires explainable AI as well as the capability to synthesize natural speech in response to questions from the trainee officer.

### TRANSFORM NATIONAL DEFENSE AND SECURITY

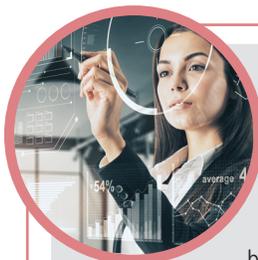

**Vignette 18**

Susan works in the emergency operations center of a smart city. There is a major hurricane bearing down, and her task is to muster hundreds of thousands of sensors, cameras, and actuators to plan and execute the safe evacuation of 10 million people. The system that she uses for this procedure has been pre-trained to estimate street capacity based on normal traffic and to predict where residents are typically located at each hour of the day based on behavioral data. From this, it can plan how to set traffic signals to facilitate rapid evacuation of the city. However, additional data must be integrated into the system in real time—weather predictions, flooding, road construction, vehicle accidents, downed utility poles—to which the plan must be adapted. In adapting to changing situations, the system can also make real-time use of residents' social media postings as they provide feedback on their own personal situations in the storm.

This sophisticated planning system has been designed to be resilient to physical damage and to surprises it cannot correct but must work around. Among other things, it can detect when the situation is beyond the conditions it can successfully handle, at which point it falls back on different systems or even hands control over to manual processes. For example, as smart sensors are damaged by the storm, missing data can be interpolated from nearby sensors, or inferred from regional models; as internet infrastructure fails in some neighborhoods, human staffers operating radios can be called in.

With the goal of instilling civil unrest, a nation state at conflict with the US seizes the opportunity to launch a coordinated campaign of disruption to interfere with the evacuation. They release malware to disrupt automated car systems and a bot-powered misinformation campaign on social media. Fortunately, safety systems in the individual cars detect inconsistencies between the user's commands and the car's physical behavior, and the city's social media analysis software identifies the bot activity and deploys communication countermeasures that have been found to be effective across many different human populations. Susan and her teammates then apply a collaborative planning system to take back control of some parts of the system, and employ the failover methods originally designed for storm resilience to route around the maliciously compromised portions of the system.

**Research Challenges.** In this scenario, the AI system combines smart detection, software countermeasures, and overall resilience to mitigate and recover from the attacks. Achieving all of this requires addressing important research challenges:

**1. Durable ML Systems.** Traffic behavior and road capacities change over time as the vehicle mix changes: e.g., the percentage of highly automated (or even autonomous) connected vehicles. Electric vehicles that run out of battery power cannot be easily recharged, which can affect these dynamics. One challenge will be to develop machine learning methods that can deal with such change in the distribution of vehicle capabilities.

A second need for managing change arises in the context of social media traffic. The behavior of people on social media is constantly changing. Any machine learning system that attempts to model the state of the city from social media posts must be prepared to deal with such changes.

A third durability requirement arises from the need to test such systems. One could accomplish this using reconstructions of historical weather evacuations, for instance. These occurred in situations very different from the current one, however, with different traffic signals and sensing capabilities, different driving behavior, and so on, and hence must be adapted to the current scenario.

**2. Trustworthy Learning.** The ubiquity of smartphones and the advent of connected vehicles provides an opportunity to coordinate the behavior of different vehicles. A central planning system could give instructions to each driver (or each vehicle), but, of course, the drivers must trust this information. Another possibility is to create a kind of market in which people could indicate their preferred destination and the planning system would coordinate traffic so that people could get to their destinations quickly. This could even be integrated with social media analysis—e.g., so that the system could ensure that family groups were all sent to the same destination in the case of an evacuation. In all cases, the system should be able to explain why it is sending people along the chosen route.

### 3.3.3 TECHNICAL CHALLENGES FOR SELF-AWARE LEARNING

#### Learning Expressive Representations

A central tenet of artificial intelligence is to build models of the world (world knowledge) and use them to both track the state of the world and formulate plans to achieve goals. While routine intelligent behaviors can be achieved via reactive (stimulus-response) mechanisms, the ability to deal with complex and novel situations is believed to require model-based reasoning. Until recently, machine learning systems were designed to map directly from an input representation (e.g., a sentence) to an output representation (e.g., a parse tree that captures its syntactic structure). The difficulty of learning is related to the size of the gap between the input and the output, and consequently this gap was kept small in order to ensure that the learning was feasible.

This section outlines three different research threads that will be required to span these gaps: automatic learning of 1) intermediate representations that act as stepping stones, 2) models that capture cause and effect, and 3) models that capture underlying mechanisms.

#### Learning Better Intermediate Representations

With the advent of deep learning—networks that use many *layers* of computational elements to map from input to output—it has been possible in some cases to learn across much larger gaps. In computer vision, our algorithms are able to map from raw image pixels to high-level image summaries (e.g., a sentence describing the image or a division of the image into its objects and background). There is some evidence that machine learning algorithms achieve this by deducing useful *intermediate representations*: ways of capturing or describing information at a level that is "between" those of the input and output. Many machine learning algorithms, for instance, leverage the notion of an *embedding space*. In word embedding, for example, each word is represented by a list of numbers. Many different natural language processing tasks can then be attacked by first mapping the words into this embedding space and then learning appropriate transformations in that space. Machine-translation systems work in this fashion, mapping the words into this embedding space prior to translating them. Similarly, systems for visual question answering proceed by combining the embedded representation of the words and the information in the image. This staged decomposition of the task has a number of advantages besides effectiveness, including robustness and extensibility; For example, systems trained on the ImageNet database of 1000 object categories can easily be "fine tuned" to work on new problems, such as medical or biological recognition. The fine tuning typically involves retraining only the final layer or two of the network (which may have 150 layers in total).

Reusable object recognition and word embeddings show that *some* forms of reusable knowledge can be learned by computer algorithms. But it is unclear how to move from these two cases to general approaches for the many problems in which reusable intermediate representations can play important roles. For example, it is widely believed that successful natural language understanding requires some event schema. Similarly in computer vision, it would be useful to build a three-dimensional model of all of the objects in an image, plus the light sources and the position and settings of the camera at the time the image was taken. In robot locomotion, there is a natural hierarchy in which the overall navigational goal ("go out the door") is the most abstract, the main trajectory—the path around the desk and through the door—is next, and at the lowest level are the individual actuator commands that direct the robot's wheels accordingly.

The main technical challenge in this endeavor is to develop such representations, either by automatic discovery from experience or by some mix of manual design and algorithmic learning. These must include transformations that bridge from the numerical representations (e.g., word embeddings, trajectory coordinates) to the symbolic representations (e.g., parse trees, movement plans) that support flexible abstract reasoning. A critical consideration in such mappings is for the AI system to know which detailed aspects of the world can be ignored within the current context. It is rare that a person needs to know the exact shape and location of every object in a room. When we are searching for our car keys, we focus on the most relevant size range for objects. In the opposite direction, the AI systems must learn how to translate abstract meanings and plans into numerical commands for generating actions (spoken language, robot motion, etc.)

**Stretch goals:** By 2040, we should have achieved a fundamental understanding of the relationships between knowledge representation, learning, and reasoning. This will enable us to create AI systems that understand natural language and engage in conversations about both the physical and the social worlds. Such systems will be able to learn from natural language instruction and in turn explain and teach via natural language. The underlying representations will allow robots to reason effectively about the physical world through direct manipulation, pointing, computer vision, and natural language. Milestones along this path include—

**5 years:** Robotics is likely to be the first field where we have good ideas about intermediate representations needed for both visual perception and physical action, making it a good starting point. Using simulated physical worlds, develop sufficient intermediate representations to enable robotic systems to learn complex, hierarchical activities through demonstration and practice. Initial work should focus on manipulation and assembly, because these will also be immediately useful.

**10 years**: Develop generalized representation-learning methodologies and use them to build effective representations of social and mental worlds in application areas where humans interact with computers (e.g., conversational agents with a broad competence). Achieve good performance in human-computer interaction by learning from examples, demonstrations, and natural language instruction.

**15 years:** Using these representations, create an AI system that can read a textbook, work the exercises, and prepare new instructional materials for customized education.

**Learning Causal Models**
Developments over the past decade in causal modeling hint at the power of these techniques in machine learning systems. Existing algorithms rely entirely on correlative patterns between the input and output variables in the training data. These correlations can be very brittle, so that even very small changes in the problem can invalidate the learned knowledge. In contrast, most scientific theories are causal, and they allow us to learn causal regularities on Earth and extrapolate far away. For example, the causal theories of physics allow us to predict with high confidence what happens inside black holes even though it is impossible for us to make observations there. The potential applications are significant. Advances in understanding causal inference could, for example, greatly advance research in biology and medicine, where these relationships are the key to designing medical interventions to fight disease.

The technical challenge here is to discover the causal relationships and assess their strength. The randomized trial (or A/B test) is the best understood method for achieving this, but it requires good data. In some settings, causal relationships can be confirmed with high confidence even in the face of noise. In some situations, one can even assess the causal impact of one variable on *multiple* result variables. Research is needed to develop a comprehensive understanding of the set of situations under which causal relationships can be inferred. This will be particularly challenging in situations where feedback loops are present, producing circular or bidirectional causal relationships. This is commonplace in real-world problems.

**Stretch goals:** By 2040, we will have a comprehensive theory of learning and inference with causal models and with models that mix causal and correlational inference. Every machine learning package and every application will include causal modeling to the extent permitted by the data and the problem. Milestones along this path include—

**5 years:** An open-source machine learning package will support general learning of causal models for supervised learning with semantically useful features. Applications in at least two domains (e.g., biology and medicine) will employ learned causal models.

**10 years:** Robotic and cyber-physical systems will learn causal models based on active exploration and experimentation in the domain. This will greatly accelerate learning in these domains.

**20 years:** Development of machine learning systems that can discover, explain, and apply new causal models for significant physical or social phenomena that rival what human experts would consider publishable in a scientific journal.

**Leveraging Mechanistic Models to Build Robust Systems**
The scientific community is increasingly considering machine learning as an alternative to hand-crafted mathematical and mechanistic models (e.g., Newton's laws of motion). These traditional scientific models are broadly powerful, but they have some major limitations, including the effort required to create them. Machine learning models, on the other hand, require little human effort, but they are "black boxes." They cannot explain their results in the language of the scientists: physical relationships, mechanistic insights, etc. There are other challenges as well. Unless they are provided with adequate information about the physical mechanisms of complex real-world processes, in the form of comprehensive training data, machine learning approaches can produce incorrect answers, false discoveries, and serious inconsistencies with known science. This reflects shortcomings in the training data, which may not be rich enough to capture the important scientific relationships of the problem at hand. It also reflects the tendency of current machine learning algorithms to focus on correlational rather than causal models (see above).

Leveraging scientific knowledge, both directly (e.g., by incorporating fundamental physical laws) and indirectly (e.g., as embodied in mechanistic models), is one way to address these issues. Incorporation of relevant physical constraints into a machine learning model, such as conservation of mass and energy, will ensure that it is both physically realistic and generalizable to situations beyond those covered in the training data. Indeed, incorporation of this kind of knowledge can reduce the amount of data required for successful training. Simulations of science-based models can even be used to generate synthetic data to "bootstrap" the training process so that only a small amount of observation data is needed to finalize the model.

Producing science-aware machine learning models will require addressing a number of challenges. Effective representation of physical processes will require development of novel abstractions and architectures that can simultaneously account for evolution at multiple spatial and temporal scales. Effective training of such models will need to consider not just accuracy (I.e., how well the output matches the specific observations) but also overall scientific correctness. This will become even more challenging when the underlying system contains disparate interacting processes that need to be represented by a collection of physical models developed by distinct scientific communities (e.g., climate science, ecology, hydrology, population dynamics, etc.).

**Stretch goals:** By 2040, we will have AI tools and methodologies that merge physical/process-based models with machine learning in a manner that effectively scaffolds the integration of multiple, disparate models that cover multiple aspects of complex systems (e.g., climate, weather, hydrology, ecology, economics, agriculture, and natural disaster mitigation and recovery). Such unified models could identify areas that are expected to be hit by hurricanes of increased intensity (due to climate change); project the path of a specific hurricane; identify areas to be impacted by flooding; create projections about damage to people, infrastructure, and biodiversity; and plan for emergency help or adaptation decisions (building a more resilient power grid, higher sea walls, etc.). Milestones along this path include—

**5 years:** Identify several use cases where the science is well understood but the physical models have limited performance. Build machine learning models that can leverage scientific knowledge to significantly exceed the performance of the associated state-of-the-art physical model.

**10 years:** Expand these efforts to a larger and more diverse set of domains with the goal of detecting commonalities in causal modeling tasks in those domains and developing a methodology that is applicable to a wide range of problems (e.g., climate, hydrological, ecological, econometric, population dynamics, crop yield, and social models).

**15 years:** Extend these tools and methodologies to handle cases where many disparate models are collectively used to solve a complex problem. Address problems of end-to-end debugging and explanation. Combine statistical and symbolic AI to help scientists come to an improved understanding of causes and effects of physical processes.

**Trustworthy Learning**

For most of its history, artificial intelligence research has focused on building systems to solve specific problems, whether it be recognizing objects in images or winning at the game of chess. But as AI is deployed in commerce, healthcare, law enforcement, and the military, it is critical that we know when we should and should not trust our AI systems. In this section, we consider the reasons why we might not trust these systems and suggest research directions to address these "threats to trust."

Every engineered system is based on a set of assumptions; if those assumptions are violated, that may be a reason to distrust the system. In machine learning, the assumptions typically include: 1) the training data capture a representative picture of the problem at hand, 2) the data have been accurately measured and correctly labeled, 3) the problem itself has been properly formulated, and 4) the solution space contains an accurate model that can be identified by the machine learning algorithm. Violations of these assumptions can arise due to bias and error introduced during the measurement, collection, and labeling of data, or when the underlying phenomena change: e.g., when a new disease spreads through the patient population. A different issue arises when the machine learning system is trained on one population (e.g., a university hospital) but then applied to a different population (e.g., a Veterans Administration hospital). A particularly extreme case of this kind of *data shift* is when the situation at hand involves categories that were not in the training data: e.g., a medical AI system that needs to recognize a disease that it has never encountered. This is known as the "open category" problem. Indeed, the classic case where the training data are fully representative of the test data is rarely encountered in practice.

A closely related threat to trust is any violation of the second two assumptions: that the problem is properly formulated and that the choice of model representation is correct. For many standard learning settings, the assumption is that each data point and query is independent of the others, which is not true if there are correlations or trends in the data. This is similar to data shift, but it is more serious, because it can require completely changing the problem formulation: e.g., considering past history, rather than just the current state, or considering other agents (humans or robots), rather than assuming the system is acting alone. To trust an AI system, we need to know what modeling assumptions it is making so that we (especially the system engineers) can check those assumptions.

A third threat to trust is adversarial attack, for which machine learning presents new challenges—e.g., modification of the training data (known as "data set poisoning"), which can cause the learning algorithm to make errors. An error in a face recognition system, for example, might allow an adversary to gain physical access to a restricted facility. If the attacker can modify the machine learning software itself, then they can also cause the system to make mistakes. An attacker may also modify examples to which the system is later applied, seeking to gain advantage. Related to adversarial models are incentive models, where the learning systems play a role in the algorithmic economy, and where participants are profit maximizing. Consider, for example, recommender systems, where some actors may seek to promote or demote particular products or methods of dynamic pricing where it is of interest to achieve collusive outcomes or strategic advantages over other learning systems.

A fourth threat to trust is the high complexity of machine learning models, such as large ensemble classifiers (e.g., boosted ensembles or random forests) and deep neural networks. These models are too complex to be directly understood either by programmers or end users.

There are four major ways to address these challenges. If the AI system can maintain complete control of the system and its data, it can avoid most of the cyberattacks and some of the sources of bias in data collection. This does not, however, address data shift or incorrect problem formulation. Moreover, it is not clear how it is possible to maintain control of data in many settings: data inherently involves interacting with the world outside of the system.

The second strategy is the *robustness* approach. In traditional engineering, the designer of a bridge or of an airplane wing can include a margin of safety so that if the assumed load on the bridge or the wing is violated by a modest amount, the system will still work. An important research challenge in AI is to discover ways to achieve a similar effect in our algorithms. A related research direction seeks to make trained models inherently less susceptible to adversarial modifications to future examples, by promoting simpler dependencies on inputs, for example. In the context of incentive considerations, an important research direction is to model costs and utility functions explicitly, and apply game theory and mechanism design to provide robustness to strategic behavior.

The third strategy is the *detect and repair* approach. In this approach, the AI system monitors the input data to verify that the underlying assumptions are not violated and takes appropriate steps if violations occur. For example, if it detects a novel category (of disease or of animal), it can apply an algorithm for one-shot learning; if it detects a change of context, it can employ transfer learning techniques to identify the ways in which the new context is similar to the old one and transfer knowledge accordingly. An important research direction is to systematize the many ways in which the data may shift or the machine learning problem may be incorrectly formulated and develop methods for diagnosing and repairing each of those errors. The field of model diagnostics in statistics provides a good starting point for this.

The fourth strategy is to *quantify, interpret, and explain*. Here, the AI system seeks to replace complex models with models that are more easily interpreted by the user. It also seeks to quantify its uncertainty over its learned knowledge and its individual predictions so that the user (or a downstream computational component) can decide whether to trust those predictions.

We now consider several topics related to creating trustworthy machine learning systems.

**Data Provenance:** Information about how the data were selected, measured, and annotated is known as data provenance. Strong data provenance principles are a determining factor in the accessibility of data and govern our ability apply machine learning to that data, but training sets with good provenance information are rare. Strong data provenance promotes meaningful labeling and metadata—data about the data—across domains. We lose the ability both to leverage this data and to track changes in an accessible and concise way.

Solutions here will require understanding how data propagates across domains and over networks, how consensus is developed around annotation and structure, and where the responsibility for defining those principles lies. Improved understanding of these principles would enable us to build and evaluate robust and flexible models in changing situations. For instance, understanding the effects of climate change on building materials and indoor environments requires integrating domain knowledge from climate scientists, materials scientists, and construction engineers. Beyond changes in statistical distribution of the data, there will be shifts in terminology associated with each of these industries over time as new materials are created and building methods are refined or improved.

**Stretch goals:** By 2040, many fields will have created shared, durable data repositories with strong provenance information. These will be continually maintained and tested so that the data can be trusted. Virtually all commercial machine learning and data mining systems employ provenance information to improve resilience and support updates and extensions to provenance standards. Milestones along this path include—

**5 years:** Provenance standards are developed in consultation with stakeholders in at least two different disciplines (e.g., pharmaceuticals and electronic health records).

**10 years**: Provenance standards are adopted by most scientific and engineering fields. At least two fields have shared, durable data repositories conforming to these standards.

**15 years:** Many fields have developed and adopted provenance standards, and these are supported by all of the leading machine learning and data mining toolkits. Most fields have shared, durable data repositories, and those repositories have well-exercised methods for making updates and extensions to the standards and testing how well machine learning and data mining systems handle updates and extensions.

**Explanation and Interpretablity**

An important technical challenge is to create interpretable, explainable machine learning methods. A model is interpretable if a person can inspect it and easily predict how it will respond to an input query or to changes in such a query. This relates also to *agency*, which is the idea that the user should retain enough understanding in order to retain control over decisions that are implied by a model. An explanation is some form of presentation (e.g., an argument, a sequence of inference steps, a visualization) that similarly allows a person to predict how the system will respond to changes in the query. In addition to being able to make query-specific predictions of model behavior, we often want an overall explanation. For example, we might want to know that a face recognition system *always* ignores the person's hair or that a pedestrian detection system is *never* fooled by shadows. Answering questions like this in the context of complex models such as deep neural networks is an open research topic. A primary reason for the success of deep neural networks is their ability to discover *intermediate representations* that bridge the gap between the inputs and the outputs. These intermediate representations, discussed in the previous section, rarely take on a form that a human can understand. An active research direction seeks to assist humans in understanding the functioning of neural networks, for example, through visualization.

Research on explanations must consider the users who are the targets of the explanations (the software engineer, the test engineer, the end user) and the possible queries in which those users might be interested (Why? Why not? What if? etc.). A major challenge here is that sometimes the explanation relies on regularities that the machine learning system has discovered about the world. For example, the system might explain why it predicts that company X2, which is located in Orlando, Florida, is not in the coal mining business by stating the general rule that there are no mining companies in Florida (a regularity that it has discovered). Such regularities may be obvious to a subject matter expert but not to end users.

**Stretch goals:** By 2040, every AI system should be able to explain its beliefs about the world and its reasoning about specific cases. The system's users (software engineers, maintenance engineers, and end users) will then be able to trust the system appropriately and understand its limitations. Milestones along this path include—

**5 years:** Algorithms that can learn high-performing, interpretable, and explainable models are developed for data with meaningful features.

**10 years:** Interpretable and/or explainable machine learning models become a requirement for all high-risk applications.

**15 years:** Algorithms are developed to explain their actions effectively. These allow AI systems to work much more successfully in open worlds.

### Quantification of Uncertainty

In many applications, it is important for an AI system to have an internal measure of its confidence. A system that can accurately assess probabilities of various outcomes and events can assist users in taking risks and identify situations in which an automated system should cede control to a user. But what if the system is not making accurate assessments of these probabilities? An accurate understanding of the sources of uncertainty can help differentiate situations that are fundamentally uncertain from those in which additional data collection would be useful.

Quantifying uncertainty is difficult. Even in the classical setting where the examples are independent and identically distributed, many existing machine learning models can be extremely confident while making incorrect predictions. Some methods do exist for quantifying uncertainty in these cases, but there are many gaps in our knowledge. For example, if the data were collected in a biased way, or if some of the attributes are missing in some of the data, how can we quantify the resulting uncertainty? When we move beyond the classical setting to consider data shift, incorrect problem formulation, and incorrect modeling assumptions, there are essentially no existing methods for uncertainty quantification. The problems of open category detection and transfer learning are particularly vexing. Although we do have some learning methods (such as meta-learning and hierarchical modeling) for one-shot learning and transfer learning—both of which are described in the previous section—the problem of uncertainty quantification in those situations is completely unexplored.

In many cases, it is important to distinguish between situations where uncertainty stems from insufficient data or an incorrect model—situations that could be remedied by additional data collection or construction of a better model—and situations in which some aspect of the world is inherently random. A related challenge has to do with accurate comparison of uncertainties across different kinds of models: When different models disagree, how should decisions be made about gathering more data or rebuilding the model?

**Stretch goals:** By 2040, every AI system should be able to build and maintain a model of its own competence and uncertainty. This should include uncertainties about the problem formulation and model choice, as well as uncertainties resulting from imperfect and non-stationary data, and uncertainties arising from the behavior of other decision makers. Every AI system should exhibit some robustness to those uncertainties, and be able to detect when that robustness will break down. Milestones along this path include—

**5 years:** Good theoretical understanding of the different sources of data and model error is achieved so that we understand what kinds of problems can be detected and repaired. Uncertainty quantification for basic cases of data shift are developed.

**10 years:** Algorithms are developed to detect most forms of flawed data. These algorithms are able to adjust the learning process to compensate for these flaws, to the extent theoretically possible. Uncertainty quantification is developed for all forms of data shift and some types of model formulation errors.

**15 years:** Algorithms are developed to detect and adapt to many forms of model formulation error. These allow AI systems to work much more successfully in open worlds.

### Machine Learning and Markets

Just as the interactions among collections of neurons lead to aggregate behavior that is intelligent at the level of whole organisms, the interactions among collections of individual decision makers leads to the phenomena of markets, which carry out tasks such as supplying all of the goods needed for a city (food, medicine, clothing, raw materials, etc.) on a daily basis, and have done so for thousands of years. Such markets work at a variety of scales and in a great variety of conditions. They can be adaptive, durable, robust and trustable—exactly the requirements outlined in this section. Moreover, the field of microeconomics provides not only conceptual tools for understanding markets—including their limitations—but also provides a theoretical framework to guide their design, e.g., through mechanism design theory. Is there a role for economic thinking in the design of trustworthy machine learning systems, and in addressing many of the major challenges that we have identified in this report?

First, consider a multi-agent view on learning systems, a *market-of-learners* view, where open platforms promote competition between learning systems (and composition of learning systems) with profits flowing to learning systems with the best functionality. Can this market-of-learners view lead to durability, with good performance over long stretches of time, offering robustness as surrounding context changes? Markets promote complements, incentivizing new products and technologies that help when used together with existing products and technologies. What is the analogy for learning systems, in the ability to promote new data, new algorithm design, and new forms of interpretability?

Second, learning systems will necessarily act in markets, leading to a *market-based AI systems* view. In domains such as commerce, transportation, finance, medicine, and education, the appropriate level of design and analysis is not individual decisions, but large numbers of coupled decisions, and the processes by which individuals are brought into contact with other individuals and with real-world resources that must be shared. We need trustworthy machine learning systems for these economic contexts. The field of multi-agent systems studies both the design of individual agents that interact with other agents (both human and artificial) and the design of the rules or mechanisms by which multi-agent systems are mediated.

The market perspective acknowledges that there may be scarcity of resources: one cannot simply offer to each individual what they want most. If a machine learning system recommends the same route to the airport to everyone, it will create congestion, and the nominally fastest route will no longer be the fastest. Learning systems need to take into account human preferences and respect these preferences. A user may be willing to go on a slower route today because she's not in such a rush, or, on the other hand, she may be willing to pay more today precisely because she's in a rush. One AI system cannot know all of these hidden facts about participants; rather, we need the participants, or AI systems representing these participants, to understand their options, how those options help them to meet their goals, and how to understand and diagnose their role in the overall system. This is what markets do. The recommendations of market-based AI systems will, like recommendation systems, be based on data and responsive to human behavior and preferences. Such blends of microeconomic principles with statistical learning will be essential for AI systems to exhibit intelligence and adaptivity, particularly in real-world domains where there is scarcity. An economic view may also inform considerations of fairness, building from theories of taste-based discrimination and statistical discrimination in an economic context.

Of course, markets have imperfections, and new kinds of imperfections will arise as we develop new kinds of markets in the context of massive data flows and machine learning algorithms. But these imperfections are properly viewed as research challenges, not as show-stoppers. Moreover, an active research direction is to use learning systems for the automatic design of the rules of markets. Moreover, the theories of normative design from economics may prove more relevant for market-based AI systems than for systems with people, as AI system come to better respect idealized rationality than people. AI must take on this research agenda of bringing together the collective intelligence of markets with the single-agent intelligence that has been the province of much of classical AI and classical machine learning.

**Stretch goals:** By 2040, we understand how to build robust AI systems that can interact in market-based contexts, making decisions that are responsive to preferences, and with market-based AI systems that achieve normative goals, and more effectively than the markets that preceded them. Market-based AI systems are adaptive to global shifts in behavior, robust against perturbations or malevolent interventions, and transparent in helping individuals to understand their role in a system. We can build robust learning systems that adjust the rules of markets and mechanisms, in order to promote objectives (efficiency, fairness, etc.). We also have vibrant, open platforms that support *economy-of-learning-systems* that promote innovation, competition, and composition of capabilities. Milestones along this path include—

**5 years:** Market-based AI systems are able to provide explanations to users about outcomes (e.g., purchases, prices), and learning systems can model other learning systems, and reason about behavior, including the way in which behavior depends on behavior of others. Learning systems can learn the preferences of individuals and build user confidence through transparency.

**10 years:** We have open platforms that support economy-of-learning-systems and succeed in driving innovation in data and algorithms, together with easy composition of capabilities. Learning systems can be employed to adapt the rules of interaction and mechanisms, in response to global changes in behavior (preferences, economic conditions, etc.), which will result in robust, market-based systems. Learning systems understand the cause-and-effect between actions and outcomes in multi-agent settings.

**15 years:** AI systems can work successfully in open, multi-agent worlds and market-based worlds, and they are effective in representing the interests of individuals and firms. Markets become more efficient, with better decisions about scarce resources, and humans spend less time making small consumption decisions. Learning systems are robust against the uncertainties that arise from the decision making of others. Learning systems can be used to solve inverse problems, going from normative goals, including those related to fairness, to mechanisms and rules of interaction.

**Durable Machine Learning Systems**

Despite tremendous progress, there remain intrinsic challenges in building machine learning systems that are robust in real-world environments. Tasks that naturally have minimal training data and situations that are observed only rarely are not well handled by current methods. Furthermore, intelligent systems today often live in a transactional setting, reacting to a single input, but without long-term evolution and accommodation of new scenarios and new distributions of data. Often learned modules are treated as endpoints, yet they should be reused and composed to exhibit greater intelligence. As a result, today's machine learning systems generally have a very short life span. They are frequently retrained from scratch, often on a daily basis. They are not durable.

These challenges cross-cut many disciplines in learning and artificial intelligence, such as computer vision, robotics, natural language processing, and speech. The ultimate goal for AI has always been general artificial intelligence: pioneers in the field envisioned general-purpose learning and reasoning agents; the recent focus on specialized systems can be seen as a step along the way but is not viewed as the pinnacle for AI research.

Consider systems that are designed to work over significant stretches of time, involving data gathered at many different places and involving many individuals and stakeholders. Systems like this arise in domains such as commerce, transportation, medicine, and security. For example, the knowledge employed in the medical treatment for a single patient can be conceptualized as a vast network involving data flows and the results of experiments, studies, and past treatments conducted at diverse locations on large numbers of individuals, against the background of changing ecologies, lifestyles and climates. Bringing all this information together requires ascertaining how relevant a given datum is to a given decision, even if the context has changed (e.g., different measurement devices, populations, etc.). It requires recognizing when something significant has changed (e.g., a mutation in a flu virus, availability of a new drug). It requires coping with the inherent uncertainty in diagnostic processes, and it requires managing biases that may creep into the way experiments are run and data are selected. It requires being robust to errors and to adversarial attack. Finally, while we wish for a certain degree of autonomy from the system—it should learn from data and not require detailed human programming or management—we must also recognize that for the foreseeable future human judgment and oversight will be required throughout the lifetime of any complex, real-world system.

In this section, we identify some of the specific research challenges associated with building durable machine learning systems, progressing toward a world of more and more general artificial intelligence.

**Dealing with Data Shift**

The knowledge acquired through machine learning is only valid as long as the regularities discovered in the training data hold true in the real world. However, the world is continually changing, and we need AI systems that can work for significant stretches of time without manual re-engineering. This is known as the problem of data shift.

It is generally acknowledged that current state-of-the-art machine learning, particularly in supervised settings, is brittle and lacks a dynamic understanding of the way data evolves over time and in response to changing environments or interactions. We need improved methods for detecting data shift, diagnosing the nature of the shift, and repairing the learned model to respond to the shift.

Detecting data shift need not be a purely statistical problem. Strong provenance information can directly tell the AI system how the data have changed and provide valuable information for responding to the shift. For example, suppose a sensor has been upgraded and the new sensor performs internal normalization. Provenance information can suggest that it is reasonable to learn how to transform the new normalized values into the old value scale so that the learned model does not need to be changed.

To build more flexible systems over time and in the event of an abrupt shift, we must be able to rapidly respond when a shift is detected and develop a resolution. Each system must ask whether it has the ability to re-train in a timely fashion and if it has easily accessible and available data to build a new model. This becomes even more important when handling real-time systems where the scale of data may make re-training prohibitively expensive. In this case, we must ensure rapid model switching and evaluation. Diagnostics, in this case, would act proactively to evaluate existing models and train parallel models that can account for whether data is anomalous and develop replacement models with varying distributions that examine ways to effectively assess new/online or unstructured data.

**Machine Learning and Memory**

The ability to create, store, and retrieve memories is of fundamental importance to the development of intelligent behavior. Just as humans depend on memory, advanced machine learning systems must be able to leverage the same kinds of capabilities. Humans also make use of external, collective memories to supplement their own, for example, through the use of libraries or, more recently, the World Wide Web. Machine learning systems must do the same through the use of external memory structures and knowledge bases.

Deep learning has become the de facto tool for analysis of many types of sequential data, such as text data for machine translation and audio signals for speech recognition. Beyond these marquee applications, deep learning is also being deployed in a broad range of areas including analyzing product demand, weather forecasting, and analyzing biological processes. Other critical areas involve systems for perception and planning in dynamic environments. Deep learning methods provide significant promise in providing flexible representational power of complex sequential data sources. However, in these areas, there are three fundamental challenges: 1) forming long-range predictions, 2) adapting to potentially rapidly changing and non-stationary environments, and 3) persisting memory of relevant past events.

Deep learning models have the potential to capture long-term dependencies; these dependencies can be critical for the task at hand, such as suggesting routes in a traffic system. However, a question is how to flexibly adapt to new settings. For example, a traffic system has to rapidly adapt to road closures and accidents. Unfortunately, most existing deep learning training procedures require batch (offline) training, and so they are unable to adapt to rapidly changing environments while persisting relevant memory. The common approach of dividing the training data into chunks (or batches) and learning a model initialized from the previous chunk (warm start) leads to catastrophic forgetting, where the model parameters adapt too quickly and overfit to the new chunk of training data. For example, when the road opens again, there is no memory from previous chunks when the road was open, so the system needs to learn about open (normally functioning) roads from scratch. Current methods for alleviating catastrophic forgetting only address one-shot classification tasks and do not address handling the temporal dependencies in sequence models. Note that there is a push-and-pull between rapid adaptation and persisting relevant information from the past, especially when one returns to previously seen settings. A related challenge is one of domain adaptation. For example, suppose the road closure was due to construction leading to a short new road segment, but an overall similar road network.

One potential avenue is to leverage new modeling techniques instead of modifying training methods. Attention mechanisms, for example, have shown promise in identifying contexts relevant for making predictions. In traffic forecasting, for example, the predicted congestion at an intersection depends on flow into and out of that intersection as well as congestion at similar types of intersections; attention can learn such relationships. As another example, memory mechanisms allow deep learning methods to store important past contexts, which can help alleviate the problem of forgetting past learning. For example, the traffic system could leverage memory to store the state of traffic dynamics prior to a road closure and recall them when the road reopens. Although attention and memory can highlight past states relevant to predicting long-term outcomes, there are many opportunities for improving long-term predictions in sequential models by learning seasonal patterns and long-term trends.

Turning to another domain where memory is critical to the next stage of advances in machine learning, the building of conversational agents has been a long-standing goal in natural language processing. Most current agents are little more than chatbots, however. Progress toward more natural, complex, and personalized utterances will hinge on agents having a large amount of knowledge about the world, how it works, and its implications for the ways in which the agent can assist the user. It will also require the agent to be able to build a personalized model of the user's goals and emotional state. Such a model will need long-term memory and lifelong learning, since it must evolve with each interaction and remember and surface information from days, weeks, or even years ago. Another related technical challenge is collecting and defining datasets and evaluation methodologies. Unlike domains such as image classification where the input can be treated as an isolated object, dialog is an ongoing interaction between the system and a user, making static datasets and discrete label sets useless. How to create good testbeds (and collect good datasets) and how to evaluate such interactive systems is an open research question.

Machine learning systems must also be designed to exploit the wealth of knowledge that has been accumulated through the ages. As we have previously noted, it should be possible to leverage the domain knowledge developed in scientific disciplines (e.g., basic principles such as conservation of mass and energy, mechanistic/process-based models) in machine learning frameworks to deal with data-sparse situations and to make machine learning algorithms generalizable to a greater range of data distributions. In the absence of adequate information about the physical mechanisms of real-world processes, machine learning approaches are prone to false discoveries and can also exhibit serious inconsistencies with known physics (the wealth of knowledge accumulated through the ages). This is because scientific problems often involve complex spaces of hypotheses with non-stationary relationships among the variables that are difficult to capture solely from raw data. Leveraging physics will be key to constraining hypothesis spaces in machine learning for small sample regimes and to produce models that can generalize to unseen data distributions.

In the natural language domain, existing recurrent neural network methods can be applied to compose word vectors into sentence and document vectors. However, this loses the structure of the text and does not capture the logical semantics of the language. By combining symbolic representations of linguistic meaning such as semantic networks, knowledge graphs, and logical forms, with continuous word embeddings, a hybrid method could integrate the strengths of both approaches. New representational formalisms and learning algorithms are needed to allow an AI system to construct such hybrid representations, reason with them efficiently, and learn to properly compose them. Such techniques are another way to exploit memory and accumulated knowledge in machine learning systems and could lead to improved document understanding, question answering, and machine translation.

Overall, new advances are necessary to form well-calibrated long-term predictions and persist relevant information from the past while performing continual, life-long learning.

**Transfer Learning**

A notion related to memory is the ability to transfer or adapt hard-earned knowledge from one setting to a different setting. Many of the most effective machine learning methods today require very large amounts of training data, which entails significant investments of time and expense in their development. However, some tasks seem so similar in their inputs and intended outputs that it ought to be possible to leverage the investments in building one system to reduce the development time for another. Imagine a classifier trained to recognize dogs in an image then being adapted to recognize cats.

Transfer learning and meta-learning (or learning to learn) have received much attention as approaches to transfer what has been learned in a source domain (e.g., with large amounts of data) to a destination domain where there is little available data. Computer vision and natural language processing are two popular application areas for transfer learning. We might ask, for example, whether classifiers trained on standard image datasets can be applied in medical imaging scenarios where much less data is available. Transfer learning is particularly valuable in the case of deep learning systems, since they require significant resources to build. Its success is based on the notion that at some layer of the network, the features learned for one problem can be effectively adapted for another problem without having to rebuild the network from scratch.

Transfer learning can be divided into *forward* transfer, which learns from one task and applies to another task, and *multi-task* transfer, which learns from a number of tasks and applies to a new task. Diversity in the training data is the key to success, but defining and capturing an effective notion of diversity remains an open research challenge. Transfer learning is often discussed in the context of few-shot, one-shot, and zero-shot learning, addressed in the next section. Meta-learning, on the other hand, is the concept of learning how to solve a new problem from previously accomplished tasks, not just from labeled training data as with traditional machine learning. Ideally, a meta-learner can acquire useful features for solving a new problem more efficiently and learn to avoid exploring alternatives that are clearly not productive. While offering intriguing possibilities, meta-learning approaches to date are sensitive to the initial task distributions they are presented with; designing effective task distributions is an open research problem.

While transfer learning has demonstrated some promising potential, more research is needed to continue to quantify its inherent uncertainty. The path forward will require innovation: clearly we should strive to be fully confident, but if there is transfer to be had from a large data source, we should be more confident than if we only had a smaller dataset to learn from.

**Learning from Very Small Amounts of Data**

As AI and deep learning models are applied to more domains, we are confronted with more and more learning scenarios that have a very limited number of labeled training examples. Furthermore, there can be continual changes in data distributions (e.g., dynamic systems) resulting in insufficient examples to retrain the model. In other scenarios, it is infeasible to obtain examples in some low-resource applications (e.g., rare diseases in health care domains). Therefore, there is a pressing need to develop learning methods that can learn effectively with fewer labels.

One promising solution is few-shot (and zero-shot) learning. These approaches typically work by first learning about various sub-parts of objects and their attributes. Then when a new object is observed for the first time, its sub-parts and attributes are compared to known sub-parts and attributes, and a recognizer for the new class is synthesized. For example, consider a computer vision system for recognizing birds. By studying multiple bird species, it can discover that important properties include the shape and length of the beak, color of the back, color of the breast, and so on. It can learn that an American robin has a black back, an orange breast, and a short light-orange beak. The first time it is shown a wood thrush, it can infer that it is similar to a robin but its back and beak are reddish-brown and its breast is white with black spots.

While many few-shot classification algorithms have achieved successes in some learning scenarios, there is still major room for improvement. First, the effectiveness of existing few-shot learning models varies. Second, it is usually assumed that the training samples (with limited examples) and the testing samples have the same distributions; while in practice there could be data shift.

Another important path to addressing low-resource learning is incorporation of domain knowledge or physical models with data-driven models (as discussed above). In many dynamical systems, such as energy systems, traffic flows, turbulent flow, aerospace and climate models, pure data-driven approaches may lead to incorrect and uninformed decisions due to limited data compared with vast state space. Models based on physical principles and domain knowledge have been developed with some initial success, although mathematical models based on these physical principles may have flaws and omit important factors. An important goal is to achieve highly accurate models by integrating physical principles with data. Research in this area will develop successful methodologies and easy-to-use tools for solving a wide range of problems in science, engineering, medicine, and manufacturing.

**Stretch goals:** Building durable machine learning systems requires developing sound principles for coping with the challenges discussed earlier in this section. By 2040, AI systems must deal with data shift, realizing that the real world is constantly changing and recognizing when the data they were trained with is no longer representative. They must find ways to exploit memory—knowledge that they or others have previously acquired. They must be able to adapt from problems they know how to solve to problems that are new through transfer learning. And they must be able to learn effectively when confronted by new and novel situations they have not encountered before. Milestones along this path include—

**5 years:** AI systems that are able to identify and correct for data shift in focused domains (e.g., a specific kind of medical diagnostic). Continued melding of data-based learning and first-principles for tasks with physical interpretations. Development of more effective paradigms for transfer learning, including increased ability to characterize task sets and their mappings into new domains. Better, more effective schemes for few-shot and zero-shot learning.

**10 years:** AI systems that begin to integrate durability along two or more of the dimensions described above, at the same time.

**20 years:** Truly durable AI systems that can simultaneously recognize and correct for data shift, exploit past knowledge and first principles in a manner reminiscent of human experts, transfer hard-earned knowledge from known problems to another related problem, and identify and perform competently when confronted by completely new tasks.

### INTEGRATING AI AND ROBOTIC SYSTEMS
#### Learning From, and For, Physical Interaction

Robots must interact with the physical world. More so, robots can be seen as a tool for automating our physical world, similar to how digital computing has automated our organization of and access to information. Through autonomous robotics, society will be able to envision and operationalize new ways to make our physical world programmable. The challenge we face, however, is that the interface of autonomous systems and the real world is filled with uncertainty and physical constraints. The convergence of robotics and AI is well suited to address these challenges and to yield next-generation autonomous systems that are robust in the context of changing real-world situations.

Robots, in order to move, must rely on physical contacts between surfaces and wheels or feet, or interactions between fluids (water, air) and wings and propellers. Locomotion often involves repeated, cyclical motions in which energy may be stored and released. Often the energy is stored mechanically, for example in springs. Currently, the mechanical and actuation components of a robot are designed first by one team; another team then develops the controllers and algorithms. Neither of these teams is able to fully model and understand the uncertainty imposed on the entire system, which limits the adaptability and robustness of the design. An exciting challenge for future work in locomotion is to replace this process with a simultaneous, collaborative design of both the hardware and the software.

Robot dexterity for grasping and manipulating objects remains a grand challenge for robotics research and development. These are crucial capabilities for robots to be useful in flexible manufacturing, logistics, and health and home-care settings, among many others. Manipulators that are currently deployed in industrial settings are designed to perform highly repetitive and exactly specified tasks. In contrast, manipulators operating in unstructured environments, such as homes, hospitals, or small-scale manufacturing sites, must handle a variety of objects, including fluids and non-rigid objects (e.g., cables and cloth). Operating in such environments requires the ability to recognize novel objects, assess their physical properties and affordances, and combine real-time feedback from visual, force, and touch sensors to constantly update those assessments and use them to generate robust controls.

While the robotics research community has developed techniques that enable manipulators to pick up rigid objects or cloth in research lab settings, fundamental advances in perception, control, and learning are necessary if these skills are to transfer to complex tasks in unconstrained environments. Consider, for example, grasping and sliding a cable through a small hole or mixing flour into cake batter, while being robust to details such as changing lighting conditions, cable stiffness, unpredictable tools and containers, and batter viscosity. Representations that capture the salient knowledge for these kinds of tasks are too complex to be manually designed and need to be learned from both real-world and simulated experiences. While recent advances have enabled progress in isolated manipulation settings, future manipulators must be able to learn far more capable representations and adapt them on the spot, without needing extensive offline learning or manual data labeling. This will require compression of high-dimensional sensor data into representations that are useful for a robot's task and allow for effective communication and collaboration with humans.

**Stretch goals:** By 2040, every robotic system should be able to safely and dexterously manipulate objects in common everyday human environments, including physically assisting people. Every system should also be continually updating models of how to perform various manipulation tasks and behaviors, helping itself and other robots to understand the form and function of objects and the behaviors of people in order to perform ever-safer and more effective physical manipulation, collaboration, and assistance. Every robot will be able to use AI systems to reason over affordances to realize goals and perform effectively in complex human environments. Every robot will have access, through the cloud, to a large repository of data, knowledge, and plans to facilitate real-time understanding of the form and function of the vast range of objects and of human behavior in the physical world. Newly created robots will be able to adapt and use these cloud repositories immediately, or through simulation-based training. Milestones along this path include—

**5 years:** Learning of representations that combine multi-sensor data to support robust operation in cluttered scenes. Learning from explicit instruction and demonstration by non-experts in day-to-day environments, one-on-one and group settings.

**10 years:** Learning of representations that support a broad set of manipulation tasks, including handling of novel rigid and deformable objects and liquids. Learning from observations of real-world environments and interactions. Safe physical manipulation of objects around people, such as in eldercare settings and homes.

**20 years:** Learning of representations that support dexterous manipulation of arbitrary objects in unstructured environments. Systems deployed in real-world human environment engaging in lifelong learning and continuous adaptation to the changing dynamics of the environment, including people. Safe physically assistive manipulation of people, such as in care for the elderly.

**Learning From Humans**

As the applications of AI and robotic systems expand, robots will increasingly be introduced to human environments, including homes, workplaces, institutions, and public venues. In these settings, they will interact and collaborate with people, which will require understanding the requests, actions, and intents of those people, and require acting in ways that do not violate both explicit and explicit human norms and expectations, including ethical standards. Such capabilities will demand designs that incorporate fundamental capabilities for interaction, collaboration, and human awareness and that can adapt to different environments and users. These capabilities are extremely complex, and information needed to achieve the necessary adaptations is often unavailable at design time, necessitating the use of learning methods.

Humans can facilitate robot learning in a number of ways, including providing explicit instruction and demonstrations through multimodal input, structuring the environment in ways that facilitate the robot but do not inconvenience people, and giving feedback to the robot as it explores a new environment to adapt its capabilities. These approaches require rich perceptual information from humans in relation to their environment, such as understanding verbal commands, including spatial references, detecting entities that are referenced through pointing or gazing, continuously tracking objects that are manipulated by the human, and inferring intent or missing information that is not directly and explicitly communicated. Making sense of this information and acting appropriately in response requires advances in sensing, recognition, prediction, and reasoning. Robots will need to not only construct models of human behavior and plan their own actions accordingly, but they must also communicate their own goals, intent, and intended actions to users in natural and informative ways.

Robots deployed in mission-critical settings (e.g., search and rescue, law enforcement, space exploration), in physical and social assistance (e.g., therapy, rehabilitation, home healthcare), and as expert tools (e.g., robot-assisted surgery) especially need to have robust and adaptive models of human input, states, and goals.

**Stretch goals:** By 2040, robotic systems will obtain new physical and social capabilities through increasingly more sophisticated ways of learning and applying learned capabilities appropriately, depending on social context and user preferences, to improve human trust and rapport. Milestones along this path include—

**5 years:** Learning from explicit instruction and demonstration by non-experts in day-to-day environments, one-on-one and group settings.

**10 years:** Learning from observations of implicit human behavior and environmental changes, adaptively engaging in active learning to improve model robustness.

**20 years:** Systems deployed in real-world human environment engaging in lifelong learning and continuous adaptation to the changing dynamics of the environment. Systems also execute learned capabilities in appropriate ways to improve human trust.

### Infrastructure and Middleware for Capable Learning Robots

Effective robot autonomy can be thought of as following a sense-plan-act pipeline involving many AI components that must work in synchrony: a computer vision system that senses the state of the world from the robot sensors, an automated planner that chooses actions to achieve a goal state, and a motion controller that carries out those planned actions. Machine learning plays important roles throughout this pipeline, drawing inferences from data to provide functionality where hand-coded solutions are infeasible.

The Robot Operating System (ROS) provides a common framework for developing AI software for robots, allowing heterogeneous communities of researchers, companies, and open-source developers to build on each other's work. Since the original robotics Roadmap in 2009, this infrastructure has served as a foundation for the rapid advancement of robotics research and development. However, these initial robot middleware efforts were geared specifically for robotics researchers. More general open-source middleware platforms, with broader functionality, are needed for prototyping new AI systems with direct uses in real-world settings involving sensing and action in the real world. A significant challenge here will be to fully understand the sense-plan-act pipeline and its interplay with robot components (software, hardware, communication) from disparate, and possibly malicious, sources. Solutions will require generalizable standards and testing suites for widely varying robot systems.

**Stretch goals:** By 2040, highly capable AI systems will be employed broadly across robotics domains, to the point that such use becomes commonplace. Just as today we take for granted that any mobile robot can map, localize, and navigate in a new environment, we will come to expect that all future robots will have the new capabilities that will be provided by forthcoming advances in AI. Reaching this level will require that AI applications can be used portably, safely, and securely in a manner that earns the trust of users. Milestones along this path include—

**5 years:** Widespread deployment of next-generation middleware for AI and robotics and accompanying testing standards for real-world conditions.

**10 years:** Industry adoption of robust security and safety standards for AI applications built on common middleware.

**20 years:** Rich ecosystem of user-centered and privacy-preserving AI applications that are portable across different robot platforms.

## 4. Major Findings

Several major findings from community discussions drive the recommendations in this Roadmap:

**I. Enabled by strong algorithmic foundations and propelled by the data and computational resources that have become available over the past decade, AI is poised to have profound positive impacts on society and the economy.** AI has become a mature science, leveraging large datasets and powerful computing resources to produce substantial progress in many areas: exploration and training of statistical models, for instance, and powerful image and video-processing techniques. Many other areas of AI might be amenable to the same dramatic leaps forward, but are starving for appropriate data. And while collection, processing, and annotation of data are key aspects of an experimental science, architectures and frameworks are also instrumental in AI solutions. Enabled by the right theoretical and applied foundations and fueled by massive datasets and growing computational power, future AI-driven successes could affect many aspects of society, including healthcare, education, business, science, government, and security (as shown in Figure 3): removing humans from harm's way in dangerous yet vital occupations; aiding in the response to public health crises and natural disasters; expanding educational opportunities for increasingly larger segments of our society; helping local, state, and federal government agencies deliver broader range of valuable services to their citizens; or providing personalized lifelong health- and wellness care accessible to each individual's changing needs.

**II. To realize the potential benefits of AI advances will require audacious AI research, along with new strategies, research models, and types of organizations for catalyzing and supporting it.** Over its first few decades, AI research was characterized by steady progress in understanding and recreating intelligent behaviors, with deployments of AI-enhanced systems in narrow application areas. Since the mid-1980s, fundamental AI research has been supported largely by short-term grant-funded projects in small single-investigator labs, limiting the types of empirical work and advances possible. In recent years, the experimental possibilities enabled by data and computationally resource-rich and generously staffed industry labs have yielded significant advances, enabling wide deployment of AI-enabled systems in many societally relevant and important venues. Cross-fertilization with other fields, ranging from social science to computer architecture, is also critical to modern AI, given the demands, breadth, and implications of its applications. To adequately address the next generation of AI challenges will require sustained effort by large, interdisciplinary teams supported by appropriate resources: massive datasets, common architectures and frameworks, shared hardware and software infrastructure, support staff, and sustained, long-term funding. This Roadmap offers new models for the resources, the critical mass, and the long-term stability that will be needed in order to enable a new era of audacious AI research that is significantly more integrative and experimental while also recognizing the need for caution regarding the impact of AI in society.

**III. The needs and roles of academia and industry, and their interactions, have critically important implications for the future of AI.** Building on the foundations of past AI research, most of which was conducted in academia, the private sector has compiled and leveraged massive resources—datasets, knowledge graphs, special-purpose computers, and large cadres of AI engineers—to propel powerful innovations. These assets, which provide major competitive advantages, are generally proprietary. Furthermore, the constraints, incentives, and timelines in these two communities are very different: Industry is largely driven by practical, near-term solutions, while academia is where many of the fundamental long-term questions are studied. Solutions to

the next generation of AI problems cannot come from either academia or industry alone. Without the right resources, academic AI research is limited; without answers to foundational questions, industrial AI innovations will be limited. The ability to explore different approaches and models through experiments on practical problems in real-world settings is limiting the ability of universities to do experimental research.

**IV. Talent and workforce issues are undergoing a sea change in AI, raising significant challenges for developing the talent pool and for ensuring adequate diversity in it.** There is great demand in industry for AI talent and the gap between the supply and the demand is likely to grow significantly over the next decades. All US universities are looking for AI faculty—and struggling to hire, particularly at senior levels and in areas relevant to industry needs. Students flock to AI course offerings and research groups. Faculty across a broad swath of fields (including humanities as well as social and physical sciences, and in such professional studies as law, medicine, and public health) seek AI collaborators, and new data-science programs continue to emerge. Drawn by the higher salaries, extensive data and infrastructure resources, and numbers of potential collaborators and support staff available in industry, many AI faculty are leaving academia or pursuing appointment arrangements where they split time between a university and a corporate research lab. While this trend creates opportunities for companies and universities to engage in innovative ways of working together to advance AI, it also has the potential to negatively affect AI education, training, and the AI research pipeline. Diversity is also a major issue. Although the number of women in STEM fields is increasing, the number of women in computer science has halved since the 1980s. In AI, the participation by women and underrepresented minorities is even lower than in computing as a whole, despite the increases in college graduation rates in these groups.

**V. The rapid deployment of AI-enabled systems is raising serious questions and societal challenges encompassing a broad range of capabilities and issues.** Novel AI technologies have been deployed quickly into market before incorporating systems engineering, safety design principles, best practices, and societal considerations. Research is needed into the incorporation into AI systems of responsibility criteria in autonomy and assistance in order to enable frameworks that are suitable for operations in critical domains, as well as a clear articulation of the limits of AI-enabled systems—settings where such systems might on balance provide more harm than good. Research is also needed into what these best practices might be and what kinds of institutional or legal scaffolding would make them most effective. It is critical for these efforts to be supported as an integral part of AI research and development, rather than as ex post facto analysis of systems after they are built. Considerations include the fairness of decisions made by these systems, as well as their potential to introduce and amplify structural patterns of inequality and distortions of the truth. The ethical ramifications of AI-assisted decision making and content generation are critical near-term issues, given that these technologies are being used as replacements of, or assistants to, human decision makers in areas as crucial to individual lives and societal well-being as criminal justice, predictive policing, credit risk systems, employment, college admissions, student and teacher evaluation, autonomous vehicles, and national security. In all of these diverse settings, AI technologies have the potential to introduce profound structural change in the way we make decisions, who gets to make them, and for whom. AI assistance is also being used extensively for content filtering and dissemination—especially on social media, mostly invisibly, and sometimes with significant consequences. It is important to ensure that these systems are designed and deployed safely and responsibly, and with adequate oversight. It is presently challenging to explain the decisions of current AI methods, for instance, leaving the user in the dark as to which factors influenced the outcomes. Addressing this constellation of challenges will require deep, extensive studies of the interactions among AI, ethics, and society, with concern for a broad range of normative questions including fairness, accountability, and transparency in AI-assisted decision making. We will also need to identify different ways of thinking about the design of these systems so that such concerns are taken into account before they are deployed and cause harm. This kind of multidisciplinary effort is naturally and best located at the nexus of the social sciences, humanities, and computing and within university environments. Locating these efforts within university environments is also important in the light of rapidly differing incentives between industry and academia that are influencing the increasing tension around the social and ethical ramifications of widely deployed AI systems.

## Substantial Long-Term Investments Drive Significant Advances

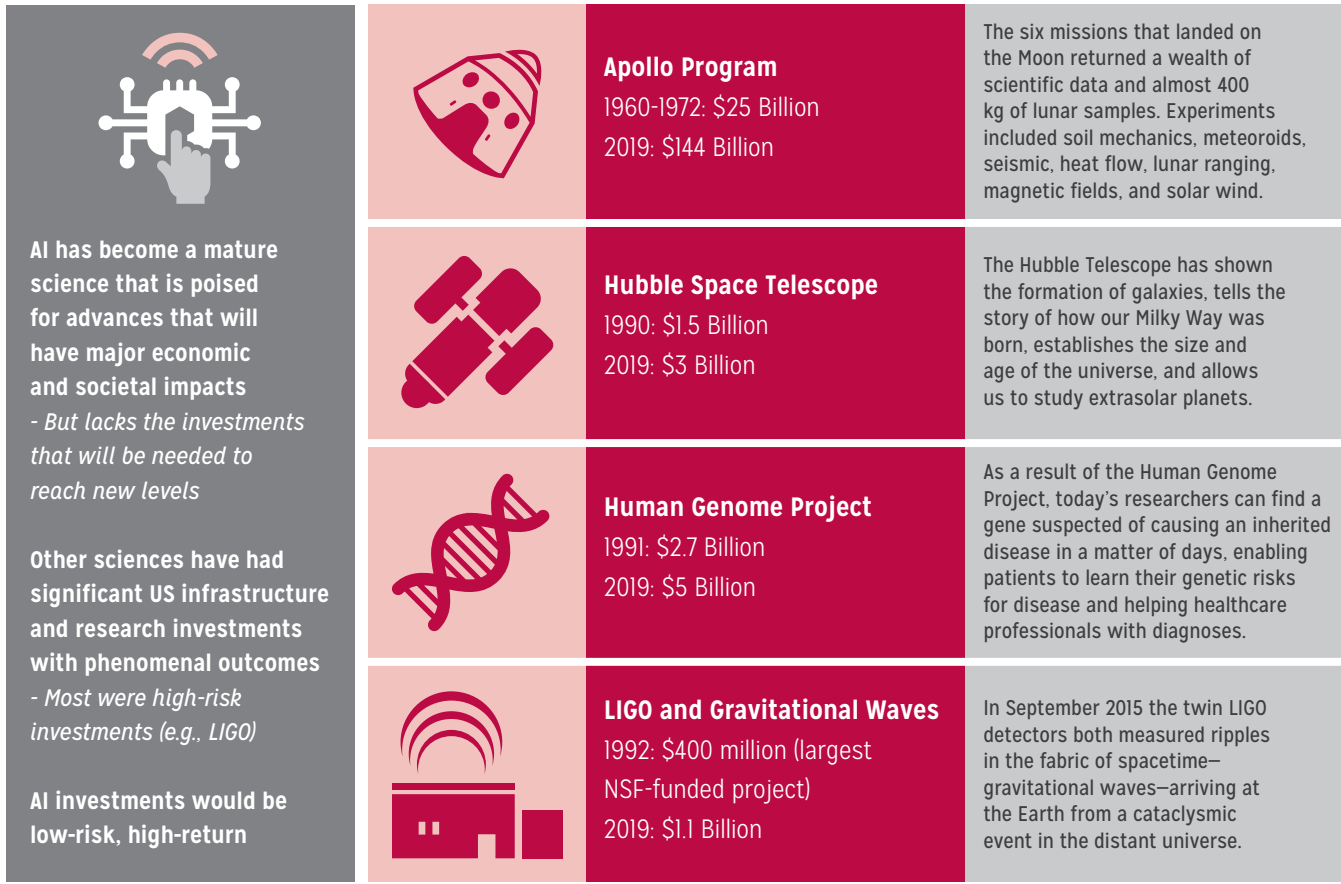

*Figure 4. Substantial Long-Term Investments Drive Significant Advances*

**VI. Significant strategic investments in AI by the United States will catalyze major scientific, technological, societal, and economic progress.** Important advances have been propelled by significant US investments in audacious projects over the past 50 years. The Laser Interferometer Gravitational Wave Observatory ($1.1B), for instance, led to the discovery of gravitational waves. The Human Genome Project ($2.7B) was the driver of major innovations in biomedical research. The Apollo program ($144B) not only accelerated space travel but also catalyzed many scientific contributions. AI is ready for similar forward leaps. Recognizing this, other major industrialized countries are already embarking on major AI research and education programs. As of the time

Figure 4 References:

Apollo Program: https://airandspace.si.edu/explore-and-learn/topics/apollo/apollo-program/

Hubble Space Telescope: https://www.nap.edu/read/11169/chapter/5#20

Human Genome Project: https://report.nih.gov/nihfactsheets/viewfactsheet.aspx?csid=45

LIGO and Gravitational Waves: https://www.ligo.caltech.edu/detection

of this writing, Germany[23] and France[24] have allocated 3B and 1.5B euros to AI, respectively. The UK[25] has pledged an investment of 1B pounds in AI, together with dedicated funding for 1,000 PhDs and 8,000 specialized teachers in AI, and has repurposed its flagship Turing Institute into a major data-driven AI research center. China[26] has announced that it will invest billions in AI over the next five years, creating at least four $50M/year AI Centers and a $1B/year National AI Research laboratory with thousands of AI researchers and engineers, and committing to training 500 instructors and 5,000 students at major universities. Significant investments in AI, structured with the guidance of this document, will allow the US to take the forefront in propelling the field into a new research era and create significant impact across all sectors of society and the economy.

## 5. Recommendations

The progress desired in integrated intelligence, meaningful interactions, and self-aware learning pose substantial research challenges that require significant investments that are sustained over time. History tells us that there is no silver bullet, no substitute for sustained effort. For example, speech recognition took four decades of research, mostly in academia, until the work was mature enough to be commercialized (e.g., the commercial Dragon speech recognition software first appeared in the market in the 1990s) and eventually mainstream (e.g., conversational assistants). Similarly, the basic ideas in neural networks were first developed in the 1980s, and it took decades of research as well as improvements in computational resources before they were able to take off. This research Roadmap is designed to prioritize investments and foster similar advances in technologies across AI over the next two decades.

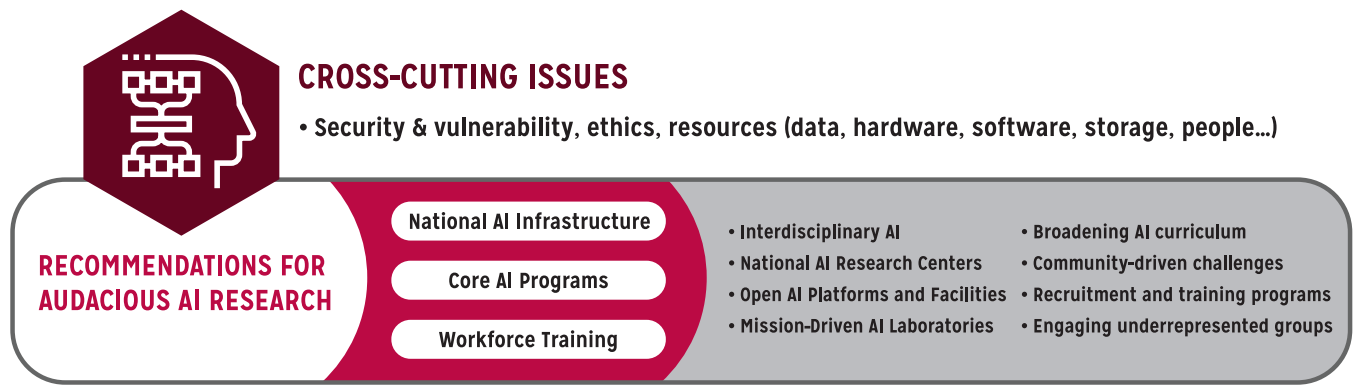

*Figure 5. Overview of the 20 Year AI Research Roadmap Recommendations: A national AI infrastructure combined with training of an all-encompassing AI workforce.*

---

[23] "Germany plans 3 billion in AI investment: government paper," Reuters, November 13, 2018, https://www.reuters.com/article/us-germany-intelligence/germany-plans-3-billion-in-ai-investment-government-paper-idUSKCN1NI1AP.

[24] "Emmanuel Macron Talks to Wired about France's AI Strategy," Wired, March 31, 2018, https://www.wired.com/story/emmanuel-macron-talks-to-wired-about-frances-ai-strategy/.

[25] "U.K. Unveils $1.4 Billion Drive Into Artificial Intelligence," Bloomberg, April 25, 2018, https://www.bloomberg.com/news/articles/2018-04-25/u-k-announces-1-4-billion-drive-into-artificial-intelligence.

[26] "Understanding China's AI Strategy," Center for a New American Security, February 6, 2019, https://www.cnas.org/publications/reports/understanding-chinas-ai-strategy.

**Achieving the audacious AI research vision put forward in this Roadmap will require a reinvention of the AI research enterprise in order to create a comprehensive national AI infrastructure and to reconceptualize AI workforce training.** The rest of this section makes specific recommendations, summarized in Figure 5. A number of important cross-cutting themes underlie all of these recommendations. Progress in AI cannot be decoupled from advances in computational resources and collection of massive datasets, nor can it ignore considerations regarding security, vulnerability, and ethics. As noted in the Code of Conduct of the Association for Computing Machinery, the field of computer science is called to "…Recognize when computer systems are becoming integrated into the infrastructure of society, and adopt an appropriate standard of care for those systems and their user." Ethics must be designed into AI systems from the start. These concerns are particularly acute in the context of high-stakes applications like healthcare, algorithmic bias, and the disparate impact of AI's application on various classes of individuals.

## 5.1 Create and Operate a National AI Infrastructure

**AI as a discipline has reached a level of maturity where significant progress can result from the availability of substantial experimental facilities.** The majority of academic AI research in the past has been driven by well-specified tasks in closed domains with short time horizons. Advancing research on AI systems that can take on open-ended tasks, carry out natural interactions, and behave appropriately under changing conditions will require living laboratories that exhibit the dynamics and unpredictability of the real world. Bringing about a new era of audacious AI research will require a National AI Infrastructure that will empower academic researchers while accelerating AI innovation and dissemination of research products.

The recommended National AI Infrastructure is not meant to replace existing AI programs, but rather complement and supplement them with significant resources.

The goal of the National AI Infrastructure is to reinvent the AI research enterprise through four interdependent capabilities: 1) open AI platforms and resources, 2) sustained community-driven challenges, 3) national AI research centers, and 4) national mission-driven AI centers. Figure 6 gives an overview of these components and their interactions. We describe each in turn below.

### 5.1.1 OPEN AI PLATFORMS AND RESOURCES

There are several key objectives for the Open AI Platforms and Resources:
- Provide the AI community with substantial experimental resources for both basic research and applications.
- Attract AI researchers to problems that are harder to tackle without significant startup resources.
- Promote collaborations across all areas of AI.
- Support multidisciplinary research by giving non-AI researchers access to AI capabilities.
- Reduce redundancy of effort and cost in the research enterprise, so research projects do not have to build up capabilities or collect new data from scratch each time.
- Reduce the cost of individual research programs to integrate relevant capabilities or to compare their work with others.
- Reduce cost of large teams of collaborators by providing already integrated or easy-to-integrate infrastructure.
- Provide the AI community with open resources that will bolster not only research in all of academia, but also in small technology companies, companies in other sectors, and government organizations.
- Provide students with practical training opportunities using operational-grade AI technologies.

◗ Attract and retain university faculty with attractive opportunities for research through access to productive ready-to-go environments for unique applications.

◗ Promote collaborations between academia, industry, and government in open research environments.

These platforms would include basic components (data repositories, software repositories, knowledge base repositories, software, services, computing and storage resources, and cloud resources), combinations of components into integrated architectures and integrated capabilities, and instrumented facilities (robotic and cyberphysical systems, instrumented homes, simulation environments, game environments, social platforms, instrumented hospitals, etc.). Important attributes of these components include:

◗ **AI-Ready:** They should offer capabilities that AI researchers can readily use with reasonable effort. This may require the implementation of services to do basic data preparation, or the integration of datasets with computing resources to facilitate algorithm testing, or APIs that enable new components to be interconnected with existing capabilities.

◗ **Open:** They should be shared under standard open licenses (including a range of recognized licenses such as unlimited reuse with attribution, no restrictions, commercial restrictions, share-alike restrictions, etc.) so that anyone can inspect the implementation and characteristics of the component and can reuse and extend it as allowed by the license agreement.

◗ **Composable:** They should offer capabilities to facilitate their integration and composition with other components. This involves adoption and development of standards, development of common APIs, mediation through ontologies and public knowledge resources, etc.

There are many efforts to promote community principles and best practices for sharing and reuse of research resources, such as the FAIR principles[27,] ontology and vocabulary standards (e.g., RDF, Common Logic), provenance standards (e.g., W3C PROV), etc. The AI research community should strive to contribute to the development of best practices and the understanding of what specific requirements are most appropriate to support AI research.

The AI community was one of the first to rely on shared data repositories to drive research. One of the first data repositories ever created was the Irvine Machine Learning Data Repository, which was started in 1987 and was instrumental in the development of effective machine learning methods. The repository still operates today. More recently, many data repositories are associated with competitions and scoreboards (Kaggle, ChiLearn, etc.). For example, ImageNet, a large dataset of labeled images and an associated classification task, helped lead to the 2012 breakthrough in computer vision using deep learning methods. That in turn has helped enable the development of major new technologies, such as self-driving cars. Research in natural language understanding received a similar boost with the creation of the Linguistic Data Consortium (LDC), which was started in 1992 and has made significant investments to create datasets for the natural language community, including datasets with sophisticated annotations such as parse trees for sentences. However, the LDC model relies on a closed-license system, while this Roadmap recommends making AI resources open.

Beyond data, AI researchers have also recognized the importance of other shared resources and community standards. Several robotics platforms have been created over the years for research, experimentation, and commercial purposes, including PR2 (from Willow Garage), AIBO (from Sony), Nao (from Aldebaran Robotics), and Cozmo (by Anki). Software toolkits have also supported more efficient and comparative research, such as ROS in robotics programming, Open CV in computer vision, and Weka in machine learning. Shared communication and representation languages have also promoted integration and comparison of approaches, including KQML, PDDL, and RDF.

---

[27] Force11. The Fair Data Principles. 2016. https://www.force11.org/group/fairgroup/fairprinciples

Industry labs have made significant contributions to the creation of shared resources for the AI research community. Google's release in 2006 of a dataset of commonly co-occurring words based on automated analysis of more than a billion sentences created a revolution in language technologies. Facebook developed PyTorch as open-source software for deep neural networks, which has allowed the research community to make great strides in machine learning.

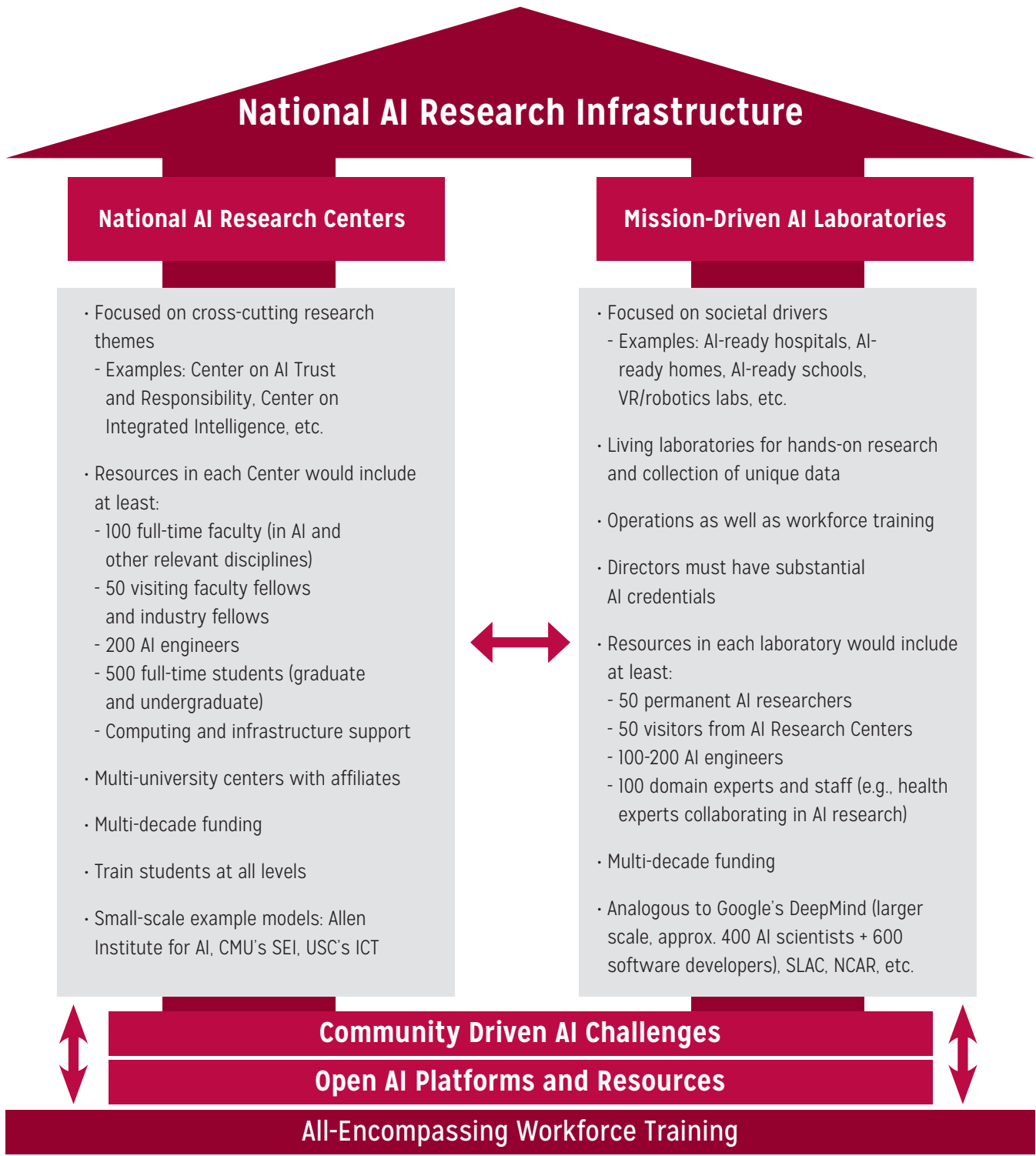

*Figure 6. Overview of the envisioned National AI Research Infrastructure.*

At the same time, many valuable resources in industry are not publicly shared and are very hard to recreate. Notable examples are the Google Knowledge Graph and AlphaGo. In the case of the Google Knowledge Graph, significant resources were invested to create this proprietary resource. In the case of AlphaGo, the deep net underlying the actual system has never been released, although it would be an amazing resource for AI researchers to study further. It is, of course, understandable that many of these AI resources developed in industry are not shared because they often provide a core competitive advantage for the companies involved.

There is a significant cost associated with maintaining public data repositories in the long term. A general challenge for any research community to sustain publicly available resources is finding a good balance of funding from government agencies, academic institutions, industry support, private donations, and foundation grants.

Unfortunately, open repositories that have been available to the research community in the past have ended up entangled in commercial interests. Figshare started as an open scientific data repository, and once it became successful it was acquired by Springer Nature. Google Open Data was a repository used by many that was shut down in a matter of months. It is important that datasets created by researchers for researchers using public funds be vigilantly protected, and that the Open AI Platforms and Facilities described in this Roadmap be created with necessary provisions for preservation for the public good.

The rest of this section gives an overview of high-priority resources envisioned by the community to be part of the Open AI Platforms and Resources, which include AI-ready data repositories, open AI software, open AI integration frameworks, an open knowledge network, and testbeds. The next sections describe each of these kinds of resources in detail.

**AI-Ready Data Repositories**
Because the collection and preparation of data often requires a significant amount of effort, the public availability of well-designed datasets are important drivers of research progress. There are many useful principles in the design of these data repositories that currently are not, but should be, widely adopted—such as proper documentation, quality control, clear problem and task statements, and sometimes incentives such as awards and scoreboards.

However, available data repositories used in the AI community tend to focus on very few tasks and very few data formats, because they are:

◗ Largely made up of labeled data for classification problems. Unfortunately, classification is a very narrow task within the much broader scope of capabilities that AI systems need.

◗ Fit for traditional formats that that are easy to characterize, use, and share. These tend to be curated compilations of images, text, and other readily available data. This is a very particular kind of data that lends itself to simple tasks (e.g., classification).

◗ Designed to demonstrate utility of algorithms, rather than utility of solutions.

◗ Collected for applications of commercial or specific research interest. Very few shared datasets are of public or societal interest.

For example, there are many data repositories for classification of text documents, or images of pets. There are few with sensor data for autonomous cars, personal task management (e.g., email exchanges, to do lists.), social interactions, etc. This makes it more challenging to address the forward-looking AI research areas outlined in this Roadmap.

Another important issue is that the many other shared data repositories are not easily usable by AI researchers, notably scientific and government datasets. Many scientific datasets would be appropriate for AI research, but are not easily findable outside the particular scientific community or are not easily usable by domain experts in the particular field of science. In some cases, scientific datasets are specifically prepared and intended for use outside the science domain but are not well known to AI researchers. Open government data offers another excellent potential source of data and concomitant opportunities, but the data are not in formats that are well documented or accessible.

Well-designed AI-ready dataset repositories will have a number of benefits:

- Allow researchers to seek and share datasets that address real-world problems and tasks, particularly in sectors and applications concerned with public and societal interests.

- Increase the opportunities for and productivity of AI researchers by making it easier to use datasets in their research, and facilitate comparing their solutions to others.

- Promote datasets that are available but not yet in a form that is conducive to research, such as government data and scientific datasets.

- Support data sharing for domains and tasks that have complex data, such as social communication, interacting behaviors, multi-criteria optimization, real-world goal accomplishment, etc.

- Promote research on managing shared datasets that contain sensitive data (e.g., personal data).

**AI Software and AI Integration Frameworks**

Open software facilitates research by providing a significant substrate to build upon. Some software libraries have come from academia (OpenCV, NLTK), others from industry (TensorFlow), and others through a combination of both (ROS). The availability of software that implements powerful approaches in an easily accessible manner spurs significant innovation. The open software movement provides many examples of the benefits of organic collaborative communities to support, and in many cases extend, software to production-grade quality.

AI needs an open collaborative ecosystem that includes a range of governance strategies for the software. Open software can be shared with minimal oversight through its publication with open licenses in open repositories with version control, a common practice today that will no doubt continue to be widespread. An alternative approach to open software is to form well-organized communities and processes organized around key software projects. For example, the framework provided by the Apache Software Foundation supports incubation of communities and establishment of governance and decision making, and has been a very successful approach to developing robust operational-grade software for both research and deployments.

Beyond software libraries, AI integration frameworks are a high priority in this research Roadmap. A wide range of approaches is possible, with varying benefits and tradeoffs. On one side of the spectrum, frameworks for integrating AI components would rest on the existence of interchange standards and the practical use of modularity and composability principles. Composing AI capabilities is challenging because they may have interacting behaviors and different assumptions that would need to be harmonized. The more heterogeneous the capabilities, the more challenging their integration. On the other side of the spectrum are frameworks that are designed in principled ways and can be adopted to develop components and capabilities that are well integrated by design. Both approaches can lead to a range of designs, depending on task constraints. For example, approaches may be based on cognitive science findings to create more human-like systems, or engineered for specific types of applications, or designed to support real-time robot platforms.

The availability of open AI integration frameworks and environments would encourage research on AI systems that combine a wide range of tasks and capabilities. This would also encourage the exploration of a diverse range of approaches and their assessment for different uses and purposes.

### 5.1.1.1 An Open Knowledge Network

A major theme in the research agenda outlined in this Roadmap is the need for AI systems to access significant amounts of knowledge about the world. The AI community has a tradition of knowledge sharing, promoted through standards (such as KQML, RDF, OWL), shared ontologies and knowledge bases (e.g., OpenCyc, WikiData), and open libraries and tools (Protégé, etc.). These traditions and practices must be complemented by a push toward generating large-scale knowledge bases, capturing both a wide range of everyday knowledge and specialist knowledge in important application areas (e.g., health, science, engineering).

Knowledge repositories have become a key competitive advantage for AI applications in industry. As tech giants push the envelope in markets such as search, question-answering, entertainment, and advertising, they have discovered the power of knowledge to make their systems ever more capable. These knowledge sources are usually captured as a graph containing entities of interest (e.g., products, organizations, notable people, events, locations, etc.) interlinked in diverse dimensions, and are highly structured to be machine-readable. Immense resources are being invested in Google's Knowledge Graph, Microsoft's Satori knowledge graph, Amazon's Product Graph, Facebook's Social Graph, and IBM Watson's Knowledge Graph. In addition to the big technology companies, many other large companies are creating knowledge graphs to drive innovative products (e.g., Thompson Reuters, Springer Nature, LinkedIn).

Although many of the technologies underlying these knowledge bases were originally developed in academia, researchers outside proprietary boundaries have limited or no access to these resources and no means to develop equivalent ones. For example, recent accounts of the Google Knowledge Graph place its size at 1 billion entities and 70 billion assertions about them, and Microsoft's at 2 billion entities and 55 billion assertions. These represent major investments that are very hard to replicate, but are all proprietary and not accessible to outside researchers. In addition, these knowledge bases are developed to support the commercial interests of the companies that created them, such as search results and ad placement. They do not necessarily contain material relevant to scientifically vital domains and other applications. Nor do they contain the broad commonsense knowledge that provides a useful substrate for the application of professional knowledge to daily life, common ground for communicating with people, and knowledge of the physical and social world needed by robots and other AI systems.

An *Open Knowledge Network (OKN)* with very large amounts of knowledge about entities, concepts, and relationships in the world would foster the next wave of knowledge-powered artificial intelligence innovations with transformative impact ranging from scientific research to the commercial sector. The OKN would enable new research in personal assistants who could better understand the world in which they operate, machine learning systems to enrich data with extensive information about the underlying objects, natural language systems to link words and sentences to meaningful descriptions, and robotics devices with context and world knowledge—among many others.

A recent report describes several key characteristics that the community considers desirable of a public resource such as the OKN.[28] First, it should be a highly distributed network of knowledge resources, with open contributions and open access to allow a diversity of uses. Second, the underlying representation should be extensible as further research is done on languages that accommodate more complex forms of knowledge. Third, the provenance of each piece of knowledge (i.e., its origins), should be explicitly represented. Fourth, the context where each piece of knowledge may be true should be qualified, whether temporal (e.g., a person is a college student only during a few years), spatial (e.g., seagulls are a common bird in coastal areas but not inland), etc.

The availability of the OKN would enable basic research in advancing the underlying technologies in order to provide AI systems with extensive knowledge about the world. Further research is needed to extend the paradigm of entities and their relationships, which is only sufficient to capture very basic knowledge. Much research is also needed in facilitating the variety of uses and application approaches that will be necessary, for example to answer new kinds of questions and to provide natural interfaces to users. New approaches to extract knowledge from text, images, videos, and interactions with people will be needed to keep the OKN up to date and to rapidly extend it for new applications and domains of interest.

The OKN should also include efforts to build knowledge resources for domains of societal interest such as science, engineering, finance, and education. Scientific research could use the OKN as a gateway to a vast trove of knowledge and data, and particularly to disseminate knowledge and data across disciplines.

---

[28] Networking and Information Technology Research and Development Program. Open Knowledge Nework: summary of the Big Data IWG Workshop. Oct. 4-5, 2017. https://www.nitrd.gov/pubs/Open-Knowledge-Network-Workshop-Report-2018.pdf

The OKN would also give industry and government organizations access to extensive world knowledge. Today, only the biggest tech companies have the resources to develop significant knowledge graphs. With an open resource such as the OKN, all companies, regardless of size and sector, would be able to use world knowledge to develop AI systems that can operate effectively in that world.

#### 5.1.1.2 AI Testbeds

AI testbeds must strive to balance tractability and real-world relevance. Many researchers choose to study simplified tasks in closed domains, rather than open-ended real-world problems, because toy tasks are more tractable for today's methods. In order to draw researcher attention, a task should admit partial progress with today's algorithms and architectures, yet not be fully solved. At the same time, overly simplified testbeds can drive illusory progress that appears impressive in the lab but does not transfer to the real world. In order to strike this balance between research tractability and real-world relevance, testbeds need to be carefully and iteratively crafted in a tight feedback loop between testbed designers and AI researchers.

AI testbeds would also help establish and maintain benchmarking standards to promote measurable research progress, allowing for principled comparisons of different approaches and facilitating reproducibility, thereby enabling the research community to make significant measurable progress.

The creation of a successful testbed should be rewarded as a valuable scientific contribution. New approaches that attain better performance on pre-defined well-scoped problems are often rewarded in the current system of publication incentives. By contrast, applications of existing algorithms to new domains, including the construction of novel testbeds that capture important real-world dynamics, are often viewed as not innovative enough for a publication in top conferences and journals. New incentives must be created to reward well-designed testbeds as valid contributions to research.

Best practices and guiding principles of AI testbed design will need to be developed. There are important research directions in determining whether a testbed captures a particular characteristic of a real-world environment, summarizing how the observations collected differ from real data, and whether those differences will affect the validity of AI systems developed on the testbed to perform well on the real environment.

The rest of this section gives examples of the kinds of testbeds that would propel AI research areas outlined in this Roadmap.

##### 5.1.1.2.1 Human-Machine Interaction Testbeds

Today, most of the progress on AI technologies is on simple perceptual tasks that require reasoning over a single interaction modality, e.g., images or text. However, many crucial real-world tasks require systems that can handle and reason over information available from multiple modalities and information sources. Furthermore, most successful AI techniques are data-driven; to build such AI systems we therefore need to develop and make widely available high-quality research ecosystems such as large datasets and software libraries. Much of the progress in AI subfields such as computer vision and natural language processing has been made possible due to openly available datasets (such as ImageNet, Gigaword Corpus, SQUAD, etc.) and software libraries (such as Tensorflow, PyTorch, AllenNLP, etc.). Similarly in robotics, the open-source Robot Operating System (ROS) has become an international standard with a massive international user and contributor community that includes academia, industry, and volunteers. Tackling future interaction research challenges requires creating large datasets that represent real-world problems and developing open-source software libraries to enable rapid prototyping and deployment at scale. Developing software libraries is especially important due to the sheer complexity and the engineering effort required to handle streams of multimodal data from many sensors.

The availability of rich interaction research platforms would enable new research to address significant real-world problems concerning human communication and human-machine interaction. They would enable the collection of data to cover the diversity and range of human communication. This diversity is not only concerned with conversation topics or communication modalities and channels, but with people of different ages (children vs. elders), different native languages, cognitive and physical abilities, speech impediments, background and technical skills, personal goals and priorities, and other attributes that affect their communication needs for AI systems. Including such diversity is crucial for generating AI systems that can operate robustly in the real world.

Another barrier to human-machine interaction research is the lack of data from realistic, and often sensitive interactions, such as those with vulnerable populations (children, elderly), private settings (doctor's office, therapist session), and group contexts (at work, school, on public transportation, etc.). The development of intelligent systems that can operate in such settings and with a diversity of users requires training data, yet such data currently take years to obtain, and are only available to a small group of researchers, due to the cost, lengthy process, and privacy constraints, which collectively inadvertently result in slow progress. Large datasets from such settings, obtained with proper consent and shared under appropriate privacy rules, would serve as major enablers for AI research into human-machine interaction and other areas of AI.

5.1.1.2.2 Life-Long and Life-Wide Personal Assistant Platforms
An exciting application area for AI with potential for high impact in society is life-long and life-wide personal assistants (LPAs). Today's personal assistant technologies are very limited, and having an open ecosystem for research in this area could lead to significant advances and have significant impact in society. Needless to say, ethical considerations (privacy, security, and accountability) are a primary concern in the design of these systems.

LPAs are AI systems that assist people in their daily lives at work and home, for leisure and for learning, for personal tasks and for social interactions. LPAs are life-long in that they will assist a person for long periods of time, possibly for their lifetime, remembering events and people that a person may not naturally recall well and adapting to a person's needs as their life changes incrementally (new preference, new daily habits) or drastically (new job, new interest). LPAs are life-wide, operating across all areas of their owners' lives to build and maintain context-sensitive models of those users.

LPAs will be not only be personalized, but also personal, in that all data shared with them will remain the property of the agent's owner. Current personalization methods typically offer minor language adjustments and content recommendations based on expressed and learned preferences. In the future, personal AI systems will learn how to best serve the interests of the people interacting with them. They will learn what sort of language is most motivating, which feedback improves a person's well-being, and which contexts are most engaging for the owner. LPAs will reason about social context and the social consequences of actions. They will also support human-machine teaming and automate tasks to achieve effective collaboration.

LPAs will learn and adapt over their users' lifetime. They will detect situations for opportunistic training and education based on modeling the user, detecting opportunities, and creating curriculum on the fly. This learning will support users both professionally and personally, serving as a cognitive multiplier for increased efficiency and learning, enhancing personal autonomy, assisting people in their daily lives, and empowering users to overcome mental and physical limitations.

LPA technology has the potential to be a game-changer across multiple sectors, including smart cities, psychology, social behavior, human body and brain function, education, workforce training and re-training, eldercare, and health. LPAs could oversee therapy, coaching, and motivation that supplement human care, and can continue long-term therapy in the home, both after hospitalization and for chronic conditions. LPAs could help people with cognitive tasks, including supporting long-term and short-term knowledge and memory.

An open research ecosystem for the development of LPAs would propel research as well as technological and commercial innovation in this area. This vision of LPAs requires innovations that cannot happen unless researchers have the ability to access and analyze data and to carry out experiments to test new approaches. Currently, personal assistants are developed and operated in the private sector, so the collected data is proprietary and the underlying platforms are not accessible for academic research. Furthermore, important privacy issues need to be addressed that can better take place in an academic open setting. Almost every AI research challenge discussed in this document is relevant to the development of LPAs. Their overall design is also a fundamental research challenge, since an LPA could be a single AI system with a diverse range of capabilities or a collection independently developed specialized systems with coordination mechanisms for sharing user history and personalization knowledge. Open platforms for continuous data collection and dynamic experimentation would significantly accelerate research, not only in this important application of AI but also in basic AI research.

#### 5.1.1.2.3 Robotics and Physical Simulation Testbeds

Physical robot development platforms are unaffordable and inaccessible to many, and simulations are only possible and available for some types of platforms and AI functions. In order to facilitate research on integrating intelligent capabilities into robots and developing robust real-world systems, accessible robotics testbeds must be created that allow experimentation and testing that bring together diverse AI components, including perception and fusion from multiple sensors, speech processing and natural language interfaces, reasoning and planning, manipulating objects in the world, navigation, interacting safely and naturally with people, and accomplishing a variety of tasks and goals. Open robotics testbeds will be critical enablers for making progress in combining AI capabilities and robotics in real-world problem contexts. To enable progress toward real-world robot deployment, robotics testbeds will also need to be embedded in realistic environments: instrumented homes, schools, and hospitals, advanced manufacturing facilities, public venues such as museums and city centers, and large-scale environments such as farms. These environments should be equipped with ambient sensors (high-fidelity cameras, microphones, etc.) and Internet of Things infrastructure, as well as virtual reality and augmented reality interfaces, to support diverse types of interactions with people. They should contain various robot platforms appropriate for their particular contexts (mobile manipulators, table-top robots, swarm-type robots, high-degrees-of-freedom arms, humanoid robots, field robots). The physical environments are critical for real-world robot development; however, testbed software infrastructure will need to include high-quality simulations to facilitate ramping up and scaling projects.

These robotics testbeds will require privacy-protected data storage and data processing facilities. They will also need to provide high-speed communications to allow for performing real-time experiments remotely. They would need to be supported with multiple technical staff who facilitate research and data collection and maintain the hardware and software infrastructure.

In addition to physical environments for robotics, AI research will also greatly benefit from high-quality, realistic virtual environments for training robotic systems. For many kinds of robotics problems, robot learning methods require significant training time (on the order of months or years of experiment time), which can only be generated in virtual environments. The creation of virtual training environments that provide a sufficiently detailed and realistic replica of the physical world is a major engineering undertaking and a critical enabler for physical machine intelligence.

### 5.1.2 SUSTAINED COMMUNITY-DRIVEN AI CHALLENGES

AI challenges with million-dollar rewards, such as the X Prize, the DARPA Robotics Challenge, and industry-driven competitions, attract significant attention and catalyze creative solutions. However, they lead to idiosyncratic non-generalizable approaches, rather than the kinds of profound new ideas that lead to transformative research in AI. Great excitement is also generated by AI competitions with common datasets and scoreboards, but they tend to target surgically delineated and unconnected problems. To move AI to the next step, we need to capitalize on the energies fostered by healthy competition, while promoting concerted progress in hard AI problems.

To that end, we recommend creating a set of sustained community-driven AI challenges with the following important features:

◗ Research-driven: The competitions would be created to drive AI research based on an explicitly articulated research Roadmap with milestones along the way.

◗ Community governance: They would involve organizational structures that enable AI researchers (rather than outsiders) to agree to the definition of a challenge and evolve the rules of the competition as research progresses.

◗ Grassroots participation: Challenges and competition rules would be designed by competition participants collaboratively.

◗ Regular updates: Competition rules would be updated based on research progress in ways that direct the community toward new advances.

- Decomposable: Ambitious challenges could be broken down into more tractable problems over a sensible timeline to formulate a feasible long-term research arch.

- Resource creation: Participants should be supported and incentivized to create shared resources, useful tools, and infrastructure that supports community participation in the competitions.

- Embedded metrics: The community should develop the necessary metrics to measure research progress based on the expertise gained by analyzing competition entries.

- Compete-collaborate cycle: Competing teams should gather after a competition to share their approaches and open-source software, to collaborate in updating the competition rules, and to regroup to undertake the next competition.

- AI ethics: These important considerations should be incorporated into the competitions at all levels.

A good model for sustained community-driven AI challenges is the RoboCup Federation. Started in 1993 to use the game of soccer to drive AI and robotics research, its ultimate goal is: "By the middle of the 21st century, a team of fully autonomous humanoid robot soccer players shall win a soccer game, complying with the official rules of FIFA, against the winner of the most recent World Cup."[29] However, it is acknowledged by the Federation that this is a challenging long-term goal, and that other research and discovery will occur along the way that will have significant societal impact. At the moment, in addition to RoboCupSoccer the federation runs other leagues under the rubrics of RoboCupRescue, RoboCup@Home, and RoboCup Industrial. The RoboCup community includes more than 1,000 people. The competition rules for each league are defined by a committee elected by the participants in that league. The rules for the competitions are designed to push the research along dimensions proposed by researchers participating in the league based on their best intuition about promising directions and anticipated topics of their funded research projects. The rules are updated each year as needed to push the research. Annual research symposia enable competition participants to share the major ideas and advances that enabled them to showcase new capabilities at the competitions. Many teams collaborate to develop research infrastructure, such as simulators or program interfaces. Companies have developed affordable platforms to support the competitions, and continue to be involved in proposing new RoboCup leagues and research challenges of interest.

The recent growth of AI for social good presents an opportunity for pursuing a sizeable variety of AI problems with the unifying theme of contributing to solving societal challenges, including human health, safety, privacy, wellness, sustainability, energy use, and many others. Applying the RoboCup model to the social good context would create a motivating opportunity for numerous AI researchers as well as students to contribute to meaningful problems.

### 5.1.3 NATIONAL AI RESEARCH CENTERS

The National AI Research Centers are intended to create unique and stable environments for large multidisciplinary and multi-university teams devoted to long-term AI research and its integration with education. While the majority of current academic research tends to be done with precisely focused, short-term, and single investigator projects that support two to three faculty, these centers will provide the expertise and critical mass needed to make significant advances in otherwise unattainable AI research goals.

Each National AI Research Center would be a catalyzer for a broad and substantially challenging AI theme that would serve as a pivot for the center's research. This research Roadmap suggests many priority areas that could serve as pivots, such as open knowledge networks, cognitive architectures, fair and trustworthy AI, representation learning, assistive robotics, LPAs, and many more.

---

[29] RoboCup. Objectives. https://www.robocup.org/objective

National AI Research Centers are envisioned to be:
- Funded through decade-long commitments, to provide stability and continuity of research

- Multi-university centers with a smaller core set of partners and a large network of affiliated educational institutions, including research universities, liberal arts colleges, community colleges, and K-12 schools.

- Multidisciplinary in the expertise involved in the research areas

- Multi-faceted in their research goals

- Inclusive hosts for visitors bringing diverse views and new ideas

- Effective dissemination vehicles for significant results

These Centers would provide:
- Practical environments for the development of open resources

- Governance and oversight for relevant Community-Driven AI Challenges

- A fertile ground for involving other in other areas of computer science and other disciplines in AI research

- Unique educational opportunities for training the next generation of AI researchers, engineers, and practitioners

Each National AI Research Center would be funded in the range of $100M/year for at least 10 years. With this level of funding, such a center would be able to support an ecosystem of roughly 100 full-time faculty (in AI and other relevant disciplines from different universities), 50 visiting fellows (faculty and industry), 200 AI engineers, and 500 students (graduate and undergraduate), and sufficient computing and infrastructure support.

There are a few examples of AI research centers that have long-term funding. The University of Maryland's Center for the Study of Language (CASL), founded in 2003 as a DoD-sponsored University Affiliated Research Center (UARC) funded by the National Security Agency, includes about 60 researchers and 70 visitors from academia and industry focused on natural language research with a defense focus. The Institute for Creative Technologies at the University of Southern California, established in 1999 as a DoD-sponsored UARC working in collaboration with the US Army Research Laboratory, has about 20 senior researchers, 30 students, and 50 support staff working on AI system and virtual environments for simulation and training. The Allen Institute for AI, which was established in 2014 and is privately funded, has more than 40 senior researchers and 60 support staff focused on machine reading and commonsense reasoning. *Although these centers have long-term funding and support staff, all are at significantly smaller scale than the ones envisioned here, and none are multi-institutional centers.*

### 5.1.4 MISSION-DRIVEN AI LABORATORIES

The Mission-Driven AI Laboratories (MAILs) are intended to be living laboratories for AI development targeting areas of great potential for societal impact with high payoff, such as healthcare, education, and science. While many aspects of academic research are driven by applications of societal interest, current work tends to be piecemeal and does not always target deployment.

MAILs are envisioned as living laboratories that would serve as real-world settings for:
- Unique collaborations between AI researchers, domain experts, and practitioners to formulate challenges and requirements for AI technologies

- Immersive environments for multidisciplinary teams to study challenging practical problems

- Exposing AI researchers to hands-on experience, allowing them to attain deep knowledge about challenging societal problems

◗ Collection of unique research-grade comprehensive data pertaining to challenging real-world problems

◗ Testing of AI prototypes for formative evaluation

◗ Experimentation with novel processes and approaches for the deployment of new AI capabilities

◗ Enabling industry and academic researchers to work in open environments on problems of mutual interest

◗ Testing the safety, security, and reliability of AI systems before broad adoption

◗ Training the next generation of AI researchers, engineers, and practitioners

◗ Exposing students and researchers from other fields to advanced AI technologies relevant to their domain and in a practical setting

MAILs would:

◗ Work closely with the National AI Research Centers to integrate and assess their most recent advances

◗ Provide sustained infrastructure, including AI engineers and facilities with associated personnel to support AI research

◗ Stress test the open AI platforms and resources

◗ Provide environments and data to formulate community-driven AI challenges

◗ Demonstrate the operational value and ethical integrity of AI technologies and advances

◗ Work closely with the National AI Research Centers to integrate and assess their most recent advances

Each MAIL would fund AI research in the range of $100M/year for several decades, with a separate budget for operations and support of the center. This would be a reasonable period of time to see significant returns on investment and transformative research in the target areas. With this level of funding, a MAIL could support an ecosystem of roughly 50 permanent AI researchers, 50 visitors from the National AI Research Centers at any given time, 100-200 AI engineers, and 100 domain experts and staff focused on supporting AI research. MAILs should be run by people with substantial AI experience and credentials in order to bring together the research community and ensure that research quality remains a priority.

*There are no existing examples of mission-centered AI laboratories.* Outside of AI, there are many such experimental laboratories in other sciences that could serve as models. The SLAC National Accelerator Laboratory, founded in 1962 as the Stanford Linear Accelerator Center, has led to four Nobel-winning results in particle physics to date. Though initially staffed by experimental physicists, it eventually became an academic department at Stanford and a national Department of Energy laboratory. SLAC is home to more than 1,500 researchers, support staff, and visitors from several hundred universities in 55 countries from both academia and industry. Its annual budget is over $430M. Other DoE laboratories are also associated with universities, including the Oak Ridge Laboratory offering an interesting model with the Oak Ridge Associated Universities (ORAU) with more than 100 affiliated universities across the US. A different model is the National Center for Atmospheric Research (NCAR) is a Federally Funded Research and Development Center (FFRDC) through National Science Foundation grants focused on atmospheric and space sciences and their impact on the environment and society. NCAR is run by the University Corporation for Atmospheric Research (UCAR), a nonprofit consortium formed by more than 100 universities. NCAR is home to more than 700 researchers and support personnel, and its annual funding surpasses $165M. Another model could be the Jet Propulsion Laboratory (JPL), a FFRDC funded through NASA awards that is administered by the California Institute of Technology with a mission focus on space and Earth sciences. Its annual budget is near $1.5B, supporting more than 6,000 employees. Interestingly, JPL has traditionally had a substantial number first-rate AI researchers developing and deploying award-winning AI technologies in space missions, with a concentration on robotics, autonomous planning and scheduling, and machine learning.

The following sections illustrate some potential target mission areas for MAILs, including AI-ready hospitals, AI-ready homes, AI-ready schools, and experimental facilities for preparation and response for natural disasters.

**AI-Ready Hospitals**

AI systems have the potential for tremendous impact in improving care in hospitals, including administrative support (e.g., bed management and staffing), care delivery (e.g., monitoring for treatment administration and early warnings of potential deterioration, assisting with integrated medical information) and decision making. Today, researchers face significant challenges in investigating effective use of AI methods in these arenas, because doing so requires access to current patient populations, not simply working with retrospective datasets. It requires significant time and effort to develop the collaborative research efforts with clinicians and build the cross-disciplinary understandings needed to make progress in these areas. The constraints on access to health data and on testing, as well as the taking of clinician practice time, make exploratory collaborations in a real setting difficult. Although this is an area of high potential impact and great interest to AI researchers, to date it has been difficult to move beyond retrospective datasets that lead to pilot studies, with full trials rare and typically only after many years.

AI-ready hospitals offer one possibility for making such research more effective. These hospitals would be designed to enable AI researchers to become part of a hospital ecosystem and easily collaborate with clinicians and hospital managers to investigate the development of novel uses of AI in healthcare settings. The IT systems for these hospitals would be designed in collaboration with AI researchers to enable the collection of rich data and easy prototyping They would collect appropriate information and support AI systems' use of that information (both adding to records and obtaining information from them). Hospital personnel, including IT engineers and healthcare professionals, would be available to collaborate with AI researchers to take care-flow constraints into account, to understand data collection and data quality, to ensure proper access of data to respect privacy, and to design appropriate user interfaces.

We recognize that the development of such hospitals raise a wide range of ethical and societal challenges related to patient participation (could a patient opt out?), location choice (could such a hospital be the only hospital in some urban neighborhood or rural area?) and clinician participation, as well as the full spectrum of privacy issues. AI-ready hospitals or healthcare systems that enable (relatively) rapid development and testing of AI approaches could be created in diverse locations, to help ensure robustness across patient populations and environments.. These could be set up as learning hospitals, where medical and nursing students would also acquire skills in advanced technologies for healthcare. In such a context, there would be a staff of clinicians, clinical trial specialists, and IT engineers whose mission is to team with AI researchers for problem selection, experimental design, data pulls, and selective access only to patients who agree to be part of a given study. This effort would be supported by HIPAA-compliant computational resources.

**AI-Ready Homes**

AI-ready homes could provide a testbed for research on AI systems that support people aging in place, improving the safety and security of homes, and making life more convenient. These heavily instrumented homes would continually provide data to researchers in agreed-upon spaces and times in and around the house. Privacy protocols would be in place governing any researcher-access to the data, including the use of identity scrubbing (e.g., automatic replacement of faces with abstract depictions, text transcripts of speech instead of audio) where feasible. Security measures would also be designed into these systems to assure their occupants appropriate control over their operations.

**AI-Ready Schools**

Much of the best work in AI for education has been done in close collaboration with in-service teachers, but establishing such collaborations is difficult, given the other demands on teachers' time and the organizational constraints of schools. AI-ready schools would encourage participation in such collaborations as one component of a teacher's job, supported and rewarded by the schools. AI-ready schools would have agreements in place for their students to participate in experiments with AI software.

Such schools would be provisioned with IT infrastructure to support a range of experimental AI systems, including machines that could be loaned to students for homework and servers for local data collection. Data gathering and processing would comply with the Family Educational Rights and Privacy Act (FERPA) using, for example, on-site anonymization, and would support educational data mining and experiment analysis without compromising student or teacher privacy. These schools would be distributed in regions that serve diverse populations, in order to ensure that the AI systems developed are robust across the full spectrum of students who need them.

**AI-Ready Science Laboratories**

Already many laboratories use software to control experimental apparatus and store data in digital form. AI-ready laboratories would go beyond this by including infrastructure for interfacing instrumentation in that field with AI systems, thereby producing data with built-in semantic description. This would enable automated analysis and experimentation and facilitate sharing and synthesis of data across experiments and laboratories. Capturing thinking and decision making processes with multimodal sensors, and enabling AI systems to participate in these conversations, would also be supported. Protocols governing data sharing that respect the privacy and priority of the domain scientists, as well as the AI scientists, would be in place.

**AI for Natural Disasters and Extreme Events**

AI can be used in many aspects of preparation for and response to natural disasters and extreme events, such as hurricane winds and storm-related flooding, which is predicted to affect 200 million Americans across 25 states, causing as much as $54B in annual economic losses.[30] Before such an event occurs, AI could be used for improving predictions, integrating geosciences models, analyzing population and asset risk, and designing robust plans in preparation for extreme events. After the onset of such an event, AI can be used in a variety of ways, from coordination and prioritization of tasks for first responders, dissemination of information and control of misinformation among the affected population, and search and rescue (e.g., via rescue robotics).

## 5.2 Re-Conceptualize and Train an All-Encompassing Workforce

Comprehensive changes need to be undertaken in order to restructure and train a diverse AI workforce to prepare highly skilled researchers and innovators. Although the demand for AI expertise is very high, the particular expertise required for any job will vary widely in terms of topics, technical skill, and experience. For example, in some cases advanced knowledge of AI algorithms will be required; in others, the demand will be for practical data analysis and user experience design skills, in still others, software engineering will be of most value. Because AI systems and supporting technologies will evolve very rapidly, training AI engineers to continuously adapt to new techniques and tools will be essential. Equally essential will be a strong focus on ethical issues around AI, including safety, privacy, control, reliability, and trust, throughout every level of this effort, in order to instill ethical thinking in all AI professionals.

This section discusses eight major recommendations: 1) Developing AI curricula at all levels, 2) Recruitment and retention programs for advanced AI degrees, 3) Engaging underrepresented and underprivileged groups, 4) Incentivizing emerging interdisciplinary AI areas, 5) AI ethics and policy 6) AI and the future of work, 7) Training highly skilled AI engineers and technicians 8) Workforce retraining. Success at all of these levels will be key for providing the general public and decision makers with general knowledge about AI that will be fundamental for everyone, both as intentional consumers or unintentional recipients of AI technologies. Professionals, as well as the public, should be informed about the high stakes of the development of AI and educated about appropriate expectations and implications.

---

[30] 30 Congressional Budget Office, Expected Costs of Damage from Hurricane Winds and Storm-Related Flooding. April 10, 2019. https://www.cbo.gov/system/files/2019-04/55019-ExpectedCostsFromWindStorm.pdf

**5.2.1 DEVELOPMENT OF AI CURRICULA AT ALL LEVELS**

We recommend comprehensive curricula for K-12, undergraduate, graduate, and postgraduate AI education. AI curricula guidelines should be developed to start at an early age to encourage interest in AI, understanding of the associated issues and implications, and curiosity to pursue careers in the field. AI curricula materials should be made freely available and teacher-training opportunities should be supported and promoted. Curricula at all levels should contain a significant part dedicated to ethical issues around AI, including safety, privacy, control, reliability, trust, etc. These issues should be integrated into the computer science curricula as a whole, rather than in separate standalone courses. To facilitate operationalization, opportunities, support, and incentives should be provided for current K-12 teachers to be trained in the use of AI curricula and course materials.

We recommend guidelines for undergraduate, graduate, and professional courses and degrees and certificates in AI, which educational institutions at all levels can employ in order to set up programs that prepare students for careers in the field. This should go well beyond computer science and engineering majors, so that programs are available to all undergraduate students to incorporate AI minors or AI concentrations as complements to their fields of study. AI undergraduate courses should not all be highly technical, so that offerings and learning opportunities are made available to and accessible for all students in all majors across a campus. Here, too, coverage of ethical issues will need to play an important role.

We recommend graduate courses and degrees that provide students with opportunities to learn about all relevant aspects of AI, particularly in practical settings, in order to preserve US leadership in AI. Courses developed jointly by faculty in different areas often lead to creative research outcomes and in many cases change the interests and career direction of students. Creating a heterogeneous ecosystem of AI and AI-related graduate studies will effectively promote much-needed spread of ideas across fields that will greatly benefit AI research and innovation. The associated course materials should be openly shared to complement curriculum guidelines, allowing every institution to offer a wide variety of AI courses. More recommendations regarding interdisciplinary studies are included below.

We also recommend developing guidelines for professional programs and certifications that will appropriately qualify individual expertise and experience in all aspects of AI, including ethics. This will enable a wide range of organizations to provide the AI education opportunities that will be needed to meet the demand for professional careers and retraining in AI. These certifications will be particularly important for jobs that require high-stakes decision making and for organizations where AI expertise is not already available.

**5.2.2 RECRUITMENT AND RETENTION PROGRAMS FOR ADVANCED AI DEGREES**

We recommend significant increases in the production of graduates with advanced degrees in AI. First, the AI education and training programs proposed in this report cannot be staffed unless significant numbers of faculty, teachers, and instructors are available with advanced AI degrees. Second, significant AI breakthroughs and innovations such as those outlined in the research plan of this report will be accomplished at speeds and quantities that are proportional to the availability of AI researchers with advanced degrees. Third, the more advanced AI expertise that is available to permeate different disciplines and areas of society, the more transformative AI technologies will be—across the board.

The creation of the AI centers described in the previous section of this report will create unique resources and stable career opportunities that will make academic careers significantly more attractive. There is such high demand for advanced AI expertise that when universities have AI job openings they find themselves competing with a vigorous marketplace. This concerns not only the hiring of junior faculty, but the retention of senior tenured faculty. It is imperative to set up competitive academic career programs for doctorate-level researchers and early-career faculty, including multi-year fellowships and career initiation grants, in order to maintain the educational programs to train the next generation. In addition, strong programs that attract students to AI from early stages (high school and undergraduate) and promote AI careers should be a key ingredient for growing a workforce with advanced AI expertise.

### 5.2.3 ENGAGING UNDERREPRESENTED AND UNDERPRIVILEGED GROUPS

We recommend significant growth of participation by diverse, underrepresented and underprivileged groups, both to ensure access to the career opportunities afforded by this growing area and to significantly broaden the talent pool. This will require incorporation of best practices into funding processes and programs, across the board, in a comprehensive approach to diversity in AI that is consistently applied across educational and professional institutions and has extensive follow-up and effective redirection when needed:

◗ Incentivizing hiring and promotion processes to increase diversity, so that all well-qualified candidates have appropriate consideration and opportunities.

◗ Ensuring the diverse role models are present at all levels of AI instruction, from K-12 to undergraduate, graduate, and professional certificate training. The absence of women and members of other underrepresented groups in the education and training process results in perpetuating under-representation.

◗ Countering undesirable effects of common cultural practices, so that any disadvantages can be addressed up front. For example, women are more likely to take time off to care for children, so childcare programs and flexible job requirements should be commonplace.

◗ Providing access to college and career opportunities, including through bridge programs, so that any individual has the awareness, the knowledge, and the means required to pursue careers in AI. For example, a disproportionate number of minority students attend high schools that do not offer computer science courses, have limited access to computers at home, and lack mentors for careers in computer science in general, let alone in AI.

◗ Broadening the sources of student application pools, especially at the graduate level, so that students originate from a variety of backgrounds and institutions beyond traditional well-connected sources.

◗ Providing free AI coding training and opportunities to contribute to open-source AI software repositories. ROS is an excellent example of such a model and a major force multiplier in robotics: It is an international open-source community where volunteers can receive training and contribute code/software, which also serves as experience on resumes. GitHub code commits have similar value and are becoming the norm in software engineering training and hiring today.

◗ Developing and deploying targeted mentoring programs, cohort programs, and special interest groups that provide exposure to the AI field, individualized advice and coaching, and peer support.

◗ Explicitly codifying best practices in ethical behavior, conduct, and inclusiveness in academic, industry, and government organizations, so that inappropriate interactions, isolation, and implicit biases are eliminated from the school and the workplace.

◗ Confronting attrition on a continuous basis, so that the core underlying reasons for underrepresented groups leaving the AI field can be addressed quickly, systematically, and thoroughly.

A number of nonprofit organizations and grassroots efforts are successfully attracting K-12 students from underrepresented groups and underprivileged areas to STEM and computer science—particularly robotics—in large numbers. Also, a number of nonprofit organizations focus on mentoring college-age students and early-career professionals belonging to underrepresented groups to retain them in the computer science field. These could provide useful models for our recommendations regarding attracting a broader population of students to pursue long-term careers in AI. Using application domains such as AI for social good will serve as compelling attractors for women and other currently underrepresented communities of AI developers.

## 5.2.4 INCENTIVIZING EMERGING INTERDISCIPLINARY AI AREAS

We recommend the creation of courses and curricula to train students to pursue careers at the intersection of AI with different disciplines. This would:

- Train a larger and more diverse workforce with AI skills

- Enhance opportunities for AI for social good

- Serve AI needs in different sectors and application areas

- Accelerate the research priorities in this Roadmap

- Stimulate the generation of transformative ideas and approaches

- Promote research and applications of AI in all areas

- Enable multidisciplinary work that would advance our understanding of hard societal challenges for AI design and deployment involving AI ethics and safety

### 5.2.4.1 Interdisciplinary AI Studies

The kinds of AI innovations and challenges envisioned in this report will require a skilled workforce with a diversity of interdisciplinary backgrounds that only exists in very limited forms today.

True interdisciplinary work requires a deep understanding of other disciplines. Learning is more efficient and effective when new concepts in a related discipline are introduced in the context of what a student already knows. Interdisciplinary courses and curricula are hard to design, and best practices should be developed to guide their creation and implementation. It will also be important to provide incentives and articulate career paths to attract students to these programs.

We recommend promoting interdisciplinary AI studies, with opportunities to combine AI research with education. There are several aspects to this interdisciplinarity that we discuss in turn: the relevance of many sciences to the study of intelligence, the close ties of AI with other areas of computer science, and the need for discipline-specific and application-specific studies of AI.

Many AI researchers have traditionally worked at the intersection of AI and other fields of science. Understanding intelligence involves many other scientific disciplines, such as psychology, cognitive science, neuroscience, philosophy, evolutionary biology, sociology, anthropology, economics, decision science, linguistics, control theory, engineering, and mathematics among others. Significantly more interdisciplinary work and training programs are needed.

AI advances will come hand in hand with advances in computer science, and educational courses and programs can not only stimulate research but also create a workforce with diverse AI-relevant skills. AI advances will drive innovations in algorithms, hardware architectures, software engineering, distributed computing, data science, sensors and computing devices, user interfaces, privacy and security, databases, computer systems and networks, and theory, among others. Interdisciplinary programs and training opportunities will be important for the pursuit of the research agenda but also to support the comprehensive deployment across sectors. Furthermore, the technology will continuously change over time, and training will need to be designed so the workforce can easily adapt to such changes.

Important application areas of AI have been pursued in the past and have vibrant communities of both research and practice. These include AI and medicine, AI and education, AI and operations research, AI and manufacturing, AI and transportation, and AI and information science to name a few. Other areas are nascent and growing, such as AI and environmental sustainability, AI and law, AI for social sciences and humanities, AI for engineering, and AI for social good, among others. Academic programs that promote the understanding and application of AI in different sectors of importance to society need to be made more ubiquitous to meet increased workforce demand.

#### 5.2.4.1.1 AI Ethics and Policy

The research priorities in this Roadmap highlight the importance of the area of AI ethics and policy, and the imperative of incorporating ethics and related responsibility principles as central elements in the design and operation of AI systems. Many of the challenges to be addressed are highly interdisciplinary, such as:

◗ **AI ethics –** concerned with responsible uses of AI, incorporating human values, respecting privacy, universal access to AI technologies, addressing AI systems bias particularly for marginalized groups, the role of AI in profiling and behavior prediction, as well as algorithmic fairness, accountability, and transparency. This research involves other disciplines such as philosophy, sociology, economics, science and technology studies, anthropology, and law.

◗ **AI privacy –** concerned with the collection and management of sensitive data about individuals and organizations, the use and protection of sensitive data in AI applications, and the use of AI for identification and tracking of people. This research involves other disciplines such as psychology and human factors, systems engineering, public policy, and law.

◗ **AI safety –** concerned with systems engineering to ensure reliable and safe operations, prevention of damage, resilience to failure, robustness, predictability, monitoring, and control. This research involves other disciplines such as industrial engineering and systems engineering.

◗ **AI policy –** concerned with legal aspects, regulation, and policy for AI deployments in different sectors and contexts. This research involves other disciplines such as public policy, government, and law.

◗ **AI trust –** concerned with high-stakes decision making, responsibility and accountability, individual's rights, and public perception and misconception of AI systems. This research involves other disciplines such as psychology, neuroscience, linguistics, and philosophy.

We recommend incentivizing the exploration of approaches that combine technology solutions, legal solutions, and policy solutions.

◗ Instrumentation of AI systems: Articulating the probes (how to instrument AI systems), indicators (what to observe), and metrics (how to measure) would enable a better quantitative characterization of key aspects of AI systems' behavior and level of autonomy to make decisions.

◗ Metrics are challenging to develop and with their effect, hard to anticipate. The domains in which AI systems are deployed are highly rich and dynamic, and a single metric can easily miss concerns that arise through complex effects.

◗ Carefully designed metrics can have a role, however, in informing a broader framework that should include legal, policy, and enforcement dimensions:
  • Legal aspects: The design of new laws that speak to concerns about agency, responsibility, and accountability, and the consequences of their violations.

  • Policy aspects: The design of new regulation and policy, as well as the characterization of desirable and permissible uses of AI in different sectors and application domains.

  • Enforcement aspects: Auditing and tracking compliance with existing policies and laws, as well as the detection of violations and undesirable outcomes.

We view that goal-driven instrumentation of AI systems, keeping in mind the larger legal, policy and enforcement goals, can contribute to the larger discussions around ethics and policy.

The mission-driven AI laboratories will provide a fertile ground for interdisciplinary research, for designing approaches that explore the complex space across these interrelated areas, and for investigating domain-specific requirements and solutions. More importantly, they will provide training and education opportunities at all levels for students and researchers in AI, public policy, legal studies, philosophy, and other areas, as well as practitioners to acquire the comprehensive background and understanding needed to tackle these challenging but crucial areas.

### 5.2.4.1.2 AI and the Future of Work

This is a key priority area of interdisciplinary research, concerned with disruption of the workforce, automation of jobs, emergence of new jobs, retraining, and other economic, social, and ethical impacts of the deployment of AI systems in the workplace. These challenges are at the intersection of AI with other disciplines such as economics, public policy, and education. Novel computing technologies often improve our lives, but they can also affect them in ways that are harmful or unjust. It is important to teach students how to think through the ethical and social implications of their work.

Great innovation and growth will result from making it easier for organizations to swiftly transform their existing processes into new ones that incorporate new technologies. AI is transforming existing jobs and creating new types of jobs, many of which will involve different forms of human-machine collaboration or assistance. A key research challenge is the characterization and management of organizational processes that combine AI systems and humans. These processes should articulate the different jobs in an organization, as well as the roles and responsibilities of people and AI systems in those jobs and processes. The characterization of organizational processes should support the identification of employee skill requirements, job reassignment, and retraining. Interdisciplinary research is needed to create proper characterizations of AI systems and human-machine systems within those processes. This will provide a useful framework to articulate the value of human roles in organizations and the limitations of automation on a task-by-task basis. It will also facilitate innovation in organizational processes and in the AI systems within them.

### 5.2.5 TRAINING HIGHLY SKILLED AI ENGINEERS AND TECHNICIANS

We recommend the training of significant numbers of AI engineers, as the demand is already high and will significantly increase over time. A significant portion of the AI workforce needed is for *AI engineers* who: 1) understand AI algorithms, data, and platforms; 2) set up and maintain AI infrastructure at scale; and 3) are able to continuously learn and adopt new technologies on the job. AI engineers combine basic expertise in AI with practical skills in software engineering, data systems, and high-end computing. These skills enable them to support the deployment of AI systems and the prototyping of AI innovations.

We also recommend training significant numbers of *AI technicians*. Companies already hire for such positions in large numbers, as these are people who can do data collection, data annotation, data cleaning, and other forms of data acquisition processes that are key for the development and improvement of AI systems. For example, AI technicians annotate failed interactions with conversational assistants, which generates data to train these AI systems to do better in the future. A workforce capability to supply skilled AI technicians could enable significant innovation across all sectors of the economy. For example, technicians could be trained to annotate sensor circuits, which could be used to train AI systems to support engineering designers. Workforce training programs and career opportunities should be created, particularly to handle complex data in diverse application areas and domains of interest.

We recommend that extensive adult education programs, distance education programs, and online courses be developed to fit personal circumstances and schedules of those interested in pursuing AI careers.

**5.2.6 WORKFORCE RETRAINING**

Workforce retraining programs to convert skilled technical workers in areas of dwindling demand into AI engineers and AI technicians will be greatly beneficial. This is particularly important for large portions of the workforce who currently have technical jobs and who will be able to repurpose their skills into AI. Other significant portions of the workforce that lack technical training will need special programs to enable them to access new jobs and opportunities in this emerging sector of the workforce.

A key contribution that AI can make in this arena is in the area of education and training. The research priorities in this Roadmap include significant aspects of AI for personalized education that broadens peoples' expertise in an area of work, retraining experts so they can repurpose their existing skills into new jobs, learning on the job to operate machines or to fit new organizational processes, and on-the-fly acquisition of skills. The fast pace of advances in AI will demand customizable, personalized, lifelong educational tools that can adapt to individual needs and incentives.

## 5.3 Core Programs for AI Research

The research priorities outlined in this Roadmap are driven by important societal drivers and require a broad range of programs. A wide variety of existing programs across agencies support AI research. However, the rate of progress in those research priorities will be slow if they continue at current funding levels and time horizons. In addition, many competitive research proposals on important areas of AI are not selected due to limited available funding. We recommend that funding for existing AI programs be increased, with particular emphasis for programs that support the research priorities in this Roadmap such as:

- Programs that support basic AI research: These programs support a single investigator (or two to three investigator collaborations) that would include students. Of particular importance are programs to support early career development and postdoctoral training.

- Programs that support application-driven AI research: These are programs that motivate AI research through the lens of practical problems.

- Interdisciplinary AI programs: These programs encourage researchers from different disciplines to pursue new approaches that combine AI, computing, engineering, social sciences, arts, and humanities. These programs should emphasize the diversity of ideas and participation, and should be for at least five years in order to reach their full potential.

- Programs representing private-public partnerships for AI: These programs would pursue unique opportunities facilitated by industry resources that can boost research in key areas of AI, such as computing and cloud resources, data resources, and other infrastructure.

- Integration of AI research and education: These programs would support the involvement of undergraduates and high school students in research projects. Programs should also be in place to incentivize their pursuit of advanced degrees.

- Outreach, diversity, and inclusion programs for AI: These programs would promote outreach and inclusion for underrepresented groups, and for institutions with limited capability in AI research and education. This would involve creating cohort groups with strong mentoring and longitudinal follow-up and building on ongoing work by nonprofit organizations and other grassroots efforts to engage broader segments of the population in AI topics.

- Education and curriculum development programs for AI: These programs would support the development of curriculum guidelines and educational materials, training opportunities for teachers and educators, online education programs for professionals, and incentives for careers in AI education.

# 6. Conclusions

This Roadmap is the result of a community activity to articulate AI research priorities for the next 20 years in a wide range of areas in AI and related disciplines. These research priorities are motivated by a detailed analysis of the potential benefits of AI for society in the domains of health, education, science, innovation, justice, and security. This Roadmap document is organized around three major priority areas: integrated intelligent systems, supporting meaningful interactions, and advancing self-aware learning. This proposed plan will bring the field to a new era of audacious AI research to tackle long-standing and multidisciplinary problems.

Along with the research priorities, a set of findings from community discussions motivate the resulting recommendations. These findings highlight significant limiting factors in infrastructure, education, and workforce capability. The document includes recommendations to overcome these challenges, with a comprehensive proposal for a new national AI infrastructure and workforce training programs that will have a profound and transformative effect on the AI R&D landscape. These investments will significantly accelerate the development and deployment of AI technologies with a profound impact across all sectors of society.

# 7. Appendices

## 7.1 Workshop Participants

**INTEGRATED INTELLIGENCE**

David Aha, US Naval Research Laboratory
Liz Bradley, U Colorado Boulder
Joyce Chai, Michigan State
Alexandra Coman, Capitol One
Vincent Conitzer, Duke University
Adnan Darwiche, UCLA
Johan de Kleer, PARC
Marie desJardins, Simmons University
Khari Douglas, CCC
Ann Schwartz Drobnis, CCC
Doug Fisher, Vanderbilt University
Ken Forbus, Northwestern University
Yolanda Gil, University of Southern California
Peter Harsha, CRA
Jim Hendler, Rensselaer Polytechnic Institute
Larry Hunter, U Colorado Denver
John Laird, U Michigan
Rick Lathrop, UC Irvine
David Leake, U Indiana
Cynthia Matuszek, U Maryland, Baltimore County
Stephanie Milani, U Maryland, Baltimore County
David McAllester, Toyota Technological Institute at Chicago
Stephanie Milani, Carnegie Mellon University
Erik Mueller, Capital One
Hector Munoz-Avila, Lehigh University
Jean Oh, Carnegie Mellon University
Paul Rosenbloom, University of Southern California
Bart Selman, Cornell University
Colleen Seifert, U Michigan
Jason Wilson, Northwestern University
Beverly Woolf, U Massachusetts Amherst

**MEANINGFUL INTERACTIONS**

Eytan Adar, U Michigan
Henny Admoni, Carnegie Mellon University
Saleema Amershi, Microsoft
Michael Bernstein, Stanford University
Jeff Bigham, Carnegie Mellon University
Gagan Bansal, U Washington
Liz Bradley, Colorado Boulder
Cynthia Breazeal, MIT
Ceren Budak, U Michigan
Carlos Busso, UT Dallas
Thomas Diettrich, Oregon State University
Ann Schwartz Drobnis, CCC
Jacob Eisenstein, Georgia Tech
Yolanda Gil, University of Southern California
Dilek Hakkani Tur, Amazon
Alon Halevy, U Washington / Megagon
Peter Harsha, CRA
Julia Hirschberg, Columbia University
Ece Kamar, Microsoft
Rao Kambhampati, Arizona State University
Benjamin Lee, U Washington
Fei-Tzin Lee, Columbia University
Blair MacIntyre, Georgia Tech
Aqueasha Martin-Hammond, Indiana University
Cynthia Matuszek, U Maryland, Baltimore County
Kathy McKeown, Columbia University
Rada Mihalcea, U Michigan
David Parkes, Harvard University
Cynthia Rudin, Duke University
Brian Scassellati, Yale University
Bart Selman, Cornell University
Milind Tambe, University of Southern California
Elsbeth Turcan, Columbia University
Dan Weld, U Washington
Helen Wright, CCC

**SELF-AWARE LEARNING**

Elias Barenboim, Purdue University
Tamara Berg, U North Carolina at Chapel Hill
Debadeepta Dey, Microsoft Research
Tom Dietterich, Oregon State University
Finale Doshi-Velez, Harvard University
Khari Douglas, CCC
Doug Downey, Northwestern University
Dieter Fox, U Washington/NVIDIA
Emily Fox, U Washington/Apple
Dileep George, Vicarious Systems
Yolanda Gil, University of Southern California
Noah Goodman, Stanford University
Kristen Grauman, UT Austin
Peter Harsha, CRA
Chad Jenkins, Michigan University
Michael Jordan, UC Berkeley
Been Kim, Google
Vipin Kumar, U Minnesota
Fei-Fei Li, Stanford University
Yan Liu, University of Southern California
Dan Lopresti, Lehigh University
Jitendra Malik, UC Berkeley
Maja Matari, University of Southern California
Cynthia Matuszek, U Maryland, Baltimore County
Ray Mooney, UT Austin
Catherine Olsson, Google
Scott Niekum, UT Austin
Devi Parikh, Georgia Tech
Brian Scasselatti, Yale University
Bart Selman, Cornell University
Reid Simmons, Carnegie Mellon University
Stefanie Tellex, Brown University
Javona White Bear, MIT Lincoln Laboratory
Helen Wright, CCC

## 7.2 Additional Contributors

Many people throughout the computing research community have contributed to the Roadmapping activity, resulting in this report. We want to especially thank workshop co-chairs Marie desJardins, Ken Forbus, Kathy McKeown, Dan Weld, Tom Dietterich, and Fei-Fei Li and all workshop participants. We acknowledge the contributions and inputs of the following people. Any opinions, findings, and conclusions or recommendations expressed in this material do not necessarily reflect the views of the individual contributors: Liz Bradley, Patti Brennan, Maya Cakmak, Tara Chklovski, Vince Conitzer, Gobi Dasu, Dieter Fox, Finale Doshi-Velez, Juliana Freire, Christina Gardner-McCune, Brian Gerkey, Carla Gomes, Robert Holte, Ayanna Howard, Charles Isbell, Chad Jenkins, Sampath Kannan, Christian Lebiere, Karen Levy, Henry Lieberman, Dan Lopresti, Travis Mandel, Maja Mataric̀, Bilge Mutlu, Hector Palacios, David Parkes, Ray Perrault, Ronitt Rubinfeld, Norman Sadeh, Aviv Segev, Andrew Selbst, Jim Spohrer, Peter Stone, Russ Tedrake, David Touretzky, Brian Van Voorst, Suresh Venkatasubramanian, Association for Computing Machinery's (ACM) US Technology Policy Committee, and the Center for Human-Compatible AI. The following individuals have provided great stewardship in advising the AI Roadmap activity: Ron Brachman, Henrik Christensen, Ed Feigenbaum, Barbara Grosz, Peter Harsha, Eric Horvitz, and Stuart Russell. Thank you to all who have helped with this activity, on behalf of the community.

Helen Wright and Khari Douglas have provided tremendous support in bringing this Roadmap to fruition and we are thankful for their contributions.

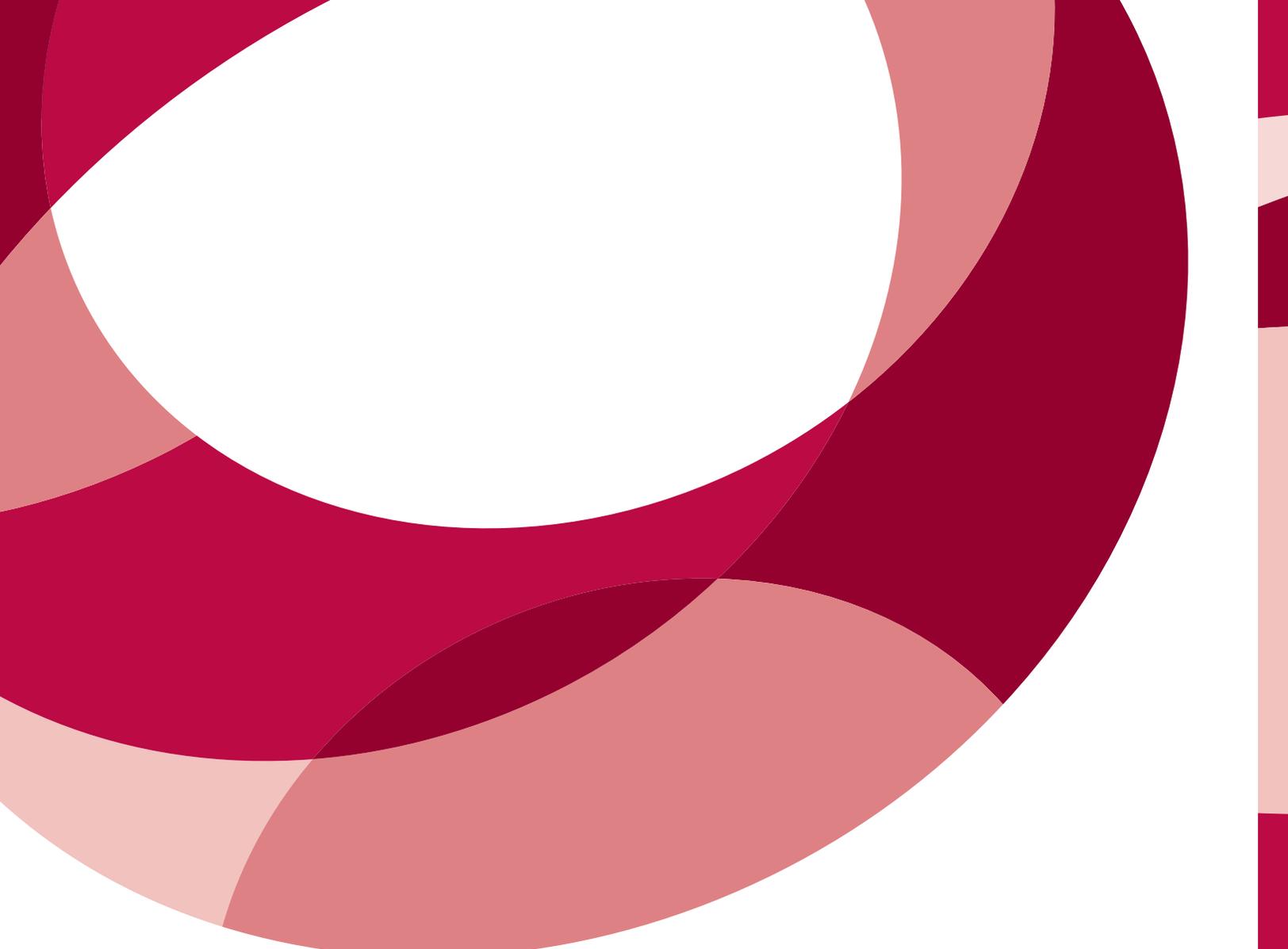
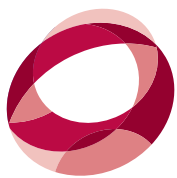

CCC
Computing Community Consortium
Catalyst